\RequirePackage{silence}
\documentclass[reprint,aps,prx,longbibliography,floatfix,superscriptaddress]{revtex4-2}
\usepackage{amsmath,amssymb,amsthm}
\usepackage{graphicx}
\graphicspath{{figures/}{./}}
\usepackage{booktabs}
\usepackage{xcolor}
\usepackage{xr}
\usepackage[hidelinks]{hyperref}
\usepackage{seqsplit}
\usepackage{tikz}
\usetikzlibrary{arrows.meta,decorations.pathreplacing}

\DeclareFontShape{OT1}{cmr}{bx}{sc}{<->ssub*cmr/bx/n}{}

\counterwithout*{equation}{section}

\newtheoremstyle{thmnoparen}{6pt}{6pt}{\itshape}{}{\bfseries}{.}{ }{\thmname{#1}\thmnumber{ #2}\thmnote{ #3}}
\theoremstyle{thmnoparen}

\newtheorem{theorem}{Theorem}

\newtheorem{proposition}{Proposition}
\newtheorem{corollary}{Corollary}

\newtheorem*{remark}{Remark}

\providecommand{\arraybackslash}{\let\\\tabularnewline}
\newcommand{\code}[1]{\texttt{\seqsplit{#1}}}

\newcommand{\Lean}{\textsc{Lean}}

\newcommand{\Ztwo}{\mathbb{Z}_2}

\newcommand{\wt}{\operatorname{wt}}

\begin{document}

\title{Exact distances and measurement allocation for logical Pauli readout}

\author{Shuoming An}
\email[Contact author: ]{anshuoming@suat-sz.edu.cn}
\affiliation{Faculty of Computility Microelectronics, Shenzhen University of Advanced Technology, Shenzhen, Guangdong 518107, China}
\affiliation{Guangdong Provincial Key Laboratory of Computility Microelectronics, Shenzhen, Guangdong 518107, China}
\affiliation{Pengxin Quantum Technology (Shenzhen) Co., Ltd., Shenzhen 518107, China}

\author{Fusheng Yang}
\email[Contact author: ]{yangfsh7@mail2.sysu.edu.cn}
\affiliation{Institute of Quantum Computing and Software, School of Computer Science and Engineering, Sun Yat-sen University, Guangzhou 510006, China}

\begin{abstract}
The minimum fault weight of a logical readout depends jointly on data-error persistence, check redundancy and time-boundary constraints.
For any binary check matrix $H$ and nonzero parity support $w$, we prove that a repeated-readout model with a zero-syndrome endpoint has exact distance $\min_{w\cdot y=1}[\operatorname{wt}(y)+q\operatorname{wt}(Hy)]$ for at least $q$ wrong readouts.
This reduces the search over fault histories to one weighted spatial minimum and gives the minimum repetition count for each adjacent-column check in square Bacon--Shor and generalized Shor readouts.
Combining those checks with redundant Gauss-check pair products yields an exact joint distance and an outcome-count optimum within the specified gauged-readout family.
At target distance $m$, one round uses $m^2$ outcomes, compared with $m(2m-1)$ for the repeated baseline in the same phenomenological model.
For a one-epoch bare-ancilla CNOT implementation, we prove a schedule-dependent distance upper bound by constructing an accepted wrong readout from one gate fault and outcome flips.
The bound identifies necessary timing conditions for retaining the target distance and gives matching upper bounds of three for the best tested schedules at both $m=3$ and $m=5$.
\end{abstract}

\maketitle

\section{Introduction}
\label{sec:intro}

Quantum error correction protects encoded information by measuring checks that detect errors without measuring the logical state~\cite{shor1995scheme,steane1996error}.
Threshold theorems establish reliable computation at sufficiently small physical error rates~\cite{aharonov2008fault}, and quantum low-density parity-check codes offer favorable rates, distances and memory overheads~\cite{panteleev2022qldpc,breuckmann2021quantum,bravyi2024highthreshold}.
Computation also needs reliable logical operations, including logical Pauli measurements.
Topological and homological codes provide the code families on which these operations were first developed~\cite{kitaev1997quantum,bombin2007homological}.
Lattice surgery implements logical measurements by merging and splitting code patches~\cite{horsman2012surgery}.
Gauging gives a systematic construction using an auxiliary graph and local checks on data and edge ancillas~\cite{williamson2026gauging}.
The product of the Gauss-check outcomes is the logical result; Fig.~\ref{fig:two-component} illustrates the construction.

A reliable readout must protect that result from both data errors and faulty measurement records.
Repeating checks and measuring redundant products can provide this protection, but their costs depend on how data faults persist and on which records are compared.
We ask how these choices set the exact minimum fault weight, and how to allocate measurements to reach a target distance.

Our main theorem reduces a multiround fault problem to one weighted spatial minimum.
For any binary check matrix $H$ and nonzero parity support $w$, the model defined in Section~\ref{sec:readout} has exact distance $\min_{w\cdot y=1}[\operatorname{wt}(y)+q\operatorname{wt}(Hy)]$ for at least $q$ wrong readouts.
Data errors persist, the checks and logical parity share a noisy record, and endpoint checks constrain the syndrome to zero.
The formula replaces a search over fault histories by a search over parity-flipping vectors $y$.
It also yields a criterion for selecting endpoint checks and the minimum number of copies of each adjacent-column check in square Bacon--Shor and generalized Shor readouts.

For a gauged readout, we combine these checks with redundant pair products of the Gauss outcomes.
The exact joint distance separates a measurement-record bottleneck from a persistent-data bottleneck.
Within the family of fixed adjacent-column checks and Gauss pair products, we minimize the outcome count over allocations and round counts.
At target distance $m$, one round uses $m^2$ outcomes, compared with $m(2m-1)$ for the repeated baseline in the same phenomenological model.
A separate circuit bound identifies low-weight accepted errors from the timing of the CNOTs and supplies a necessary condition for preserving distance $m$ with bare ancillas.

These results build on data-syndrome coding.
Ashikhmin, Lai and Brun encode syndrome bits through redundant stabilizer products and include repeated measurements as a repetition-code special case~\cite{ashikhmin2020ds}; Fujiwara also studies protection against data and syndrome errors~\cite{fujiwara2014self}.
Generalized data-syndrome codes represent accumulating data faults and noisy syndrome records across multiple rounds~\cite{kuo2025generalized}.
Here the quantity optimized is a selected logical parity under an explicit acceptance and readout predicate.
The resulting contributions are the exact threshold-dependent spatial reduction, the minimum Shor-family check allocation, and the joint outcome-count optimum for the specified gauged-readout family.
The CNOT bound then tests whether a chosen schedule can preserve that phenomenological distance.

Williamson and Yoder establish the gauging construction and its spatial, temporal and spacetime bounds under explicit construction and boundary assumptions~\cite{williamson2026gauging}.
The present readout and circuit results have direct proofs; the Supplemental Material also provides formalized algebraic results and code examples that separate graph, boundary and check-redundancy effects.

Section~\ref{sec:readout} proves the general distance formula and check allocation, Section~\ref{sec:gauged-allocation} gives the joint gauged-readout distance and measurement count, and Section~\ref{sec:circuit-bound} establishes the circuit scheduling bound.
The Supplemental Material~\cite{supplemental} gives the proofs, verification scope and reproduction commands.

\section{Exact distances and check allocation}
\label{sec:readout}

We first solve a repeated-readout problem for an arbitrary binary matrix.
The same formula then determines which checks to repeat in a Shor-family readout.

\subsection{Fault model and readout predicate}
We call the model with accumulating data errors the \emph{persistent-fault readout model}.
Rounds are indexed by $t=0,\ldots,T-1$, with $T\ge1$.
For the general model, let $H$ be a binary check matrix with $n$ columns and let $w\in\mathbb F_2^n$ be a nonzero parity support.
In each round, $d_t$ is a new data error, $\mu_t$ is a readout-record error, and $\gamma_t$ is a check-result error; each nonzero bit is one independent fault location.
The data errors persist, so $x_t=\sum_{s=0}^t d_s$.
The checks and logical parity use the same noisy record $x_t+\mu_t$, with independent check-result flips added afterward:
\begin{equation}
\sigma_t=H(x_t+\mu_t)+\gamma_t,
\qquad r_t=w\cdot(x_t+\mu_t).
\label{eq:readout-model}
\end{equation}
Here $\sigma_t$ collects the check-result error bits, and $r_t=1$ means that the round readout differs from the ideal logical value.
This shared record explains why $\mu_t$ enters both expressions.
Equation~\eqref{eq:readout-model} defines a phenomenological model; a circuit realization must specify observables and fault propagation that produce these effects.
With a boundary constraint, the detectors require $\sigma_0=0$ and $\sigma_t+\sigma_{t-1}=0$ for $t\ge1$.
Without it, they impose only the adjacent comparisons.
Let $D(f)=0$ mean that a fault $f$ fires no detector.
The distance for an undetected strict-majority error is
\begin{equation}
d_{\mathrm{maj}}(T)=\min\{\wt(f):D(f)=0,\ 2|\{t:r_t=1\}|>T\}.
\label{eq:majority-distance}
\end{equation}
Ties are outside this predicate.
To compute a decoder failure distance, one must specify the decoder's observed inputs and its decision or rejection rule, including ties.

\subsection{A spatial minimum for a multiround fault problem}

The exact distance has a closed form for any threshold $q$ of wrong readouts.
For $1\le q\le T$, let $d_q(H,w;T)$ minimize the number of fault locations over detector-free patterns with at least $q$ wrong readouts.
With the model in Eq.~\eqref{eq:readout-model} and the zero-syndrome endpoint constraint, we prove
\begin{theorem}[Exact repeated-readout distance]
\label{thm:weighted-readout}
For every binary matrix $H$, nonzero $w$, positive $T$, and $1\le q\le T$,
\begin{equation}
d_q(H,w;T)=\Gamma_q(H,w)
=\min_{w\cdot y=1}\bigl[\wt(y)+q\wt(Hy)\bigr].
\label{eq:weighted-readout-distance}
\end{equation}
In particular, strict majority uses $q=\lfloor T/2\rfloor+1$, so $d_{\mathrm{maj}}(T)=\Gamma_{\lfloor T/2\rfloor+1}(H,w)$.
\end{theorem}
To see the lower bound, choose the wrong round with the fewest transient faults among the at least $q$ wrong rounds.
If $y=x_t+\mu_t$ in that round, zero syndrome gives $Hy=\gamma_t$ and a wrong readout gives $w\cdot y=1$.
If $k_t=\wt(\mu_t)+\wt(\gamma_t)$, then $q k_t$ is no greater than the total transient-fault cost in the wrong rounds.
Also, $\wt(y)\le\wt(x_t)+\wt(\mu_t)$ and $\wt(x_t)$ is at most the total data-fault cost.
Together these inequalities give $\wt(y)+q\wt(Hy)$ no greater than the total fault weight.
For the upper bound, introduce one minimizing data error $y$ at the start of the final $q$-round block and cancel $Hy$ with check-result faults in those rounds.
The full proof is in the Supplemental Material, Supplemental Sec.~\ref*{sec:sm-readout-allocation}.

The theorem removes the time history from the minimization.
The vector $y$ specifies a parity-flipping error in one round, and $q\wt(Hy)$ is the cost of hiding its syndrome in $q$ wrong rounds.
No longer history can give a cheaper undetected pattern.
At $q=1$, the expression is the parity-restricted binary data-and-syndrome minimum: the undetected pair is $(y,Hy)$~\cite{ashikhmin2020ds}.
Equation~\eqref{eq:weighted-readout-distance} identifies when the multiround problem reduces exactly to that weighted spatial minimum.

\subsection{Which endpoint checks are needed?}

Removing the boundary syndrome constraint reduces the distance to one for every $T\ge1$.
One data fault in the logical support, introduced in the first round, persists through all later rounds.
Every round then has the same syndrome, so adjacent-round comparisons detect nothing while every readout is flipped.
More generally, let $H$ be any check matrix and let $w$ be a nonzero linear parity readout on the data.
Choose a coordinate $j$ with $w_j=1$ and introduce the unit error $e_j$ in the first round.
This weight-one fault is undetected by adjacent comparisons and flips every readout.
Since a zero-weight fault cannot flip a linear readout, the distance is exactly one.
Thus boundary collapse follows from error accumulation and comparison-only detection, for any original code distance.

The zero-syndrome constraint can be imposed at either endpoint.
Adjacent comparisons make all syndromes equal, so requiring either the first or the last syndrome to vanish forces all of them to vanish.
Requiring a zero syndrome at the second endpoint gives no further restriction on undetected faults in this model.

The endpoint constraint acts on the total syndrome, so fault contributions may cancel there.
The source convention and its surrounding error correction are described under C3 in Supplemental Sec.~\ref*{sec:decision}.

The endpoint need not constrain every syndrome bit.
Suppose adjacent comparisons still use the full syndrome $\sigma_t$, but an endpoint checks only $B\sigma_t=0$ for a binary linear map $B$.
Define $\delta_B=\min\{\wt(y):BHy=0,\ w\cdot y=1\}$ and assume this set is nonempty.
For strict majority, the resulting distance obeys
\begin{equation}
\min\{\delta_B,q+1\}\le d_{\mathrm{maj}}^B(T)\le\delta_B,
\qquad q=\lfloor T/2\rfloor+1.
\label{eq:selected-endpoint-distance}
\end{equation}
It equals $\delta_B$ when $q\ge\delta_B-1$.
Thus, for a target $d$, the endpoint map must detect every parity-flipping vector below weight $d$; this condition is also sufficient once $q\ge d-1$.
The proof and the cases where the bound is not tight are given in Supplemental Sec.~\ref*{sec:sm-readout-allocation}.

\subsection{Minimum repetitions for Shor-family checks}

For the square $m\times m$ Bacon--Shor and generalized Shor families with $m\ge2$, the exact formula also determines how often to measure each adjacent-column check.
Let $v_j$ be the parity of the data error in column $j$, so the logical readout is $v_0$ and check $j$ measures $v_{j-1}+v_j$ for $1\le j<m$.
If check $j$ has $c_j\ge1$ independent copies per round, then
\begin{equation}
\begin{aligned}
d_q(H_c,w;T)&=\min\left\{m,\ \min_{1\le j<m}(j+q c_j)\right\},\\
c_j^*&=\left\lceil\frac{m-j}{q}\right\rceil
\qquad (d_q=m).
\end{aligned}
\label{eq:path-check-allocation}
\end{equation}
The copy counts $c_j^*$ are the componentwise minimum that gives distance $m$ within this fixed family of checks.
They allocate more repetitions to checks closer to the selected readout column.
The derivation appears in Supplemental Sec.~\ref*{sec:sm-readout-allocation}.

\section{Measurement allocation for gauged readout}
\label{sec:gauged-allocation}

Gauging measures a logical Pauli operator through the product of local Gauss-check outcomes~\cite{williamson2026gauging}.
We combine those outcomes with the repeated-check formula to protect both the data and the measurement records.

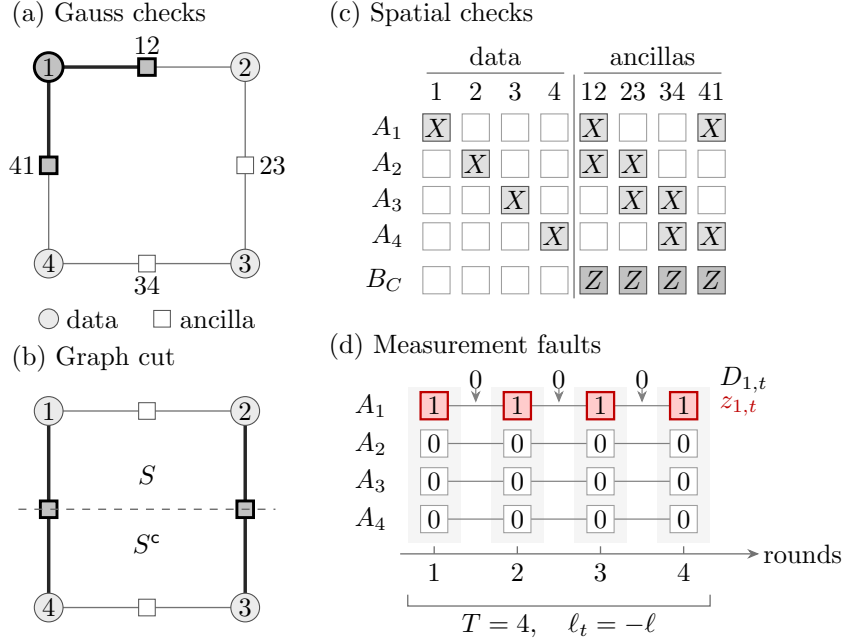
\begin{figure*}[t]
\centering
\begin{tikzpicture}[
  font=\normalsize,
  >={Stealth[length=1.6mm,width=1.3mm]},
  title/.style={anchor=west,inner sep=0pt},
  tag/.style={inner sep=0pt},
  data/.style={circle,draw=black!70,fill=black!8,minimum size=0.38cm,inner sep=0pt},
  active/.style={data,draw=black,line width=1.1pt,fill=black!24},
  anc/.style={rectangle,draw=black!65,fill=white,minimum size=0.22cm,inner sep=0pt},
  activeanc/.style={anc,draw=black,line width=1.1pt,fill=black!24},
  edge/.style={draw=black!55,line width=0.5pt},
  heavy/.style={draw=black!85,line width=1.3pt},
  cut/.style={draw=black!60,dashed,line width=0.5pt},
  cell/.style={rectangle,draw=black!38,fill=white,minimum size=0.36cm,inner sep=0pt},
  xcell/.style={cell,draw=black!65,fill=black!12},
  zcell/.style={cell,draw=black!65,fill=black!24},
  fcell/.style={cell,draw=red!75!black,fill=red!20,line width=0.85pt},
  detector/.style={draw=black!50,line width=0.45pt},
  band/.style={fill=black!4,draw=none}
]

\begin{scope}[shift={(0,4.75)}]
\node[title] at (0,3.75) {(a) Gauss checks};
\draw[edge] (0.50,3.05) -- (3.10,3.05) -- (3.10,0.45) -- (0.50,0.45) -- cycle;
\draw[heavy] (0.50,3.05) -- (1.80,3.05);
\draw[heavy] (0.50,3.05) -- (0.50,1.75);
\node[active] at (0.50,3.05) {$1$};
\node[data] at (3.10,3.05) {$2$};
\node[data] at (3.10,0.45) {$3$};
\node[data] at (0.50,0.45) {$4$};
\node[activeanc] at (1.80,3.05) {};
\node[anc] at (3.10,1.75) {};
\node[anc] at (1.80,0.45) {};
\node[activeanc] at (0.50,1.75) {};
\node[tag,anchor=south] at (1.80,3.21) {$12$};
\node[tag,anchor=west] at (3.28,1.75) {$23$};
\node[tag,anchor=north] at (1.80,0.28) {$34$};
\node[tag,anchor=east] at (0.33,1.75) {$41$};
\node[data,minimum size=0.26cm] at (0.50,-0.28) {};
\node[tag,anchor=west] at (0.73,-0.28) {data};
\node[anc] at (2.00,-0.28) {};
\node[tag,anchor=west] at (2.23,-0.28) {ancilla};
\end{scope}

\begin{scope}[shift={(0,0.20)}]
\node[title] at (0,3.75) {(b) Graph cut};
\draw[edge] (0.50,3.05) -- (3.10,3.05) -- (3.10,0.45) -- (0.50,0.45) -- cycle;
\draw[heavy] (0.50,3.05) -- (0.50,0.45);
\draw[heavy] (3.10,3.05) -- (3.10,0.45);
\node[data] at (0.50,3.05) {$1$};
\node[data] at (3.10,3.05) {$2$};
\node[data] at (3.10,0.45) {$3$};
\node[data] at (0.50,0.45) {$4$};
\node[anc] at (1.80,3.05) {};
\node[activeanc] at (3.10,1.75) {};
\node[anc] at (1.80,0.45) {};
\node[activeanc] at (0.50,1.75) {};
\draw[cut] (0.10,1.75) -- (3.50,1.75);
\node[tag] at (1.80,2.21) {$S$};
\node[tag] at (1.80,1.29) {$S^{\mathsf c}$};
\end{scope}

\begin{scope}[shift={(4.20,3.85)}]
\node[title] at (0,4.65) {(c) Spatial checks};
\node[tag] at (2.21,4.10) {data};
\node[tag] at (4.29,4.10) {ancillas};
\draw[edge] (1.30,3.87) -- (3.12,3.87);
\draw[edge] (3.38,3.87) -- (5.20,3.87);
\foreach \label/\x in {1/1.43,2/1.95,3/2.47,4/2.99,12/3.51,23/4.03,34/4.55,41/5.07}
  {\node[tag] at (\x,3.63) {$\label$};}
\foreach \label/\y in {A_1/3.15,A_2/2.67,A_3/2.19,A_4/1.71,B_C/1.13}
  {\node[tag,anchor=east] at (1.00,\y) {$\label$};}
\foreach \y in {3.15,2.67,2.19,1.71,1.13}
  {\foreach \x in {1.43,1.95,2.47,2.99,3.51,4.03,4.55,5.07}
    {\node[cell] at (\x,\y) {};}}
\foreach \x/\y in {1.43/3.15,3.51/3.15,5.07/3.15,1.95/2.67,3.51/2.67,4.03/2.67,2.47/2.19,4.03/2.19,4.55/2.19,2.99/1.71,4.55/1.71,5.07/1.71}
  {\node[xcell] at (\x,\y) {$X$};}
\foreach \x in {3.51,4.03,4.55,5.07}
  {\node[zcell] at (\x,1.13) {$Z$};}
\draw[edge] (3.25,3.85) -- (3.25,0.89);
\end{scope}

\begin{scope}[shift={(4.20,0)}]
\node[title] at (0,4.10) {(d) Measurement faults};
\foreach \x in {1.40,2.50,3.60,4.70}
  {\fill[band] (\x-0.35,1.51) rectangle (\x+0.35,3.57);}
\foreach \label/\y in {A_1/3.32,A_2/2.82,A_3/2.32,A_4/1.82}
  {\node[tag,anchor=east] at (0.78,\y) {$\label$};
   \foreach \x in {1.58,2.68,3.78}
     {\draw[detector] (\x,\y) -- ++(0.74,0);}}
\foreach \y in {2.82,2.32,1.82}
  {\foreach \x in {1.40,2.50,3.60,4.70}
    {\node[cell] at (\x,\y) {$0$};}}
\foreach \x in {1.40,2.50,3.60,4.70}
  {\node[fcell] at (\x,3.32) {$1$};}
\foreach \x in {1.95,3.05,4.15}
  {\node[tag] at (\x,3.67) {$0$};
   \draw[->,detector] (\x,3.54) -- (\x,3.36);}
\node[tag,anchor=west] at (5.17,3.67) {$D_{1,t}$};
\node[tag,anchor=west,text=red!75!black] at (5.17,3.32) {$z_{1,t}$};
\draw[->,edge] (0.95,1.36) -- (5.70,1.36);
\foreach \label/\x in {1/1.40,2/2.50,3/3.60,4/4.70}
  {\draw[edge] (\x,1.36) -- (\x,1.46);
   \node[tag,anchor=north] at (\x,1.24) {$\label$};}
\node[tag,anchor=west] at (5.75,1.36) {rounds};
\draw[edge] (1.05,0.82) -- (1.05,0.67) -- (5.05,0.67) -- (5.05,0.82);
\node[tag,anchor=north] at (3.05,0.58) {$T=4,\quad\ell_t=-\ell$};
\end{scope}

\end{tikzpicture}
\caption{
Gauging the weight-four logical operator $L=X_1X_2X_3X_4$.
(a) Circles are support data; squares are edge ancillas initialized in $|0\rangle$.
Dark markers show $A_1=X_1X_{12}X_{41}$; the outcome product of all four Gauss checks measures $L$.
(b) The dashed line separates the vertex set $S$ from its complement $S^{\mathsf c}$, cutting two edges.
C1 requires $|\partial S|\ge\min(|S|,|S^{\mathsf c}|)$ for every $S$.
(c) Displayed Gauss checks and the cycle check $B_C=Z_{12}Z_{23}Z_{34}Z_{41}$; blank cells denote $I$.
Original checks, their deformed partners and other data qubits are omitted.
(d) Four measurement-outcome flips on $A_1$ pass every adjacent-round comparison and reverse every logical readout.
Here $\ell$ is the ideal logical outcome and $\ell_t$ is the observed outcome in round $t$.
Here $z_{v,t}$ marks a flipped check outcome and $D_{v,t}=z_{v,t}\oplus z_{v,t+1}$ is its comparison detector.
Fault weight is four in this measurement-only example.
OpenAI Codex assisted the drawing code and layout; see Section~\ref{sec:verification}.
}

\label{fig:two-component}
\end{figure*}

An $X$-type gauging graph $G$ places a vertex on each data qubit in the measured logical support and an edge ancilla in $|0\rangle$ on each edge.
Its Gauss check is $A_v=X_v\prod_{e\ni v}X_e$.
Multiplying all Gauss checks cancels every edge factor, leaving the original logical operator.
These are checks on the enlarged system; measuring the original stabilizers alone does not measure that logical operator.
Figure~\ref{fig:two-component} shows the check supports and a constant measurement-error history.
The complete gauging construction also includes the deformed original checks and cycle checks~\cite{williamson2026gauging}.
For its spatial guarantee, the source uses the sufficient graph condition C1, $|\partial S|\ge\min(|S|,|S^{\mathsf c}|)$ for every vertex set $S$, together with the check-completion hypotheses.
The readout model below specifies the original-check and Gauss records explicitly; Supplemental Sec.~\ref*{sec:setting} gives the full construction background.

Let $A_0,\ldots,A_{m-1}$ be the commuting Gauss checks whose product is the logical operator $L$.
Use a separate simple graph $K$ to choose additional pair products $A_uA_v$ for measurement.
This graph $K$ is independent of the auxiliary graph $G$ used to construct the gauged code.
If $a\in\mathbb F_2^m$ is the vector of original Gauss outcomes, the added records are the pair parities $a_u+a_v$ for edges $\{u,v\}\in E(K)$.
Same-round detectors check each pair result against its two original outcomes.

With independent outcome flips and fault-free data, an undetected logical flip is a set $S$ of original outcomes and the pair outcomes on its cut $\partial_K S$.
Its weight is $|S|+|\partial_K S|$, and it changes the logical result exactly when $|S|$ is odd.
The minimum measurement-only distance is
\begin{equation}
d_{\mathrm{pair}}(K)=\min_{|S|\,\mathrm{odd}}\bigl(|S|+|\partial_K S|\bigr)
=1+\min_v\deg_K(v).
\label{eq:pair-product-distance}
\end{equation}
For $T$ repeated rounds with adjacent comparisons of the original Gauss outcomes, the distance is $T d_{\mathrm{pair}}(K)$.
Consequently, reaching single-round distance $\delta\le m$ requires at least $\lceil m(\delta-1)/2\rceil$ pair-product measurements; a simple graph attaining this count exists for every such $m$ and $\delta$.
The proof and graph constructions are in Supplemental Sec.~\ref*{sec:sm-readout-allocation}.

We now combine these records with persistent data and auxiliary-edge errors.
Let $P$ restrict the original data error to the $m$ qubits in the measured support, and let $J_G$ be the vertex-edge incidence matrix of the gauging graph $G$.
Thus $\mathbf1^{\mathsf T}P=w^{\mathsf T}$ and $\mathbf1^{\mathsf T}J_G=0$: the Gauss product reads the selected parity, and each edge ancilla touches two vertices.
Let $H_c$ be the original $X$-check matrix with the chosen repeated adjacent-column checks, and let $R_K$ encode the original Gauss outcomes and pair products.
Here $R_K a=(a,(a_u+a_v)_{\{u,v\}\in E(K)})$; $x_t$ and $e_t$ are the cumulative $Z$-type errors on the original data and edge ancillas through round $t$.
Each round also has independent result errors $\eta_t$ on original checks and independent result errors $\nu_t$ on the encoded Gauss records.
Each new data or edge error bit and each result-error bit counts as one fault location.
All syndrome and Gauss-record vectors in this model are flips relative to the fault-free measurement history.
\begin{samepage}
\begin{proposition}[Exact joint gauged-readout distance]
\label{prop:joint-readout}
In the persistent-fault model above, let $T\ge1$ and use a zero-syndrome endpoint check, adjacent-round syndrome and Gauss-record comparisons, and same-round pair consistency.
An accepted wrong readout has one fixed vector $z$ with $\mathbf1\cdot z=1$ in every round, and its exact minimum fault weight is
\begin{equation}
\begin{aligned}
\sigma_t&=H_c x_t+\eta_t=0,\\
b_t&=R_K(Px_t+J_G e_t)+\nu_t\\
&=R_K z,\\
d_{\mathrm{gauge}}(T;H_c,w,R_K)
&=\min\left\{T\delta_{R_K},\ \Gamma_T(H_c,w)\right\},\\
\delta_{R_K}&=\min_{\mathbf1\cdot u=1}\operatorname{wt}(R_Ku).
\end{aligned}
\label{eq:joint-gauged-readout-distance}
\end{equation}
\end{proposition}
\end{samepage}
Here $b_t=R_Kz$ means the encoded Gauss records are mutually consistent and unchanged between rounds, and a wrong logical result has $\mathbf1\cdot z=1$.
The first term is the cost of repeating an odd-parity Gauss-result error; the second is the cost of hiding a persistent data error from the original checks in every round.
The proof appears in Supplemental Sec.~\ref*{sec:sm-joint-distance}.

For the square Shor-family instance, the exact distance gives a resource optimum within the family that repeats fixed adjacent-column checks and adds pair products.
\begin{samepage}
\begin{corollary}[Minimum outcome count]
\label{cor:outcome-count}
For the square $m\times m$ Shor-family readout with $m\ge1$, use the joint model of Proposition~\ref{prop:joint-readout} and the same measurement allocation in every round.
Allow the $m$ Gauss checks, at least one independent copy of each fixed adjacent-column check, and pair products selected by a simple graph $K$.
At target distance $m$ and $1\le T\le m$, the minimum total outcome count is
\begin{equation}
\begin{aligned}
N_T^*&=T\Bigl[m+\left\lceil\frac{m(\lceil m/T\rceil-1)}{2}\right\rceil\\
&\qquad+\sum_{s=1}^{m-1}\left\lceil\frac{s}{T}\right\rceil\Bigr]\\
&\ge m^2+\frac{m(T-1)}{2}.
\end{aligned}
\label{eq:gauged-resource-count}
\end{equation}
It is attained by $c_j=\lceil(m-j)/T\rceil$ and a pair graph with minimum degree $\lceil m/T\rceil-1$ and $\lceil m(\lceil m/T\rceil-1)/2\rceil$ edges.
Over all positive round counts, one round uniquely minimizes the total at $m^2$ outcomes.
\end{corollary}
\end{samepage}
For $T>m$, the $m$ Gauss outcomes per round already cost $Tm>m^2$.
The proof of the minimum for $1\le T\le m$, the attaining allocations and the bound in Eq.~\eqref{eq:gauged-resource-count} appears in Supplemental Sec.~\ref*{sec:sm-outcome-count}.
The repeated baseline with $T=m$ measures the $m$ Gauss checks and all $m-1$ adjacent-column checks in every round, for $m(2m-1)$ outcomes.
For $m=5$, these counts are 25 and 45 at the same phenomenological distance five.
The quantity compared here is the number of measurement outcomes in the stated model; gate, ancilla, deformed-check and surrounding-round resources are outside this count.
The next section gives a circuit scheduling bound for a bare-ancilla implementation of this measurement family.

\section{A circuit scheduling bound}
\label{sec:circuit-bound}

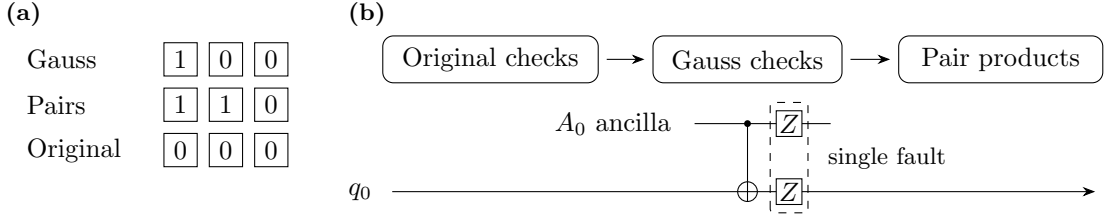
\begin{figure*}[t]
\centering
\begin{tikzpicture}[x=1cm,y=1cm,>=Stealth,font=\normalsize]
\node[anchor=west,font=\bfseries] at (-0.28,1.85) {(a)};
\node[anchor=west] at (0,1.25) {Gauss};
\node[anchor=west] at (0,0.65) {Pairs};
\node[anchor=west] at (0,0.05) {Original};
\foreach \x/\a/\b/\c in {2.15/1/1/0,2.75/0/1/0,3.35/0/0/0} {
  \node[draw,minimum size=0.44cm,inner sep=0pt] at (\x,1.25) {$\a$};
  \node[draw,minimum size=0.44cm,inner sep=0pt] at (\x,0.65) {$\b$};
  \node[draw,minimum size=0.44cm,inner sep=0pt] at (\x,0.05) {$\c$};
}
\node[anchor=west,font=\bfseries] at (4.25,1.85) {(b)};
\node[draw,rounded corners,minimum width=2.85cm,minimum height=0.62cm,inner xsep=6pt] (originalstage) at (6.25,1.25) {Original checks};
\node[draw,rounded corners,minimum width=2.50cm,minimum height=0.62cm,inner xsep=6pt] (gaussstage) at (9.65,1.25) {Gauss checks};
\node[draw,rounded corners,minimum width=2.70cm,minimum height=0.62cm,inner xsep=6pt] (pairstage) at (13.00,1.25) {Pair products};
\draw[->,shorten <=3pt,shorten >=3pt] (originalstage.east) -- (gaussstage.west);
\draw[->,shorten <=3pt,shorten >=3pt] (gaussstage.east) -- (pairstage.west);
\draw[->] (4.95,-0.50) -- (14.25,-0.50);
\draw (8.95,0.40) -- (10.75,0.40);
\draw (9.65,0.40) -- (9.65,-0.50);
\fill (9.65,0.40) circle (0.045);
\draw[fill=white] (9.65,-0.50) circle (0.13);
\draw (9.52,-0.50) -- (9.78,-0.50);
\draw (9.65,-0.63) -- (9.65,-0.37);
\draw[dashed,rounded corners=1.5pt] (9.95,-0.78) rectangle (10.45,0.68);
\node[draw,fill=white,minimum size=0.34cm,inner sep=0pt] at (10.20,0.40) {$Z$};
\node[draw,fill=white,minimum size=0.34cm,inner sep=0pt] at (10.20,-0.50) {$Z$};
\node[anchor=west,font=\small] at (10.60,-0.05) {single fault};
\node[anchor=east] at (4.80,-0.50) {$q_0$};
\node[anchor=east] at (8.75,0.40) {$A_0$ ancilla};
\end{tikzpicture}
\caption{A consistent wrong record caused by one gate fault.
(a) For three Gauss checks and their three pair products, the displayed vectors are flips relative to the ideal records, with pair order $(01,02,12)$.
The Gauss errors $(1,0,0)$ and pair errors $(1,1,0)$ satisfy every pair-consistency relation but flip the logical product; the three original-check records have zero error.
(b) A serial schedule measures all original $X$ checks, then the Gauss checks, then the pair products.
The drawing shows one Gauss-ancilla CNOT on the selected support qubit $q_0$; the blocks denote complete stages, with other interactions omitted.
A $Z$ on both CNOT qubits immediately after the gate flips $A_0$ and propagates to the later pair-product controls.
There is no later original check and no earlier pair check on $q_0$, so $\alpha=\beta=0$ and the circuit distance is one by Proposition~\ref{prop:circuit-bound}.
For the distance-three Shor allocation, the same measured-check family has phenomenological distance three.
OpenAI Codex assisted the drawing code and layout; see Section~\ref{sec:verification}.
}
\label{fig:serial-bound}
\end{figure*}

Figure~\ref{fig:serial-bound} illustrates an accepted one-gate fault in a serial readout with three Gauss checks.

The outcome-count optimum treats data and result faults as independent locations.
In a circuit, one gate fault can create both, and the order of the checks matters.
We now give a bound for a specified one-epoch implementation of the same measurement family.
The fault-free input is an original-code state with the gauging edge ancillas initialized in $|0\rangle$.

Each original $X$ check, Gauss check $A_v$ and selected pair product $A_uA_v$ is measured once using its own bare ancilla.
Every ancilla starts in $|+\rangle$, controls one CNOT to each qubit in its check support, and is measured in $X$ after its last CNOT.
The CNOTs may be interleaved in any order that uses each qubit at most once per time step; data and edge qubits are always targets, and no other gates or error-correction rounds are inserted.
The detectors are the original-check results and the same-round pair-consistency relations.
We count a Pauli fault after a CNOT as one location and each independent outcome flip as one location.
The allowed gate faults include $Z$ on the target and $Z$ on both control and target.
The numerical search also includes initial Pauli faults on original-data and edge qubits and preparation faults on extraction ancillas; each faulty input qubit or preparation counts as one location.
An accepted wrong readout passes these detectors but reverses the product of the Gauss outcomes relative to the fault-free circuit.

For each support qubit $q_v$, consider a CNOT $g$ on $q_v$ no later than its CNOT with the $A_v$ ancilla.
Let $\alpha_v(g)$ count original-check CNOTs on $q_v$ strictly after $g$, and let $\beta_v(g)$ count pair-product CNOTs on $q_v$ strictly before $g$.
Independent copies of an original check are counted separately.

\begin{proposition}[Schedule-dependent circuit bound]
\label{prop:circuit-bound}
For any valid schedule $\mathcal S$ in the circuit family above, its accepted-wrong readout distance obeys
\begin{equation}
d_{\mathrm{bare}}(\mathcal S)
\le 1+\min_{v,g}\bigl[\alpha_v(g)+\beta_v(g)\bigr].
\label{eq:circuit-schedule-bound}
\end{equation}
\end{proposition}

To construct the fault, insert target $Z$ after $g$; also insert control $Z$ at the same location if $g$ belongs to a Gauss or pair-product check.
This flips the $A_v$ result and every later check on $q_v$.
If $g$ is a pair-product CNOT, the control $Z$ also flips that pair result.
Cancel the $\alpha_v(g)$ affected original-check results and flip the $\beta_v(g)$ earlier pair results.
The Gauss record now differs by the unit vector at $v$, and every pair record agrees with it.
All detectors vanish and the logical product is wrong, with the weight in Eq.~\eqref{eq:circuit-schedule-bound}.
The Supplemental Material gives the full Pauli-propagation proof.

This bound gives a necessary condition on a distance-$m$ implementation: $\alpha_v(g)+\beta_v(g)\ge m-1$ at every eligible gate.
For example, a serial schedule that measures all original checks, then all Gauss checks, then all pair products has $\alpha_v(g)=\beta_v(g)=0$ at a Gauss CNOT.
Its distance is exactly one, even when the independent-outcome model has distance $m$.
Thus the measured-check family alone does not determine the circuit distance; the checks must also detect errors introduced during extraction.

Backward Pauli propagation gives exact distance three for the best tested $m=3$ and $m=5$ schedules among 1,024 schedules of each size.
Equation~\eqref{eq:circuit-schedule-bound} also gives three for both schedules, so explicit three-fault patterns attain the bound.
At $m=3$, flipping all three Gauss outcomes leaves every pair relation unchanged and reverses the logical product.
This witness applies to every schedule in the circuit family, so the saved distance-three schedule is optimal within that family.
At $m=5$, one final-outcome flip and two target-$Z$ gate faults give the accepted wrong readout listed in Supplemental Sec.~\ref*{sec:sm-schedule-witnesses}.
The value three is the largest found in the finite test set, below the target distance five; the complete schedules and fault locations are given in Supplemental Sec.~\ref*{sec:sm-schedule-data}.
Deformed $Z$ checks, cycle checks, flagged ancillas and surrounding error-correction rounds require their own detector and fault analysis.

\section{Examples and design implications}
\label{sec:examples}

The general formulas distinguish the effects of the readout predicate, the check allocation and the auxiliary graph.
We illustrate these effects with a nine-qubit readout, a distance-five allocation, and two gauged codes on the same measured support.

\subsection{A nine-qubit readout with persistent faults}

We specialize the persistent-fault model of Section~\ref{sec:readout} to the nine-qubit Bacon--Shor readout.
The Bacon--Shor distance three refers to its protected subsystem; this readout tests the selected logical parity using two $X$-type stabilizer checks.
The nine-qubit Bacon--Shor instance used in Table~\ref{tab:models} has $n=9$, two check rows and qubit labels $j=3a+b$ on a three-by-three grid.
Its two rows of $H$ have supports $\{0,1,3,4,6,7\}$ and $\{1,2,4,5,7,8\}$, and $w=\mathbf 1_{\{0,3,6\}}$.

In the measurement-fault model, each fault flips an independent measurement outcome, and the detectors compare consecutive logical parities.

The persistent-fault model also permits ties at two rounds.
At two rounds, a final-round data fault on qubit zero and a canceling check-result fault give $(r_0,r_1)=(0,1)$ with zero syndromes and weight two.
A first-round readout fault and its canceling check-result fault instead give $(1,0)$.
Table~\ref{tab:models} collects the four model choices and their independently reproduced values.

\begin{table*}[t]
\caption{Readout models and exact distances at $T=1,2,3,4$ rounds.
Data faults persist; check-result and readout faults act in their own round.
The boundary constraint is $\sigma_0=0$; syndrome comparisons impose $\sigma_t=\sigma_{t-1}$.
Logical comparisons instead require consecutive logical readouts to agree.
The first three rows' twelve values are certified in Lean; the last row is a numerical control.
}
\label{tab:models}
\begin{ruledtabular}
\begin{tabular}{p{82pt}p{82pt}p{122pt}p{87pt}l}
Model & Fault locations & Detectors & Error criterion & Distances \\
\hline
Measurement only & Measurement outcomes & Logical comparisons & Constant logical flip & $1,2,3,4$ \\
Persistent faults & Data, check results, readout & Boundary and syndrome comparisons & Strict majority & $2,3,3,3$ \\
No boundary constraint & Data, check results, readout & Syndrome comparisons & Strict majority & $1,1,1,1$ \\
Last-round control & Data, check results, readout & Boundary and syndrome comparisons & Last readout wrong & $2,2,2,2$ \\
\end{tabular}
\end{ruledtabular}
\end{table*}

For the original two checks, Eq.~\eqref{eq:path-check-allocation} at $m=3$ gives $d_q=\min\{3,1+q\}$.
Strict majority therefore has distance two at $T=1$ and three for every $T\ge2$.
At one round, a data error on qubit zero changes the logical parity and one check; a check-result flip hides that syndrome, giving weight two.
Duplicating the first check changes the allocation to $(c_1,c_2)=(2,1)$ and makes $d_1=3$.
An error on the readout column must then be hidden in both copies, while a weight-three logical data error gives the matching upper bound.
This raises the one-round distance by adding one independent check-result location, without changing the static check row space.

\begin{figure}[!htbp]
\centering
\includegraphics[width=0.792\columnwidth]{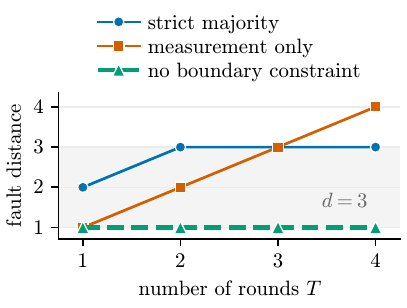}
\caption{
Exact distances for the three Bacon--Shor $[[9,1,3]]$ models in Table~\ref{tab:models}.
The blue circles show $d_{\mathrm{maj}}$: two at one round and three for $T\ge2$.
Deleting the endpoint syndrome constraint gives the green triangles at one for every positive $T$.
Orange squares show the measurement-only distance $T$.
The dotted line is the static code distance $d=3$; shading spans the region down to the no-boundary curve.
Independent enumeration reproduces all twelve values, each certified in Lean.
OpenAI Codex assisted the drawing code and layout; see Section~\ref{sec:verification}.
}

\label{fig:sharpness}
\end{figure}

Figure~\ref{fig:sharpness} compares this readout with two other models.
The measurement-only model compares consecutive logical parities and requires a wrong answer in every round, giving distance $T$.
The persistent-fault model uses syndrome comparisons and the strict-majority predicate, so its relevant threshold is $q=\lfloor T/2\rfloor+1$.
Removing its endpoint constraint admits a constant nonzero syndrome and gives distance one.
A second implementation recomputes the figure, and all twelve of its points are exactly the proved ones, four on each curve.
The circuit and joint gauged-readout analyses use their separately stated record-consistency predicates; their distances should be evaluated with those predicates.

\subsection{Allocating measurements at distance five}

For $m=5$, Table~\ref{tab:allocation-five} makes the outcome-count optimum concrete.
At $T=1$, the original adjacent-column checks have copy counts $(4,3,2,1)$, and the pair graph is complete.
The ten original-check outcomes, ten pair outcomes and five Gauss outcomes give 25 in total.
At $T=2$, a five-cycle pair graph gives $\delta_{R_K}=3$, while the original-check allocation is $(2,2,1,1)$.
This uses 16 outcomes per round and 32 overall.
At $T=5$, pair products are unnecessary and each original check is measured once per round, giving the repeated baseline of 45.
Each entry reaches the same target distance five in the joint phenomenological model.
The comparison counts these measured records; Proposition~\ref{prop:circuit-bound} separately tests the timing of a bare-ancilla implementation.

\begin{table}[t]
\caption{Minimum outcome counts at target distance five within the fixed adjacent-column and Gauss-pair measurement family. $T$ is the round count, $\delta_{R_K}$ is the Gauss-record distance per round, $c$ lists the adjacent-column check copy counts, and $N_T^*$ is the total number of outcomes. The five Gauss outcomes per round are included in every total.}
\label{tab:allocation-five}
\begin{ruledtabular}
\begin{tabular}{ccccc}
$T$ & $\delta_{R_K}$ & $c$ & Pairs/round & $N_T^*$ \\
\hline
1 & 5 & $(4,3,2,1)$ & 10 & 25 \\
2 & 3 & $(2,2,1,1)$ & 5 & 32 \\
3 & 2 & $(2,1,1,1)$ & 3 & 39 \\
4 & 2 & $(1,1,1,1)$ & 3 & 48 \\
5 & 1 & $(1,1,1,1)$ & 0 & 45 \\
\end{tabular}
\end{ruledtabular}
\end{table}

\subsection{Spatial distance and readout distance}

The auxiliary graph $G$ defines the deformed quantum code, whereas $K$ chooses the redundant Gauss-result measurements.
Their distance statements concern different fault sets.
The spatial distance tests all nontrivial logical operators of the deformed code; the readout distance tests faults that change the selected result while passing its detectors.

\begin{figure}[!htbp]
\centering
\begin{tikzpicture}[
    every node/.style={inner sep=1.5pt},
    vert/.style={circle, draw=black!75, fill=black!10, minimum size=5.5pt,
                 inner sep=0pt},
    ed/.style={draw=black!75, line width=0.9pt},
    unshared/.style={draw=black!75, line width=0.9pt},
    slab/.style={font=\normalsize, text=black!80},
    rdg/.style={font=\normalsize, text=black!85},
  ]

  \def\ax{0.0}\def\ay{1.15}
  \def\bx{0.0}\def\by{0.0}
  \def\cx{1.15}\def\cy{1.15}
  \def\dx{1.15}\def\dy{0.0}

  \begin{scope}[shift={(0,0)}]
    \draw[ed] (\ax,\ay) -- (\cx,\cy);
    \draw[ed] (\cx,\cy) -- (\dx,\dy);
    \draw[unshared] (\ax,\ay) -- (\bx,\by);
    \node[vert] at (\ax,\ay) {};
    \node[vert] at (\bx,\by) {};
    \node[vert] at (\cx,\cy) {};
    \node[vert] at (\dx,\dy) {};
    \node[slab, anchor=south east, xshift=-2.5pt, yshift=1.5pt] at (\ax,\ay) {$0$};
    \node[slab, anchor=north east, xshift=-2.5pt, yshift=-2.0pt] at (\bx,\by) {$2$};
    \node[slab, anchor=south west, xshift=2.5pt, yshift=1.5pt] at (\cx,\cy) {$4$};
    \node[slab, anchor=north west, xshift=2.5pt, yshift=-2.0pt] at (\dx,\dy) {$5$};
    \draw[decorate, decoration={brace, amplitude=3.5pt, mirror}, black!60]
      (\bx-0.06,\by-0.62) -- (\dx+0.06,\dy-0.62);
    \node[rdg, anchor=north] at (0.575,-0.80) {$d_X^{\mathrm{phys}}=1$};
  \end{scope}

  \begin{scope}[shift={(2.85,0)}]
    \draw[ed] (\ax,\ay) -- (\cx,\cy);
    \draw[ed] (\cx,\cy) -- (\dx,\dy);
    \draw[unshared] (\bx,\by) -- (\dx,\dy);
    \node[vert] at (\ax,\ay) {};
    \node[vert] at (\bx,\by) {};
    \node[vert] at (\cx,\cy) {};
    \node[vert] at (\dx,\dy) {};
    \node[slab, anchor=south east, xshift=-2.5pt, yshift=1.5pt] at (\ax,\ay) {$0$};
    \node[slab, anchor=north east, xshift=-2.5pt, yshift=-2.0pt] at (\bx,\by) {$2$};
    \node[slab, anchor=south west, xshift=2.5pt, yshift=1.5pt] at (\cx,\cy) {$4$};
    \node[slab, anchor=north west, xshift=2.5pt, yshift=-2.0pt] at (\dx,\dy) {$5$};
    \draw[decorate, decoration={brace, amplitude=3.5pt, mirror}, black!60]
      (\bx-0.06,\by-0.62) -- (\dx+0.06,\dy-0.62);
    \node[rdg, anchor=north] at (0.575,-0.80) {$d_X^{\mathrm{phys}}=2$};
  \end{scope}

\end{tikzpicture}
\caption{
Two paths on the same weight-four support of the toric code $[[8,2,2]]$ give different deformed-code distances.
Vertex numbers label data qubits; each edge adds one ancilla.
Both paths fail C1 at a two-vertex cut with one outgoing edge.
Braces show the physical $X$-type distances: one and two.
The corresponding physical $Z$-type distances are five and at least two, so the total distances are one and two.
For the left path, a zero $Z$-check column and the corresponding unit vector outside the $X$-check row space give the weight-one logical operator.
OpenAI Codex assisted the drawing code and layout; see Section~\ref{sec:verification}.
}

\label{fig:c1pair}
\end{figure}
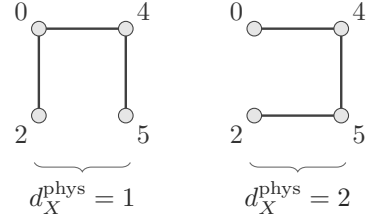

For a stabilizer code, $[[n,k,d]]$ denotes $n$ physical qubits, $k$ encoded qubits and distance $d$.
Two paths on the same weight-four logical support of the toric code $[[8,2,2]]$ illustrate the spatial distinction (Fig.~\ref{fig:c1pair}).
Both paths violate C1, because a two-vertex set has only one outgoing edge.
One path gives physical $X$-type distance one and physical $Z$-type distance five.
The other has physical $X$-type distance two and no weight-one logical on either side, so its total distance is two.
The source's sufficient expansion condition does not determine the exact distance of each construction below its threshold.
Supplemental Sec.~\ref*{sec:sharp} gives the check completions and finite census.

For the distance-one path, a zero column of the deformed $Z$-check matrix and a unit vector outside the $X$-check row space identify the weight-one logical operator.
The zero column means that no $Z$ check detects it; the row-space condition means that it is not a product of $X$ stabilizers.
This matrix test concerns the deformed static code.
The weighted formula and circuit bound additionally account for record faults and the order of extraction.

\section{Verification and reproducibility}
\label{sec:verification}

The general readout, allocation and circuit statements have direct proofs in the Supplemental Material.
Independent dynamic programs check the readout formulas on finite instances, and backward Pauli propagation checks the circuit witnesses and saved schedules.
Supplemental Sec.~\ref*{sec:sm-schedule-data} specifies the circuit fault locations, complete schedules and minimum-weight witnesses.
The separate Lean arithmetic project verifies five lemmas used in the resource count.
The algebraic and finite code results in the supplement are supported by the main Lean development; Supplemental Sec.~\ref*{sec:sm-audit} reports its axiom audit.
The general matrix theorems and circuit proposition are established by the direct proofs, rather than by that formal audit.

\section{Conclusion}
\label{sec:conclusion}

The weighted minimum $\Gamma_q(H,w)$ gives the exact distance for a threshold of wrong readouts in the persistent-fault model.
It turns measurement allocation into a spatial problem: an error costs its data weight plus the cost of hiding its syndrome in the required wrong rounds.
The endpoint criterion and Shor-family allocation make this relation explicit.

For gauged readout, the joint distance separates persistent-data faults from measurement-record faults and gives the one-round outcome-count optimum within the stated measurement family.
The circuit scheduling bound identifies how a gate fault can connect those two mechanisms during extraction.
Together, the exact formula, allocation rule and explicit circuit witnesses give conditions for designing and auditing logical Pauli readouts at the model and circuit levels.

\begin{acknowledgments}
This work was supported by the National Natural Science Foundation of China under Grant
No.~12674609 and by the Guangdong Provincial Key Laboratory of Computility Microelectronics
under Grant No.~2024B1212010007.
The authors declare no competing interests.
The authors used DeepSeek-V4.1-Flash for text polishing, the Lean derivations, and the debugging of individual numerical values.
They also used OpenAI Codex (GPT-6) for manuscript revision, literature checks, numerical cross-checks, formal-proof exploration and figure-code preparation.
The authors are responsible for all of its content.
\end{acknowledgments}

\section*{Author Contributions}

All authors contributed equally to this work.

\section*{Data availability}

The formal development and baseline numerical data are archived on Zenodo~\cite{gaugedarchive}, together with a last-round-readout control and its regression tests. The readout-analysis scripts, selected CNOT schedules and fault witnesses are provided as Supplemental data and code. The Supplemental Material~\cite{supplemental} identifies the formal statement supporting each claim and explains how to reproduce every numerical value. The formal layer beneath these results is released as a separate package~\cite{qeccertificates}.

\nocite{*}
\bibliography{refs}

\begin{thebibliography}{44}%
\makeatletter
\providecommand \@ifxundefined [1]{%
 \@ifx{#1\undefined}
}%
\providecommand \@ifnum [1]{%
 \ifnum #1\expandafter \@firstoftwo
 \else \expandafter \@secondoftwo
 \fi
}%
\providecommand \@ifx [1]{%
 \ifx #1\expandafter \@firstoftwo
 \else \expandafter \@secondoftwo
 \fi
}%
\providecommand \natexlab [1]{#1}%
\providecommand \enquote  [1]{``#1''}%
\providecommand \bibnamefont  [1]{#1}%
\providecommand \bibfnamefont [1]{#1}%
\providecommand \citenamefont [1]{#1}%
\providecommand \href@noop [0]{\@secondoftwo}%
\providecommand \href [0]{\begingroup \@sanitize@url \@href}%
\providecommand \@href[1]{\@@startlink{#1}\@@href}%
\providecommand \@@href[1]{\endgroup#1\@@endlink}%
\providecommand \@sanitize@url [0]{\catcode `\\12\catcode `\$12\catcode
  `\&12\catcode `\#12\catcode `\^12\catcode `\_12\catcode `\%12\relax}%
\providecommand \@@startlink[1]{}%
\providecommand \@@endlink[0]{}%
\providecommand \url  [0]{\begingroup\@sanitize@url \@url }%
\providecommand \@url [1]{\endgroup\@href {#1}{\urlprefix }}%
\providecommand \urlprefix  [0]{URL }%
\providecommand \Eprint [0]{\href }%
\providecommand \doibase [0]{https://doi.org/}%
\providecommand \selectlanguage [0]{\@gobble}%
\providecommand \bibinfo  [0]{\@secondoftwo}%
\providecommand \bibfield  [0]{\@secondoftwo}%
\providecommand \translation [1]{[#1]}%
\providecommand \BibitemOpen [0]{}%
\providecommand \bibitemStop [0]{}%
\providecommand \bibitemNoStop [0]{.\EOS\space}%
\providecommand \EOS [0]{\spacefactor3000\relax}%
\providecommand \BibitemShut  [1]{\csname bibitem#1\endcsname}%
\let\auto@bib@innerbib\@empty
\bibitem [{\citenamefont {Shor}(1995)}]{shor1995scheme}%
  \BibitemOpen
  \bibfield  {author} {\bibinfo {author} {\bibfnamefont {P.~W.}\ \bibnamefont
  {Shor}},\ }\bibfield  {title} {\bibinfo {title} {Scheme for reducing
  decoherence in quantum computer memory},\ }\href
  {https://doi.org/10.1103/PhysRevA.52.R2493} {\bibfield  {journal} {\bibinfo
  {journal} {Phys. Rev. A}\ }\textbf {\bibinfo {volume} {52}},\ \bibinfo
  {pages} {R2493} (\bibinfo {year} {1995})}\BibitemShut {NoStop}%
\bibitem [{\citenamefont {Steane}(1996)}]{steane1996error}%
  \BibitemOpen
  \bibfield  {author} {\bibinfo {author} {\bibfnamefont {A.~M.}\ \bibnamefont
  {Steane}},\ }\bibfield  {title} {\bibinfo {title} {Error correcting codes in
  quantum theory},\ }\href {https://doi.org/10.1103/PhysRevLett.77.793}
  {\bibfield  {journal} {\bibinfo  {journal} {Phys. Rev. Lett.}\ }\textbf
  {\bibinfo {volume} {77}},\ \bibinfo {pages} {793} (\bibinfo {year}
  {1996})}\BibitemShut {NoStop}%
\bibitem [{\citenamefont {Aharonov}\ and\ \citenamefont
  {Ben-Or}(2008)}]{aharonov2008fault}%
  \BibitemOpen
  \bibfield  {author} {\bibinfo {author} {\bibfnamefont {D.}~\bibnamefont
  {Aharonov}}\ and\ \bibinfo {author} {\bibfnamefont {M.}~\bibnamefont
  {Ben-Or}},\ }\bibfield  {title} {\bibinfo {title} {Fault-tolerant quantum
  computation with constant error rate},\ }\href
  {https://doi.org/10.1137/S0097539799359385} {\bibfield  {journal} {\bibinfo
  {journal} {SIAM J. Comput.}\ }\textbf {\bibinfo {volume} {38}},\ \bibinfo
  {pages} {1207} (\bibinfo {year} {2008})}\BibitemShut {NoStop}%
\bibitem [{\citenamefont {Panteleev}\ and\ \citenamefont
  {Kalachev}(2022)}]{panteleev2022qldpc}%
  \BibitemOpen
  \bibfield  {author} {\bibinfo {author} {\bibfnamefont {P.}~\bibnamefont
  {Panteleev}}\ and\ \bibinfo {author} {\bibfnamefont {G.}~\bibnamefont
  {Kalachev}},\ }\bibfield  {title} {\bibinfo {title} {Quantum {LDPC} codes
  with almost linear minimum distance},\ }\href
  {https://doi.org/10.1109/TIT.2021.3119384} {\bibfield  {journal} {\bibinfo
  {journal} {IEEE Trans. Inf. Theory}\ }\textbf {\bibinfo {volume} {68}},\
  \bibinfo {pages} {213} (\bibinfo {year} {2022})}\BibitemShut {NoStop}%
\bibitem [{\citenamefont {Breuckmann}\ and\ \citenamefont
  {Eberhardt}(2021)}]{breuckmann2021quantum}%
  \BibitemOpen
  \bibfield  {author} {\bibinfo {author} {\bibfnamefont {N.~P.}\ \bibnamefont
  {Breuckmann}}\ and\ \bibinfo {author} {\bibfnamefont {J.~N.}\ \bibnamefont
  {Eberhardt}},\ }\bibfield  {title} {\bibinfo {title} {Quantum low-density
  parity-check codes},\ }\href {https://doi.org/10.1103/PRXQuantum.2.040101}
  {\bibfield  {journal} {\bibinfo  {journal} {PRX Quantum}\ }\textbf {\bibinfo
  {volume} {2}},\ \bibinfo {pages} {040101} (\bibinfo {year}
  {2021})}\BibitemShut {NoStop}%
\bibitem [{\citenamefont {Bravyi}\ \emph {et~al.}(2024)\citenamefont {Bravyi},
  \citenamefont {Cross}, \citenamefont {Gambetta}, \citenamefont {Maslov},
  \citenamefont {Rall},\ and\ \citenamefont {Yoder}}]{bravyi2024highthreshold}%
  \BibitemOpen
  \bibfield  {author} {\bibinfo {author} {\bibfnamefont {S.}~\bibnamefont
  {Bravyi}}, \bibinfo {author} {\bibfnamefont {A.~W.}\ \bibnamefont {Cross}},
  \bibinfo {author} {\bibfnamefont {J.~M.}\ \bibnamefont {Gambetta}}, \bibinfo
  {author} {\bibfnamefont {D.}~\bibnamefont {Maslov}}, \bibinfo {author}
  {\bibfnamefont {P.}~\bibnamefont {Rall}},\ and\ \bibinfo {author}
  {\bibfnamefont {T.~J.}\ \bibnamefont {Yoder}},\ }\bibfield  {title} {\bibinfo
  {title} {High-threshold and low-overhead fault-tolerant quantum memory},\
  }\href {https://doi.org/10.1038/s41586-024-07107-7} {\bibfield  {journal}
  {\bibinfo  {journal} {Nature}\ }\textbf {\bibinfo {volume} {627}},\ \bibinfo
  {pages} {778} (\bibinfo {year} {2024})}\BibitemShut {NoStop}%
\bibitem [{\citenamefont {Kitaev}(1997)}]{kitaev1997quantum}%
  \BibitemOpen
  \bibfield  {author} {\bibinfo {author} {\bibfnamefont {A.~Y.}\ \bibnamefont
  {Kitaev}},\ }\bibfield  {title} {\bibinfo {title} {Quantum computations:
  algorithms and error correction},\ }\href
  {https://doi.org/10.1070/RM1997v052n06ABEH002155} {\bibfield  {journal}
  {\bibinfo  {journal} {Russ. Math. Surveys}\ }\textbf {\bibinfo {volume}
  {52}},\ \bibinfo {pages} {1191} (\bibinfo {year} {1997})}\BibitemShut
  {NoStop}%
\bibitem [{\citenamefont {Bomb{\'i}n}\ and\ \citenamefont
  {Martin-Delgado}(2007)}]{bombin2007homological}%
  \BibitemOpen
  \bibfield  {author} {\bibinfo {author} {\bibfnamefont {H.}~\bibnamefont
  {Bomb{\'i}n}}\ and\ \bibinfo {author} {\bibfnamefont {M.~A.}\ \bibnamefont
  {Martin-Delgado}},\ }\bibfield  {title} {\bibinfo {title} {Homological error
  correction: Classical and quantum codes},\ }\href
  {https://doi.org/10.1063/1.2731356} {\bibfield  {journal} {\bibinfo
  {journal} {J. Math. Phys.}\ }\textbf {\bibinfo {volume} {48}},\ \bibinfo
  {pages} {052105} (\bibinfo {year} {2007})}\BibitemShut {NoStop}%
\bibitem [{\citenamefont {Horsman}\ \emph {et~al.}(2012)\citenamefont
  {Horsman}, \citenamefont {Fowler}, \citenamefont {Devitt},\ and\
  \citenamefont {Van~Meter}}]{horsman2012surgery}%
  \BibitemOpen
  \bibfield  {author} {\bibinfo {author} {\bibfnamefont {D.}~\bibnamefont
  {Horsman}}, \bibinfo {author} {\bibfnamefont {A.~G.}\ \bibnamefont {Fowler}},
  \bibinfo {author} {\bibfnamefont {S.}~\bibnamefont {Devitt}},\ and\ \bibinfo
  {author} {\bibfnamefont {R.}~\bibnamefont {Van~Meter}},\ }\bibfield  {title}
  {\bibinfo {title} {Surface code quantum computing by lattice surgery},\
  }\href {https://doi.org/10.1088/1367-2630/14/12/123011} {\bibfield  {journal}
  {\bibinfo  {journal} {New J. Phys.}\ }\textbf {\bibinfo {volume} {14}},\
  \bibinfo {pages} {123011} (\bibinfo {year} {2012})}\BibitemShut {NoStop}%
\bibitem [{\citenamefont {Williamson}\ and\ \citenamefont
  {Yoder}(2026)}]{williamson2026gauging}%
  \BibitemOpen
  \bibfield  {author} {\bibinfo {author} {\bibfnamefont {D.~J.}\ \bibnamefont
  {Williamson}}\ and\ \bibinfo {author} {\bibfnamefont {T.~J.}\ \bibnamefont
  {Yoder}},\ }\bibfield  {title} {\bibinfo {title} {Low-overhead fault-tolerant
  quantum computation by gauging logical operators},\ }\href
  {https://doi.org/10.1038/s41567-026-03220-8} {\bibfield  {journal} {\bibinfo
  {journal} {Nat. Phys.}\ }\textbf {\bibinfo {volume} {22}},\ \bibinfo {pages}
  {598} (\bibinfo {year} {2026})}\BibitemShut {NoStop}%
\bibitem [{\citenamefont {Ashikhmin}\ \emph {et~al.}(2020)\citenamefont
  {Ashikhmin}, \citenamefont {Lai},\ and\ \citenamefont
  {Brun}}]{ashikhmin2020ds}%
  \BibitemOpen
  \bibfield  {author} {\bibinfo {author} {\bibfnamefont {A.}~\bibnamefont
  {Ashikhmin}}, \bibinfo {author} {\bibfnamefont {C.-Y.}\ \bibnamefont {Lai}},\
  and\ \bibinfo {author} {\bibfnamefont {T.~A.}\ \bibnamefont {Brun}},\
  }\bibfield  {title} {\bibinfo {title} {Quantum data-syndrome codes},\ }\href
  {https://doi.org/10.1109/JSAC.2020.2968997} {\bibfield  {journal} {\bibinfo
  {journal} {IEEE Journal on Selected Areas in Communications}\ }\textbf
  {\bibinfo {volume} {38}},\ \bibinfo {pages} {449} (\bibinfo {year}
  {2020})}\BibitemShut {NoStop}%
\bibitem [{\citenamefont {Fujiwara}(2014)}]{fujiwara2014self}%
  \BibitemOpen
  \bibfield  {author} {\bibinfo {author} {\bibfnamefont {Y.}~\bibnamefont
  {Fujiwara}},\ }\bibfield  {title} {\bibinfo {title} {Ability of stabilizer
  quantum error correction to protect itself from its own imperfection},\
  }\href {https://doi.org/10.1103/PhysRevA.90.062304} {\bibfield  {journal}
  {\bibinfo  {journal} {Physical Review A}\ }\textbf {\bibinfo {volume} {90}},\
  \bibinfo {pages} {062304} (\bibinfo {year} {2014})}\BibitemShut {NoStop}%
\bibitem [{\citenamefont {Kuo}\ and\ \citenamefont
  {Lai}(2025)}]{kuo2025generalized}%
  \BibitemOpen
  \bibfield  {author} {\bibinfo {author} {\bibfnamefont {K.-Y.}\ \bibnamefont
  {Kuo}}\ and\ \bibinfo {author} {\bibfnamefont {C.-Y.}\ \bibnamefont {Lai}},\
  }\bibfield  {title} {\bibinfo {title} {Generalized quantum data-syndrome
  codes and belief propagation decoding for phenomenological noise},\ }\href
  {https://doi.org/10.1109/TIT.2025.3529773} {\bibfield  {journal} {\bibinfo
  {journal} {IEEE Transactions on Information Theory}\ }\textbf {\bibinfo
  {volume} {71}},\ \bibinfo {pages} {1824} (\bibinfo {year}
  {2025})}\BibitemShut {NoStop}%
\bibitem [{sup()}]{supplemental}%
  \BibitemOpen
  \href@noop {} {\bibinfo {title} {Supplemental material}},\ \bibinfo {note}
  {see Supplemental Material at [URL will be inserted by publisher] for the
  module map, the conventions the numbers depend on, the axiom audit, and the
  commands that reproduce every number quoted here}\BibitemShut {NoStop}%
\bibitem [{\citenamefont {An}\ and\ \citenamefont
  {Yang}(2026)}]{gaugedarchive}%
  \BibitemOpen
  \bibfield  {author} {\bibinfo {author} {\bibfnamefont {S.}~\bibnamefont
  {An}}\ and\ \bibinfo {author} {\bibfnamefont {F.}~\bibnamefont {Yang}},\
  }\href {https://doi.org/10.5281/zenodo.22998740} {\bibinfo {title} {The cost
  of each side condition in a gauged logical measurement}} (\bibinfo {year}
  {2026}),\ \bibinfo {note} {version 0.3, formal development and numerical
  data}\BibitemShut {NoStop}%
\bibitem [{\citenamefont {An}(2026)}]{qeccertificates}%
  \BibitemOpen
  \bibfield  {author} {\bibinfo {author} {\bibfnamefont {S.}~\bibnamefont
  {An}},\ }\href {https://doi.org/10.5281/zenodo.23056679} {\bibinfo {title}
  {{QECCertificates}: checkable distance certificates for quantum error
  correction}} (\bibinfo {year} {2026})\BibitemShut {NoStop}%
\bibitem [{\citenamefont {Lee}(2026)}]{gaugingspacetime2026}%
  \BibitemOpen
  \bibfield  {author} {\bibinfo {author} {\bibfnamefont {G.}~\bibnamefont
  {Lee}},\ }\href {https://doi.org/10.48550/arXiv.2606.05664} {\bibinfo {title}
  {Gauging the spacetime code}} (\bibinfo {year} {2026}),\ \Eprint
  {https://arxiv.org/abs/2606.05664} {arXiv:2606.05664} \BibitemShut {NoStop}%
\bibitem [{\citenamefont {Cowtan}\ \emph {et~al.}(2025)\citenamefont {Cowtan},
  \citenamefont {He}, \citenamefont {Williamson},\ and\ \citenamefont
  {Yoder}}]{logicalmeasurements2026}%
  \BibitemOpen
  \bibfield  {author} {\bibinfo {author} {\bibfnamefont {A.}~\bibnamefont
  {Cowtan}}, \bibinfo {author} {\bibfnamefont {Z.}~\bibnamefont {He}}, \bibinfo
  {author} {\bibfnamefont {D.~J.}\ \bibnamefont {Williamson}},\ and\ \bibinfo
  {author} {\bibfnamefont {T.~J.}\ \bibnamefont {Yoder}},\ }\href
  {https://doi.org/10.48550/arXiv.2510.14895} {\bibinfo {title} {Fast and
  fault-tolerant logical measurements: auxiliary hypergraphs and transversal
  surgery}} (\bibinfo {year} {2025}),\ \Eprint
  {https://arxiv.org/abs/2510.14895} {arXiv:2510.14895} \BibitemShut {NoStop}%
\bibitem [{\citenamefont {Bacon}(2006)}]{bacon2006operator}%
  \BibitemOpen
  \bibfield  {author} {\bibinfo {author} {\bibfnamefont {D.}~\bibnamefont
  {Bacon}},\ }\bibfield  {title} {\bibinfo {title} {Operator quantum
  error-correcting subsystems for self-correcting quantum memories},\ }\href
  {https://doi.org/10.1103/PhysRevA.73.012340} {\bibfield  {journal} {\bibinfo
  {journal} {Phys. Rev. A}\ }\textbf {\bibinfo {volume} {73}},\ \bibinfo
  {pages} {012340} (\bibinfo {year} {2006})}\BibitemShut {NoStop}%
\bibitem [{\citenamefont {Aliferis}\ and\ \citenamefont
  {Cross}(2007)}]{aliferis2007subsystem}%
  \BibitemOpen
  \bibfield  {author} {\bibinfo {author} {\bibfnamefont {P.}~\bibnamefont
  {Aliferis}}\ and\ \bibinfo {author} {\bibfnamefont {A.~W.}\ \bibnamefont
  {Cross}},\ }\bibfield  {title} {\bibinfo {title} {Subsystem fault tolerance
  with the {B}acon-{S}hor code},\ }\href
  {https://doi.org/10.1103/PhysRevLett.98.220502} {\bibfield  {journal}
  {\bibinfo  {journal} {Phys. Rev. Lett.}\ }\textbf {\bibinfo {volume} {98}},\
  \bibinfo {pages} {220502} (\bibinfo {year} {2007})}\BibitemShut {NoStop}%
\bibitem [{\citenamefont {Dennis}\ \emph {et~al.}(2002)\citenamefont {Dennis},
  \citenamefont {Kitaev}, \citenamefont {Landahl},\ and\ \citenamefont
  {Preskill}}]{dennis2002topological}%
  \BibitemOpen
  \bibfield  {author} {\bibinfo {author} {\bibfnamefont {E.}~\bibnamefont
  {Dennis}}, \bibinfo {author} {\bibfnamefont {A.}~\bibnamefont {Kitaev}},
  \bibinfo {author} {\bibfnamefont {A.}~\bibnamefont {Landahl}},\ and\ \bibinfo
  {author} {\bibfnamefont {J.}~\bibnamefont {Preskill}},\ }\bibfield  {title}
  {\bibinfo {title} {Topological quantum memory},\ }\href
  {https://doi.org/10.1063/1.1499754} {\bibfield  {journal} {\bibinfo
  {journal} {J. Math. Phys.}\ }\textbf {\bibinfo {volume} {43}},\ \bibinfo
  {pages} {4452} (\bibinfo {year} {2002})}\BibitemShut {NoStop}%
\bibitem [{\citenamefont {Bravyi}\ and\ \citenamefont
  {Hastings}(2014)}]{bravyi2014homological}%
  \BibitemOpen
  \bibfield  {author} {\bibinfo {author} {\bibfnamefont {S.}~\bibnamefont
  {Bravyi}}\ and\ \bibinfo {author} {\bibfnamefont {M.~B.}\ \bibnamefont
  {Hastings}},\ }\bibfield  {title} {\bibinfo {title} {Homological product
  codes},\ }in\ \href {https://doi.org/10.1145/2591796.2591870} {\emph
  {\bibinfo {booktitle} {Proceedings of the 46th Annual ACM Symposium on Theory
  of Computing (STOC)}}}\ (\bibinfo {year} {2014})\ pp.\ \bibinfo {pages}
  {273--282}\BibitemShut {NoStop}%
\bibitem [{\citenamefont {Ehatamm}\ \emph {et~al.}(2026)\citenamefont
  {Ehatamm}, \citenamefont {Lee}, \citenamefont {Wu},\ and\ \citenamefont
  {Tao}}]{ehatamm2026leanqec}%
  \BibitemOpen
  \bibfield  {author} {\bibinfo {author} {\bibfnamefont {M.}~\bibnamefont
  {Ehatamm}}, \bibinfo {author} {\bibfnamefont {Y.}~\bibnamefont {Lee}},
  \bibinfo {author} {\bibfnamefont {X.}~\bibnamefont {Wu}},\ and\ \bibinfo
  {author} {\bibfnamefont {R.}~\bibnamefont {Tao}},\ }\href
  {https://doi.org/10.48550/arXiv.2605.16523} {\bibinfo {title} {End-to-end
  formalization of quantum error correction}} (\bibinfo {year} {2026}),\
  \bibinfo {note} {code: \url{https://github.com/VerifiedQC/Lean-QEC}},\
  \Eprint {https://arxiv.org/abs/2605.16523} {arXiv:2605.16523} \BibitemShut
  {NoStop}%
\bibitem [{\citenamefont {Jain}(2025)}]{jain2025qeclean}%
  \BibitemOpen
  \bibfield  {author} {\bibinfo {author} {\bibfnamefont {S.}~\bibnamefont
  {Jain}},\ }\href {https://github.com/Stavan-Jain/QECLean} {\bibinfo {title}
  {{QECLean}: Formalized quantum error correction}},\ \bibinfo {howpublished}
  {GitHub repository} (\bibinfo {year} {2025}),\ \bibinfo {note} {apache-2.0
  license}\BibitemShut {NoStop}%
\bibitem [{\citenamefont {Hillmann}\ \emph {et~al.}(2025)\citenamefont
  {Hillmann}, \citenamefont {Dauphinais}, \citenamefont {Tzitrin},\ and\
  \citenamefont {Vasmer}}]{hillmann2025faultcomplexes}%
  \BibitemOpen
  \bibfield  {author} {\bibinfo {author} {\bibfnamefont {T.}~\bibnamefont
  {Hillmann}}, \bibinfo {author} {\bibfnamefont {G.}~\bibnamefont
  {Dauphinais}}, \bibinfo {author} {\bibfnamefont {I.}~\bibnamefont
  {Tzitrin}},\ and\ \bibinfo {author} {\bibfnamefont {M.}~\bibnamefont
  {Vasmer}},\ }\bibfield  {title} {\bibinfo {title} {Single-shot and
  measurement-based quantum error correction via fault complexes},\ }\href
  {https://doi.org/10.1103/cjb4-l57n} {\bibfield  {journal} {\bibinfo
  {journal} {Phys. Rev. A}\ }\textbf {\bibinfo {volume} {112}},\ \bibinfo
  {pages} {L040401} (\bibinfo {year} {2025})}\BibitemShut {NoStop}%
\bibitem [{\citenamefont {Yuan}\ \emph {et~al.}(2026)\citenamefont {Yuan},
  \citenamefont {Cowtan}, \citenamefont {He}, \citenamefont {Lin},\ and\
  \citenamefont {Williamson}}]{yuan2026parsimonious}%
  \BibitemOpen
  \bibfield  {author} {\bibinfo {author} {\bibfnamefont {A.~C.}\ \bibnamefont
  {Yuan}}, \bibinfo {author} {\bibfnamefont {A.}~\bibnamefont {Cowtan}},
  \bibinfo {author} {\bibfnamefont {Z.}~\bibnamefont {He}}, \bibinfo {author}
  {\bibfnamefont {T.-C.}\ \bibnamefont {Lin}},\ and\ \bibinfo {author}
  {\bibfnamefont {D.~J.}\ \bibnamefont {Williamson}},\ }\href
  {https://doi.org/10.48550/arXiv.2603.05082} {\bibinfo {title} {Parsimonious
  quantum low-density parity-check code surgery}} (\bibinfo {year} {2026}),\
  \Eprint {https://arxiv.org/abs/2603.05082} {arXiv:2603.05082} \BibitemShut
  {NoStop}%
\bibitem [{\citenamefont {Cowtan}\ \emph {et~al.}(2026)\citenamefont {Cowtan},
  \citenamefont {He}, \citenamefont {Williamson},\ and\ \citenamefont
  {Yoder}}]{cowtan2026parallel}%
  \BibitemOpen
  \bibfield  {author} {\bibinfo {author} {\bibfnamefont {A.}~\bibnamefont
  {Cowtan}}, \bibinfo {author} {\bibfnamefont {Z.}~\bibnamefont {He}}, \bibinfo
  {author} {\bibfnamefont {D.~J.}\ \bibnamefont {Williamson}},\ and\ \bibinfo
  {author} {\bibfnamefont {T.~J.}\ \bibnamefont {Yoder}},\ }\bibfield  {title}
  {\bibinfo {title} {Parallel logical measurements via quantum code surgery},\
  }\href {https://doi.org/10.1103/gj8x-n5gg} {\bibfield  {journal} {\bibinfo
  {journal} {PRX Quantum}\ }\textbf {\bibinfo {volume} {7}},\ \bibinfo {pages}
  {020325} (\bibinfo {year} {2026})}\BibitemShut {NoStop}%
\bibitem [{\citenamefont {Swaroop}\ \emph {et~al.}(2026)\citenamefont
  {Swaroop}, \citenamefont {Jochym-O'Connor},\ and\ \citenamefont
  {Yoder}}]{swaroop2026adapters}%
  \BibitemOpen
  \bibfield  {author} {\bibinfo {author} {\bibfnamefont {E.}~\bibnamefont
  {Swaroop}}, \bibinfo {author} {\bibfnamefont {T.}~\bibnamefont
  {Jochym-O'Connor}},\ and\ \bibinfo {author} {\bibfnamefont {T.~J.}\
  \bibnamefont {Yoder}},\ }\bibfield  {title} {\bibinfo {title} {Universal
  adapters between quantum low-density parity check codes},\ }\href
  {https://doi.org/10.1103/1g44-jp62} {\bibfield  {journal} {\bibinfo
  {journal} {PRX Quantum}\ }\textbf {\bibinfo {volume} {7}},\ \bibinfo {pages}
  {010324} (\bibinfo {year} {2026})}\BibitemShut {NoStop}%
\bibitem [{\citenamefont {Williamson}(2026)}]{williamson2026higherform}%
  \BibitemOpen
  \bibfield  {author} {\bibinfo {author} {\bibfnamefont {D.~J.}\ \bibnamefont
  {Williamson}},\ }\href {https://doi.org/10.48550/arXiv.2601.22939} {\bibinfo
  {title} {Fast magic state preparation by gauging higher-form transversal
  gates in parallel}} (\bibinfo {year} {2026}),\ \Eprint
  {https://arxiv.org/abs/2601.22939} {arXiv:2601.22939} \BibitemShut {NoStop}%
\bibitem [{\citenamefont {Delfosse}\ and\ \citenamefont
  {Paetznick}(2023)}]{delfosse2023spacetime}%
  \BibitemOpen
  \bibfield  {author} {\bibinfo {author} {\bibfnamefont {N.}~\bibnamefont
  {Delfosse}}\ and\ \bibinfo {author} {\bibfnamefont {A.}~\bibnamefont
  {Paetznick}},\ }\href {https://doi.org/10.48550/arXiv.2304.05943} {\bibinfo
  {title} {Spacetime codes of {C}lifford circuits}} (\bibinfo {year} {2023}),\
  \Eprint {https://arxiv.org/abs/2304.05943} {arXiv:2304.05943} \BibitemShut
  {NoStop}%
\bibitem [{\citenamefont {Bravyi}\ and\ \citenamefont
  {Kitaev}(1998)}]{bravyi1998quantum}%
  \BibitemOpen
  \bibfield  {author} {\bibinfo {author} {\bibfnamefont {S.~B.}\ \bibnamefont
  {Bravyi}}\ and\ \bibinfo {author} {\bibfnamefont {A.~Y.}\ \bibnamefont
  {Kitaev}},\ }\href {https://doi.org/10.48550/arXiv.quant-ph/9811052}
  {\bibinfo {title} {Quantum codes on a lattice with boundary}} (\bibinfo
  {year} {1998}),\ \Eprint {https://arxiv.org/abs/quant-ph/9811052}
  {arXiv:quant-ph/9811052} \BibitemShut {NoStop}%
\bibitem [{\citenamefont {Fowler}\ \emph {et~al.}(2012)\citenamefont {Fowler},
  \citenamefont {Mariantoni}, \citenamefont {Martinis},\ and\ \citenamefont
  {Cleland}}]{fowler2012surface}%
  \BibitemOpen
  \bibfield  {author} {\bibinfo {author} {\bibfnamefont {A.~G.}\ \bibnamefont
  {Fowler}}, \bibinfo {author} {\bibfnamefont {M.}~\bibnamefont {Mariantoni}},
  \bibinfo {author} {\bibfnamefont {J.~M.}\ \bibnamefont {Martinis}},\ and\
  \bibinfo {author} {\bibfnamefont {A.~N.}\ \bibnamefont {Cleland}},\
  }\bibfield  {title} {\bibinfo {title} {Surface codes: Towards practical
  large-scale quantum computation},\ }\href
  {https://doi.org/10.1103/PhysRevA.86.032324} {\bibfield  {journal} {\bibinfo
  {journal} {Phys. Rev. A}\ }\textbf {\bibinfo {volume} {86}},\ \bibinfo
  {pages} {032324} (\bibinfo {year} {2012})}\BibitemShut {NoStop}%
\bibitem [{\citenamefont {Bolt}\ \emph {et~al.}(2016)\citenamefont {Bolt},
  \citenamefont {Duclos-Cianci}, \citenamefont {Poulin},\ and\ \citenamefont
  {Stace}}]{bolt2016foliated}%
  \BibitemOpen
  \bibfield  {author} {\bibinfo {author} {\bibfnamefont {A.}~\bibnamefont
  {Bolt}}, \bibinfo {author} {\bibfnamefont {G.}~\bibnamefont {Duclos-Cianci}},
  \bibinfo {author} {\bibfnamefont {D.}~\bibnamefont {Poulin}},\ and\ \bibinfo
  {author} {\bibfnamefont {T.~M.}\ \bibnamefont {Stace}},\ }\bibfield  {title}
  {\bibinfo {title} {Foliated quantum error-correcting codes},\ }\href
  {https://doi.org/10.1103/PhysRevLett.117.070501} {\bibfield  {journal}
  {\bibinfo  {journal} {Phys. Rev. Lett.}\ }\textbf {\bibinfo {volume} {117}},\
  \bibinfo {pages} {070501} (\bibinfo {year} {2016})}\BibitemShut {NoStop}%
\bibitem [{\citenamefont {Xu}\ \emph {et~al.}(2026)\citenamefont {Xu},
  \citenamefont {Wang},\ and\ \citenamefont {Liu}}]{xu2026lifting}%
  \BibitemOpen
  \bibfield  {author} {\bibinfo {author} {\bibfnamefont {Y.}~\bibnamefont
  {Xu}}, \bibinfo {author} {\bibfnamefont {Y.}~\bibnamefont {Wang}},\ and\
  \bibinfo {author} {\bibfnamefont {Z.-W.}\ \bibnamefont {Liu}},\ }\href
  {https://doi.org/10.48550/arXiv.2606.06365} {\bibinfo {title} {A framework
  for low-overhead quantum fault tolerance via spacetime lifting}} (\bibinfo
  {year} {2026}),\ \Eprint {https://arxiv.org/abs/2606.06365}
  {arXiv:2606.06365} \BibitemShut {NoStop}%
\bibitem [{\citenamefont {Het\'enyi}\ \emph {et~al.}(2026)\citenamefont
  {Het\'enyi}, \citenamefont {Brown},\ and\ \citenamefont
  {Williamson}}]{hetenyi2026cultivation}%
  \BibitemOpen
  \bibfield  {author} {\bibinfo {author} {\bibfnamefont {B.}~\bibnamefont
  {Het\'enyi}}, \bibinfo {author} {\bibfnamefont {B.~J.}\ \bibnamefont
  {Brown}},\ and\ \bibinfo {author} {\bibfnamefont {D.~J.}\ \bibnamefont
  {Williamson}},\ }\bibfield  {title} {\bibinfo {title} {Constant-depth magic
  state cultivation with {C}lifford measurements by gauging},\ }\href
  {https://doi.org/10.1103/f7fz-6zsy} {\bibfield  {journal} {\bibinfo
  {journal} {Phys. Rev. Lett.}\ }\textbf {\bibinfo {volume} {137}},\ \bibinfo
  {pages} {140601} (\bibinfo {year} {2026})}\BibitemShut {NoStop}%
\bibitem [{\citenamefont {Evra}\ \emph {et~al.}(2024)\citenamefont {Evra},
  \citenamefont {Kaufman},\ and\ \citenamefont {Z\'emor}}]{evra2022decodable}%
  \BibitemOpen
  \bibfield  {author} {\bibinfo {author} {\bibfnamefont {S.}~\bibnamefont
  {Evra}}, \bibinfo {author} {\bibfnamefont {T.}~\bibnamefont {Kaufman}},\ and\
  \bibinfo {author} {\bibfnamefont {G.}~\bibnamefont {Z\'emor}},\ }\bibfield
  {title} {\bibinfo {title} {Decodable quantum {LDPC} codes beyond the
  $\sqrt{n}$ distance barrier using high-dimensional expanders},\ }\href
  {https://doi.org/10.1137/20M1383689} {\bibfield  {journal} {\bibinfo
  {journal} {SIAM J. Comput.}\ }\textbf {\bibinfo {volume} {53}},\ \bibinfo
  {eid} {FOCS20-276} (\bibinfo {year} {2024})}\BibitemShut {NoStop}%
\bibitem [{\citenamefont {Kaufman}\ and\ \citenamefont
  {Oppenheim}(2020)}]{kaufman2020highorder}%
  \BibitemOpen
  \bibfield  {author} {\bibinfo {author} {\bibfnamefont {T.}~\bibnamefont
  {Kaufman}}\ and\ \bibinfo {author} {\bibfnamefont {I.}~\bibnamefont
  {Oppenheim}},\ }\bibfield  {title} {\bibinfo {title} {High order random
  walks: Beyond spectral gap},\ }\href
  {https://doi.org/10.1007/s00493-019-3847-0} {\bibfield  {journal} {\bibinfo
  {journal} {Combinatorica}\ }\textbf {\bibinfo {volume} {40}},\ \bibinfo
  {pages} {245} (\bibinfo {year} {2020})}\BibitemShut {NoStop}%
\bibitem [{\citenamefont {Lubotzky}(2019)}]{lubotzky2017expanders}%
  \BibitemOpen
  \bibfield  {author} {\bibinfo {author} {\bibfnamefont {A.}~\bibnamefont
  {Lubotzky}},\ }\bibfield  {title} {\bibinfo {title} {High dimensional
  expanders},\ }in\ \href {https://doi.org/10.1142/9789813272880_0027} {\emph
  {\bibinfo {booktitle} {Proceedings of the International Congress of
  Mathematicians (ICM 2018)}}}\ (\bibinfo  {publisher} {World Scientific},\
  \bibinfo {year} {2019})\ pp.\ \bibinfo {pages} {705--730}\BibitemShut
  {NoStop}%
\bibitem [{\citenamefont {Dikstein}\ and\ \citenamefont
  {Dinur}(2024)}]{dikstein2024coboundary}%
  \BibitemOpen
  \bibfield  {author} {\bibinfo {author} {\bibfnamefont {Y.}~\bibnamefont
  {Dikstein}}\ and\ \bibinfo {author} {\bibfnamefont {I.}~\bibnamefont
  {Dinur}},\ }\bibfield  {title} {\bibinfo {title} {Coboundary and cosystolic
  expansion without dependence on dimension or degree},\ }in\ \href
  {https://doi.org/10.4230/LIPIcs.APPROX/RANDOM.2024.62} {\emph {\bibinfo
  {booktitle} {Approximation, Randomization, and Combinatorial Optimization.
  Algorithms and Techniques (APPROX/RANDOM 2024)}}},\ \bibinfo {series}
  {LIPIcs}, Vol.\ \bibinfo {volume} {317}\ (\bibinfo  {publisher} {Schloss
  Dagstuhl -- Leibniz-Zentrum f{\"u}r Informatik},\ \bibinfo {year} {2024})\
  pp.\ \bibinfo {pages} {62:1--62:24}\BibitemShut {NoStop}%
\bibitem [{\citenamefont {Vardy}(1997)}]{vardy1997intractability}%
  \BibitemOpen
  \bibfield  {author} {\bibinfo {author} {\bibfnamefont {A.}~\bibnamefont
  {Vardy}},\ }\bibfield  {title} {\bibinfo {title} {The intractability of
  computing the minimum distance of a code},\ }\href
  {https://doi.org/10.1109/18.641542} {\bibfield  {journal} {\bibinfo
  {journal} {IEEE Trans. Inf. Theory}\ }\textbf {\bibinfo {volume} {43}},\
  \bibinfo {pages} {1757} (\bibinfo {year} {1997})}\BibitemShut {NoStop}%
\bibitem [{\citenamefont {Wu}\ \emph {et~al.}(2026)\citenamefont {Wu},
  \citenamefont {Zhong}, \citenamefont {Brun},\ and\ \citenamefont
  {Lidar}}]{wu2026switch}%
  \BibitemOpen
  \bibfield  {author} {\bibinfo {author} {\bibfnamefont {S.}~\bibnamefont
  {Wu}}, \bibinfo {author} {\bibfnamefont {D.}~\bibnamefont {Zhong}}, \bibinfo
  {author} {\bibfnamefont {T.~A.}\ \bibnamefont {Brun}},\ and\ \bibinfo
  {author} {\bibfnamefont {D.~A.}\ \bibnamefont {Lidar}},\ }\bibfield  {title}
  {\bibinfo {title} {Universal weakly fault-tolerant quantum computation via
  code switching in the $[[8,3,2]]$ code},\ }\href
  {https://doi.org/10.1103/4gwn-6cnx} {\bibfield  {journal} {\bibinfo
  {journal} {Phys. Rev. Research}\ }\textbf {\bibinfo {volume} {8}},\ \bibinfo
  {pages} {033380} (\bibinfo {year} {2026})}\BibitemShut {NoStop}%
\bibitem [{\citenamefont {Cross}\ \emph {et~al.}(2024)\citenamefont {Cross},
  \citenamefont {He}, \citenamefont {Rall},\ and\ \citenamefont
  {Yoder}}]{cross2024improved}%
  \BibitemOpen
  \bibfield  {author} {\bibinfo {author} {\bibfnamefont {A.~W.}\ \bibnamefont
  {Cross}}, \bibinfo {author} {\bibfnamefont {Z.}~\bibnamefont {He}}, \bibinfo
  {author} {\bibfnamefont {P.~J.}\ \bibnamefont {Rall}},\ and\ \bibinfo
  {author} {\bibfnamefont {T.~J.}\ \bibnamefont {Yoder}},\ }\href
  {https://doi.org/10.48550/arXiv.2407.18393} {\bibinfo {title} {Improved
  {QLDPC} surgery: logical measurements and bridging codes}} (\bibinfo {year}
  {2024}),\ \Eprint {https://arxiv.org/abs/2407.18393} {arXiv:2407.18393}
  \BibitemShut {NoStop}%
\bibitem [{\citenamefont {Ide}\ \emph {et~al.}(2025)\citenamefont {Ide},
  \citenamefont {Gowda}, \citenamefont {Nadkarni},\ and\ \citenamefont
  {Dauphinais}}]{ide2025homological}%
  \BibitemOpen
  \bibfield  {author} {\bibinfo {author} {\bibfnamefont {B.}~\bibnamefont
  {Ide}}, \bibinfo {author} {\bibfnamefont {M.~G.}\ \bibnamefont {Gowda}},
  \bibinfo {author} {\bibfnamefont {P.~J.}\ \bibnamefont {Nadkarni}},\ and\
  \bibinfo {author} {\bibfnamefont {G.}~\bibnamefont {Dauphinais}},\ }\bibfield
   {title} {\bibinfo {title} {Fault-tolerant logical measurements via
  homological measurement},\ }\href
  {https://doi.org/10.1103/PhysRevX.15.021088} {\bibfield  {journal} {\bibinfo
  {journal} {Phys. Rev. X}\ }\textbf {\bibinfo {volume} {15}},\ \bibinfo
  {pages} {021088} (\bibinfo {year} {2025})}\BibitemShut {NoStop}%
\bibitem [{\citenamefont {Berent}\ \emph {et~al.}(2026)\citenamefont {Berent},
  \citenamefont {Cohen},\ and\ \citenamefont {Quintavalle}}]{berent2026lifted}%
  \BibitemOpen
  \bibfield  {author} {\bibinfo {author} {\bibfnamefont {L.}~\bibnamefont
  {Berent}}, \bibinfo {author} {\bibfnamefont {L.~Z.}\ \bibnamefont {Cohen}},\
  and\ \bibinfo {author} {\bibfnamefont {A.~O.}\ \bibnamefont {Quintavalle}},\
  }\href {https://doi.org/10.48550/arXiv.2609.11723} {\bibinfo {title} {Lifted
  surgery: Fast processing with qldpc codes}} (\bibinfo {year} {2026}),\
  \Eprint {https://arxiv.org/abs/2609.11723} {arXiv:2609.11723} \BibitemShut
  {NoStop}%
\end{thebibliography}%

\clearpage
\setcounter{equation}{0}
\setcounter{table}{0}
\setcounter{figure}{0}

\section*{Supplemental material: Exact distances and measurement allocation for logical Pauli readout}
\label{sec:sm}
\newcounter{smsection}
\renewcommand{\thesmsection}{S\arabic{smsection}}
\renewcommand{\thetable}{S\arabic{table}}
\renewcommand{\thefigure}{S\arabic{figure}}
\renewcommand{\theequation}{S\arabic{equation}}

\refstepcounter{smsection}\section*{\thesmsection\quad How to read this supplement}

All references cited here are listed in the bibliography of the main article.

The first proof section establishes the general readout formula, endpoint bounds, measurement allocations and circuit scheduling bound. The next section specifies the circuit fault model, full best-tested schedules and explicit accepted-error witnesses.
The remaining sections identify the Lean statements supporting the algebraic and finite results, give the cosystolic-distance and surgery analysis, and present step-by-step protocol derivations.
The final sections state the conventions, report the axiom audit and provide reproduction commands.

\refstepcounter{smsection}\section*{\thesmsection\quad Readout-distance and allocation proofs}
\label{sec:sm-readout-allocation}

This section proves the weighted repeated-readout formula, the selected-endpoint bound, the Shor-family check allocation and the pair-product results used in the main text. These statements concern the fault models defined there. The general matrix formulas are proved directly here; the separate Lean project certifies arithmetic lemmas used in the resource bound, not the full matrix theorems.

\subsection{Exact distance for a threshold of wrong readouts}

Let $A=\sum_t\operatorname{wt}(d_t)$ be the total persistent data-fault cost and let $b_t=\operatorname{wt}(\mu_t)+\operatorname{wt}(\gamma_t)$ be the transient cost in round $t$. Consider a detector-free fault pattern with at least $q$ wrong readouts. Among those rounds choose $t$ with minimum $b_t$, so $q b_t\le\sum_s b_s$. Set $y=x_t+\mu_t$. The zero-syndrome condition gives $Hy=\gamma_t$, and the wrong readout gives $w\cdot y=1$. Also, $\operatorname{wt}(y)\le\operatorname{wt}(x_t)+\operatorname{wt}(\mu_t)\le A+\operatorname{wt}(\mu_t)$. Therefore, by the definition of $\Gamma_q$, we have $\Gamma_q(H,w)\le\operatorname{wt}(y)+q\operatorname{wt}(Hy)\le A+\operatorname{wt}(\mu_t)+q\operatorname{wt}(\gamma_t)\le A+q b_t\le A+\sum_s b_s$, where the penultimate inequality uses $q\ge1$.
This is the lower bound on total fault weight. For the matching upper bound, choose $y$ minimizing $\operatorname{wt}(y)+q\operatorname{wt}(Hy)$ subject to $w\cdot y=1$, introduce one data error $d_t=y$ at the start of the final $q$-round block, and set $\gamma_t=Hy$ in those rounds. Set all other faults to zero. Then exactly those $q$ readouts are wrong, every syndrome is zero, and the fault weight is $\operatorname{wt}(y)+q\operatorname{wt}(Hy)$. This proves Eq.~(\ref*{eq:weighted-readout-distance}) for all binary $H$, nonzero $w$, and $1\le q\le T$.

\subsection{Selected endpoint checks}

Suppose adjacent comparisons constrain the full syndrome, while one endpoint checks only $B\sigma_t=0$. Since the comparisons make $\sigma_t$ constant, the accepted syndrome lies in $\ker B$. Let $\delta_B$ be the minimum weight of a data vector $y$ satisfying $BHy=0$ and $w\cdot y=1$, and assume this set is nonempty. An initial data error attaining $\delta_B$ has a constant syndrome in $\ker B$ and flips all $T$ readouts, so $d_{\mathrm{maj}}^B(T)\le\delta_B$.

For the lower bound, suppose a detector-free fault has total weight $K<\min\{\delta_B,q+1\}$ and at least $q$ wrong readouts. If it contains a data fault, then every wrong round must also contain at least one transient fault. Indeed, a wrong round with $\mu_t=\gamma_t=0$ would have $B H x_t=0$, $w\cdot x_t=1$, and $\operatorname{wt}(x_t)\le K<\delta_B$, a contradiction. The data fault and the required transient faults in $q$ distinct rounds would give $K\ge q+1$, also a contradiction.

If there are no data faults, then a wrong round cannot be produced by a single check-result fault, since $\gamma_t$ does not enter $r_t$. A single readout fault would have to satisfy $BH\mu_t=0$ and $w\cdot\mu_t=1$, which would imply $\delta_B=1$; this is incompatible with $K<\delta_B$ for any nonzero fault. Thus every wrong round costs at least two transient faults. The $q$ wrong rounds then cost at least $2q\ge q+1>K$. This contradiction proves $d_{\mathrm{maj}}^B(T)\ge\min\{\delta_B,q+1\}$. The equality when $q\ge\delta_B-1$ follows by combining the two bounds. For a target $d$, the condition $\delta_B\ge d$ is necessary; together with $q\ge d-1$, it is sufficient.

The lower bound need not be tight. At $T=q=1$, the Shor allocation $c_j=m-j$ with $B=I$ has $\delta_B=m$ and exact distance $m$, whereas the bound $\min\{\delta_B,q+1\}$ is only two for $m>2$. The exact value follows from the allocation formula proved next.

\subsection{Adjacent-column check allocation}

For the square Shor-family readout with $m\ge2$, write $v_j$ for the parity of the data-error bits in column $j$. The logical readout error is $v_0$, and check $j$ has syndrome $v_{j-1}+v_j$. If this check is measured in $c_j$ independent copies per round, its row appears $c_j$ times in $H_c$, with an independent check-result fault bit for each copy.

Consider any data error with $v_0=1$. If every $v_j=1$, at least one data bit is needed in each of the $m$ columns, so its weight is at least $m$. Otherwise, let $j$ be the first column with $v_j=0$. The first $j$ column parities are odd, requiring at least $j$ data bits, and check $j$ has nonzero syndrome in all $c_j$ copies. To hide this syndrome in at least $q$ wrong rounds requires at least $q c_j$ check-result faults. Hence the weighted cost is at least $j+q c_j$. Conversely, one data error in each of the first $j$ columns and none in later columns makes check $j$ the only nonzero adjacent-column syndrome; canceling its $c_j$ copies in the final $q$ rounds attains $j+q c_j$. Taking the least of these witnesses and the all-odd-column witness proves Eq.~(\ref*{eq:path-check-allocation}).

The condition that this distance be at least $m$ is exactly $j+q c_j\ge m$ for every $j$. Since $c_j$ must be a positive integer, the componentwise smallest allocation is $c_j=\lceil(m-j)/q\rceil$. Summing over $j$ gives $\sum_{s=1}^{m-1}\lceil s/q\rceil$ check-result locations per round. This is optimal only within the family that repeats these fixed adjacent-column checks.

\subsection{Pair products and their distance}

Let $K$ be a simple graph on the $m$ Gauss-check labels. For a set $S$ of original outcomes flipped in one round, consistency of the pair records forces exactly the outcomes on the cut $\partial_K S$ to flip as well. The logical product is wrong if and only if $|S|$ is odd. Thus the one-round distance is the minimum of $|S|+|\partial_K S|$ over odd sets.

Write $r=\min_v\deg_K(v)$ and $k=|S|$. The singleton consisting of a minimum-degree vertex has cost $r+1$. For the converse, when $k\le r+1$, the sum of degrees in $S$ is at least $kr$, while at most $k(k-1)$ of this sum can come from internal edges. Therefore $|\partial_K S|\ge kr-k(k-1)$ and the same count gives $|S|+|\partial_K S|\ge k(r-k+2)\ge r+1$,
where the last inequality follows from $(k-1)(r+1-k)\ge0$. When $k\ge r+1$, the $k$ original outcome flips already cost at least $r+1$. This proves Eq.~(\ref*{eq:pair-product-distance}).

For target distance $\delta\le m$, the graph must have minimum degree at least $\delta-1$, so the degree sum gives $|E(K)|\ge\lceil m(\delta-1)/2\rceil$. This lower bound is attainable. Put $r=\delta-1$. If $mr$ is even, an $r$-regular circulant graph exists: connect each vertex to its $r/2$ nearest neighbors on each side when $r$ is even, and add the antipodal matching when $r$ is odd and $m$ is even. If both $m$ and $r$ are odd, first form the $(r+1)$-regular circulant graph on the odd cycle and remove a matching covering all but one vertex; the unmatched vertex has degree $r+1$ and all others have degree $r$. The result has the required minimum degree and exactly $\lceil mr/2\rceil$ edges.

Under repeated adjacent comparisons, a detector-free encoded outcome must be the same $R_K z$ in every round. If $\mathbf1\cdot z=1$, each round costs at least $\delta_{R_K}=\min_{\mathbf1\cdot u=1}\operatorname{wt}(R_Ku)$, and repeating a minimum-weight odd $u$ attains the bound. The measurement-only distance is therefore $T\delta_{R_K}=T d_{\mathrm{pair}}(K)$.

\subsection{Joint gauged-readout distance}
\label{sec:sm-joint-distance}

This subsection proves Proposition~\ref*{prop:joint-readout} of the main text.

Write $H=H_c$ for the original-check matrix with its repeated rows.

Use $A$ for the total number of persistent data and edge-ancilla fault bits, and let $k_t=\operatorname{wt}(\eta_t)+\operatorname{wt}(\nu_t)$ be the result-fault cost in round $t$. Consider a detector-free pattern with odd output vector $z$. In a round with $w\cdot x_t=1$, the original-check constraint gives $Hx_t=\eta_t$, so the repeated-readout formula implies $\Gamma_T(H,w)\le\operatorname{wt}(x_t)+T\operatorname{wt}(\eta_t)\le A+T k_t$.

In a round with $w\cdot x_t=0$, the vector $u=z+Px_t+J_Ge_t$ has odd parity: $\mathbf1\cdot z=1$, $\mathbf1^{\mathsf T}P x_t=w\cdot x_t=0$, and $\mathbf1^{\mathsf T}J_G=0$. Since $b_t=R_K(Px_t+J_Ge_t)+\nu_t=R_Kz$, we have $\nu_t=R_Ku$, and therefore $\operatorname{wt}(\nu_t)\ge\delta_{R_K}$. In either case, $\min\{T\delta_{R_K},\Gamma_T(H,w)\}\le A+T k_t$. Averaging over all $T$ rounds gives the lower bound by the total fault weight $A+\sum_t k_t$.

Both terms are attainable. For $T\delta_{R_K}$, repeat in every round an outcome-fault vector $R_Ku$ attaining $\delta_{R_K}$ for odd $u$, with no data faults. For $\Gamma_T(H,w)$, introduce a minimizing data error $y$ in the first round, set $\eta_t=Hy$ in every round, and set the edge and encoded Gauss-result faults to zero. The output is then the constant odd vector $Py$. These witnesses prove Eq.~(\ref*{eq:joint-gauged-readout-distance}).

\subsection{Outcome-count optimum}
\label{sec:sm-outcome-count}

This subsection proves Corollary~\ref*{cor:outcome-count} of the main text within the stated measurement family.

At $m=1$, there is one Gauss check and no original or pair checks, so target distance one is reached at a cost of $T$ outcomes. We henceforth take $m\ge2$.

For target distance $m$ and $1\le T\le m$, the Gauss-result code must have $\delta_{R_K}\ge\lceil m/T\rceil$. The graph bound above shows that the fewest pair outcomes are obtained with $\lceil m(\lceil m/T\rceil-1)/2\rceil$ edges. The path formula shows that the fewest adjacent-column check outcomes per round are obtained with $c_j=\lceil(m-j)/T\rceil$, whose sum is $\sum_{s=1}^{m-1}\lceil s/T\rceil$. Adding the $m$ original Gauss outcomes per round proves the exact count in Eq.~(\ref*{eq:gauged-resource-count}).

For the lower bound on this count, $T\lceil m(\lceil m/T\rceil-1)/2\rceil\ge m(m-T)/2$, and $T\sum_{s=1}^{m-1}\lceil s/T\rceil\ge m(m-1)/2$. Thus $N_T^*\ge Tm+m(m-T)/2+m(m-1)/2=m^2+m(T-1)/2$. The value $m^2$ is attained at $T=1$; for $T\ge2$ the lower bound is strictly larger. At $T=m$, the construction has no pair-product edges, one copy of each adjacent-column check, and $m(2m-1)$ total outcomes.

These outcome counts do not include every gate, ancilla, deformed check or surrounding error-correction round. A commuting $X$-type check can be measured with a $|+\rangle$ ancilla, controlled CNOTs to its support, and an $X$ measurement of the ancilla. Measuring a pair product adds a consistency result without changing the joint eigenspaces already fixed by the individual commuting Gauss checks. State-vector and projector comparisons verified this ideal measurement map for the tested four-cycle examples.

For $T>m$, the required $m$ original Gauss outcomes per round give a count at least $Tm>m^2$ before any other checks are added.
Together with the bound for $1\le T\le m$, this proves the one-round optimum over all positive round counts within the specified family.

\subsection{Circuit scheduling bound}
\label{sec:sm-circuit-bound}

We use exactly the one-epoch circuit family and detectors defined in main-text Section~\ref*{sec:circuit-bound}.
All extraction ancillas are controls, all original and edge qubits are targets, and each check uses a distinct ancilla prepared in $|+\rangle$ and measured in $X$.
There is one CNOT for each nontrivial check-support entry.
For a readout-support qubit $q_v$, its incident checks are $A_v$, the pair products indexed by edges of $K$ incident to $v$, and the original $X$ checks whose supports contain $q_v$.

Fix an eligible CNOT $g$ on $q_v$ at or before its $A_v$ interaction.
Insert $Z$ on $q_v$ immediately after $g$.
If $g$ belongs to $A_v$ or a pair-product check, include $Z$ on its control at the same gate-fault location; if $g$ belongs to an original check, leave the control unchanged.
Under a later CNOT with control $a$ and target $q_v$, conjugation maps $Z_{q_v}$ to $Z_a Z_{q_v}$.
The accumulated $Z_a$ flips that ancilla's final $X$ result, and it commutes with the ancilla's other CNOTs.
The fault therefore flips every subsequent check interaction on $q_v$.
If $g$ precedes the Gauss interaction, that later interaction flips $A_v$; if $g$ is the Gauss interaction itself, the control $Z$ flips $A_v$.
No other original Gauss outcome flips.

For each pair product incident to $v$, an interaction after $g$ flips its result through propagation.
If $g$ is itself a pair interaction, the control $Z$ flips that result.
The only missing pair flips are thus the $\beta_v(g)$ pair interactions strictly before $g$.
Insert an independent outcome flip at each of those measurements.
The Gauss and pair record is then exactly $R_K e_v$, so every pair-consistency detector is zero and the logical product is wrong.
The $\alpha_v(g)$ original checks strictly after $g$ are the only original-check results changed by propagation; cancel each by one outcome flip.
An original check at $g$ itself has no control $Z$ and remains correct.
All original-check detectors now vanish as well.
The pattern occupies one CNOT location and $\alpha_v(g)+\beta_v(g)$ distinct measurement locations.
Minimizing over eligible gates and support vertices proves Eq.~(\ref*{eq:circuit-schedule-bound}).

This construction is valid for arbitrary interleavings in the stated family, because each original or edge qubit is a target and each check has its own control ancilla.
It proves an upper bound, not a converse: satisfying its necessary timing inequalities does not guarantee the target distance.
In the serial original-check/Gauss/pair schedule, a Gauss interaction has no later original check and no earlier pair check.
The constructed one-location fault is accepted and wrong; with fault-free input, zero faults cannot change the ideal logical product, so the distance is exactly one.

We also searched one-epoch CNOT circuits for $m=3$ and $m=5$, including the original $X$ checks, pair products, input Pauli faults, CNOT Pauli faults and outcome flips. The generated test set contains 1,024 schedules at each size. The largest accepted-wrong readout distance found was three at both sizes. The best $m=3$ schedule therefore retains its phenomenological target distance, while the best tested $m=5$ schedule has distance three rather than five. An independent backward Pauli calculation reproduced the fault signatures and minimum-weight witnesses for the best schedules. The witnesses from Eq.~(\ref*{eq:circuit-schedule-bound}) attain weight three on both best schedules. The schedule bound applies to the entire defined circuit family, while the reported best distances refer to this finite test set. The tested fault model includes the original $X$ checks, pair products and listed Pauli and outcome faults; it leaves out the deformed $Z$ checks, cycle checks and surrounding error-correction rounds.

Finite dynamic programs test the repeated-readout, partial-endpoint and joint gauged-model formulas on 29,232, 9,198 and 4,608 instances, respectively; the general statements follow from the proofs above. The separate Lean development verifies five arithmetic lemmas in the resource analysis. The matrix theorems and circuit scheduling bound have direct proofs above; the optimum over tested schedules refers to the stated finite search.

\refstepcounter{smsection}\section*{\thesmsection\quad Circuit fault model, schedules and witnesses}
\label{sec:sm-schedule-data}

\subsection{Fault locations and detector signatures}

The numerical search uses the one-epoch circuit family in main-text Sec.~\ref*{sec:circuit-bound}, with complete graphs for both the gauging graph $G$ and the pair-product graph $K$ at $m=3,5$.
Original data qubits are indexed by $q=mi+b$ for $0\le i,b<m$, and the readout support is column $b=0$.
Edge-ancilla indices start at $m^2$ and follow lexicographic order of the unordered pairs of row labels.
Extraction-ancilla indices $a$ enumerate the measured checks: the $m$ Gauss checks first, the pair products in lexicographic order next, and then $m-j$ copies of adjacent-column check $j$ for $1\le j<m$.
All tabulated indices are zero based; a layer index is one based.

Fault locations are initial original-data or edge-qubit Pauli errors, extraction-ancilla preparation errors, Pauli errors immediately after a CNOT, and independent flips of a final ancilla outcome.
Each faulty preparation, input qubit, CNOT or outcome is one location; a two-qubit Pauli after a CNOT still counts as one location.
No idle faults, deformed $Z$ checks, cycle checks or surrounding error-correction rounds are included in this numerical model.
For the final $X$ records, the $X$ components of a Pauli commute with the backward-propagated $X$ observables.
The record signatures therefore depend only on the $Z$ components.
Control $Z$ is equivalent to a single flip of that ancilla's final outcome; target $Z$ flips all later extraction controls acting on the target; their product gives the third nonzero gate signature.
An extraction preparation error has the same relevant signature as an outcome flip or has zero signature.
Thus the input-$Z$, outcome, target-$Z$ and control-target-$ZZ$ channels cover the record effects of all allowed Pauli faults.

The detector signature consists of all original-check result errors and the pair-relation errors $a_u+a_v+p_{uv}$.
The logical signature is the parity of the original Gauss-result error vector.
An accepted wrong record has zero detector signature and logical signature one.
The minimum-weight search operates on these signatures; the general schedule bound is proved independently in Sec.~\ref{sec:sm-circuit-bound}.

\subsection{Explicit witnesses and exactness checks}
\label{sec:sm-schedule-witnesses}

For $m=3$, the three outcome faults $M_0,M_1,M_2$ flip every Gauss record.
Each pair parity remains unchanged, and the odd number of Gauss flips reverses the logical product.
This weight-three outcome witness applies to every $m=3$ schedule in the defined family, so a schedule with exact distance three also maximizes the distance over that family.
For $m=5$, the target-$Z$ fault $(1,16,15;Z_{\rm target})$ flips the later $A_3$ record, its four pair records, and original-check records 15, 17 and 18.
The target-$Z$ fault $(8,21,11;Z_{\rm target})$ flips original-check records 17 and 18, canceling those two errors.
The outcome fault $M_{15}$ cancels the remaining original-check error.
Both witnesses therefore have weight three and give a consistent odd Gauss record with no detector response.

Independent backward Pauli propagation checks every stored channel.
For either saved schedule, neither one channel nor any pair of channels has the accepted-wrong signature.
Together with the explicit witnesses, these tests establish the exact distances of the saved schedules in the stated fault model.
The bound of Proposition~\ref*{prop:circuit-bound} gives matching upper bounds on both schedules; satisfying its necessary inequalities alone would not establish either lower bound.

\subsection{Complete best-tested schedules}

The generated test set contains 1,024 valid schedules per size, obtained by decomposing the padded regular bipartite check-support incidence graph into perfect matchings.
The search uses Julia 1.12.7 with MersenneTwister seeds 1 through 1,024 at each size.
The largest exact distance within each set is three.
The saved $m=3$ schedule (seed 145) uses 12 input qubits, 9 checks, 6 CNOT layers and 39 CNOTs.
Its exact distance is three, attained by the outcome faults $M_0,M_1,M_2$.
The saved $m=5$ schedule (seed 2) uses 35 input qubits, 25 checks, 10 layers and 205 CNOTs.
Its exact distance is also three, attained by $M_{15}$ and the two target-$Z$ gate faults $(8,21,11;Z_{\rm target})$ and $(1,16,15;Z_{\rm target})$.
The schedules below list every CNOT as a target qubit with its execution layer, grouped by extraction ancilla.
This grouping specifies the full schedule without relying on the seed or its random-number generator.
The machine-readable check supports, layers and fault signatures are included in the Supplemental data and code as \texttt{gate\_schedule\_m3.json} and \texttt{gate\_schedule\_m5.json}.
Table~\ref{tab:sm-schedule-3} and Table~\ref{tab:sm-schedule-5} give all CNOT layers for the two saved schedules.

\begin{table*}[t]
\caption{Complete CNOT schedule for $m=3$. For each extraction ancilla $a$, $q@t$ means a CNOT with control $a$ and target $q$ in layer $t$. Gauss labels are $A_v$, pair labels are $A_uA_v$, and $C_j^{(c)}$ is copy $c$ of adjacent-column check $j$.}
\label{tab:sm-schedule-3}
\begin{ruledtabular}
\begin{tabular}{clp{345pt}}
$a$ & Measured check & Target qubit $q$ @ layer $t$ \\
\hline
0 & $A_0$ & $0@6$, $9@1$, $10@3$ \\
1 & $A_1$ & $3@2$, $9@4$, $11@6$ \\
2 & $A_2$ & $6@1$, $10@5$, $11@2$ \\
3 & $A_0A_1$ & $0@2$, $3@6$, $10@4$, $11@1$ \\
4 & $A_0A_2$ & $0@1$, $6@4$, $9@6$, $11@5$ \\
5 & $A_1A_2$ & $3@1$, $6@5$, $9@3$, $10@2$ \\
6 & $C_1^{(1)}$ & $0@5$, $1@6$, $3@4$, $4@3$, $6@2$, $7@1$ \\
7 & $C_1^{(2)}$ & $0@3$, $1@4$, $3@5$, $4@1$, $6@6$, $7@2$ \\
8 & $C_2^{(1)}$ & $1@5$, $2@6$, $4@4$, $5@2$, $7@3$, $8@1$ \\
\end{tabular}
\end{ruledtabular}
\end{table*}
\begin{table*}[t]
\caption{Complete CNOT schedule for $m=5$. For each extraction ancilla $a$, $q@t$ means a CNOT with control $a$ and target $q$ in layer $t$. Gauss labels are $A_v$, pair labels are $A_uA_v$, and $C_j^{(c)}$ is copy $c$ of adjacent-column check $j$.}
\label{tab:sm-schedule-5}
\begin{ruledtabular}
\begin{tabular}{clp{345pt}}
$a$ & Measured check & Target qubit $q$ @ layer $t$ \\
\hline
0 & $A_0$ & $0@10$, $25@5$, $26@8$, $27@1$, $28@4$ \\
1 & $A_1$ & $5@2$, $25@3$, $29@5$, $30@7$, $31@8$ \\
2 & $A_2$ & $10@3$, $26@5$, $29@6$, $32@1$, $33@10$ \\
3 & $A_3$ & $15@2$, $27@8$, $30@1$, $32@4$, $34@5$ \\
4 & $A_4$ & $20@10$, $28@3$, $31@6$, $33@5$, $34@1$ \\
5 & $A_0A_1$ & $0@6$, $5@8$, $26@3$, $27@10$, $28@2$, $29@1$, $30@4$, $31@7$ \\
6 & $A_0A_2$ & $0@3$, $10@8$, $25@2$, $27@9$, $28@5$, $29@10$, $32@7$, $33@4$ \\
7 & $A_0A_3$ & $0@4$, $15@5$, $25@7$, $26@9$, $28@6$, $30@2$, $32@10$, $34@3$ \\
8 & $A_0A_4$ & $0@7$, $20@1$, $25@8$, $26@6$, $27@4$, $31@9$, $33@3$, $34@10$ \\
9 & $A_1A_2$ & $5@10$, $10@2$, $25@6$, $26@7$, $30@9$, $31@3$, $32@5$, $33@1$ \\
10 & $A_1A_3$ & $5@5$, $15@3$, $25@4$, $27@6$, $29@8$, $31@10$, $32@2$, $34@7$ \\
11 & $A_1A_4$ & $5@6$, $20@8$, $25@1$, $28@7$, $29@2$, $30@3$, $33@9$, $34@4$ \\
12 & $A_2A_3$ & $10@1$, $15@6$, $26@10$, $27@7$, $29@4$, $30@8$, $33@2$, $34@9$ \\
13 & $A_2A_4$ & $10@5$, $20@7$, $26@4$, $28@10$, $29@3$, $31@2$, $32@9$, $34@8$ \\
14 & $A_3A_4$ & $15@9$, $20@2$, $27@3$, $28@1$, $30@6$, $31@5$, $32@8$, $33@7$ \\
15 & $C_1^{(1)}$ & $0@2$, $1@9$, $5@7$, $6@5$, $10@10$, $11@3$, $15@8$, $16@1$, $20@4$, $21@6$ \\
16 & $C_1^{(2)}$ & $0@5$, $1@7$, $5@3$, $6@8$, $10@9$, $11@2$, $15@1$, $16@4$, $20@6$, $21@10$ \\
17 & $C_1^{(3)}$ & $0@8$, $1@1$, $5@4$, $6@7$, $10@6$, $11@9$, $15@10$, $16@2$, $20@5$, $21@3$ \\
18 & $C_1^{(4)}$ & $0@1$, $1@6$, $5@9$, $6@2$, $10@7$, $11@10$, $15@4$, $16@5$, $20@3$, $21@8$ \\
19 & $C_2^{(1)}$ & $1@3$, $2@9$, $6@6$, $7@1$, $11@5$, $12@2$, $16@8$, $17@4$, $21@7$, $22@10$ \\
20 & $C_2^{(2)}$ & $1@10$, $2@1$, $6@9$, $7@4$, $11@6$, $12@7$, $16@3$, $17@2$, $21@5$, $22@8$ \\
21 & $C_2^{(3)}$ & $1@2$, $2@10$, $6@3$, $7@5$, $11@8$, $12@4$, $16@7$, $17@6$, $21@9$, $22@1$ \\
22 & $C_3^{(1)}$ & $2@8$, $3@9$, $7@2$, $8@3$, $12@6$, $13@7$, $17@1$, $18@10$, $22@5$, $23@4$ \\
23 & $C_3^{(2)}$ & $2@2$, $3@4$, $7@8$, $8@9$, $12@5$, $13@10$, $17@7$, $18@3$, $22@6$, $23@1$ \\
24 & $C_4^{(1)}$ & $3@6$, $4@8$, $8@2$, $9@7$, $13@3$, $14@9$, $18@5$, $19@1$, $23@10$, $24@4$ \\
\end{tabular}
\end{ruledtabular}
\end{table*}

\refstepcounter{smsection}\section*{\thesmsection\quad Setting: gauging a logical measurement}
\label{sec:setting}

\subsection{Related constructions and formal developments}

Related constructions include auxiliary-graph surgery~\cite{cross2024improved}, homological measurement~\cite{ide2025homological}, adapters between code blocks~\cite{swaroop2026adapters}, and auxiliary-hypergraph surgery~\cite{logicalmeasurements2026}.
Parallel measurements~\cite{cowtan2026parallel}, parsimonious surgery~\cite{yuan2026parsimonious} and lifted surgery~\cite{berent2026lifted} address concurrency, ancilla overhead and fast logical operations.
Gauging also supports higher-form transversal gates~\cite{williamson2026higherform} and magic-state preparation~\cite{hetenyi2026cultivation}; gauge measurements enable code switching~\cite{wu2026switch}.
Spacetime codes describe circuit faults and detectors~\cite{dennis2002topological,delfosse2023spacetime}, and gauging has been applied directly to them~\cite{gaugingspacetime2026}.
QECLean provides Lean formalizations of foundational quantum-error-correction constructions~\cite{jain2025qeclean}, while Lean-QEC formalizes stabilizer-code theory and machine-checked distance certificates~\cite{ehatamm2026leanqec}.
Our development also treats repeated readouts, persistent faults and boundary detectors.
The distance statements provide inputs to fault-tolerance analyses with specified noise models and decoders~\cite{fowler2012surface,bravyi2024highthreshold,aliferis2007subsystem}.

\subsection{Check matrices and the two components}

A logical measurement has two useful descriptions: an effective code on the original data qubits and an enlarged code containing the gauging ancillas.
We write $[[n,k,d]]$ for a code with $n$ physical qubits, $k$ logical qubits and distance $d$.
For the effective description, take a CSS code with binary check matrices $H_X,H_Z$ satisfying $H_XH_Z^{\mathsf T}=0$.
We use physical Pauli labels $d_X^{\mathrm{phys}}$ and $d_Z^{\mathrm{phys}}$ for the minimum weights in $\ker H_Z\setminus\operatorname{rowsp}H_X$ and $\ker H_X\setminus\operatorname{rowsp}H_Z$, respectively.
The formal library names these quantities $d_Z$ and $d_X$, indexing them by the kernel matrix; Section~\ref{app:conventions} records this conversion.
An $X$-type logical operator is a vector $L\in\ker H_Z$, meaning $H_ZL=0$, that cannot be written as a binary sum of the rows of $H_X$.
Conditioning on the measured eigenvalue adds $L$ to the effective $X$ checks.
The new check commutes with the $Z$ checks and is independent of the original $X$ rows, so the resulting CSS code has one fewer logical qubit.
For example, the length-three repetition code has two adjacent-pair $Z$ checks and no $X$ check.
Adding its all-ones logical vector as an $X$ check leaves no logical qubit.

The physical gauging construction realizes this measurement by adding ancilla qubits and local checks.
Its auxiliary graph has one vertex per qubit in the support of $L$ and one ancilla per edge.
Each vertex has a Gauss-law check $A_v=X_v\prod_{e\ni v}X_e$, acting on that vertex and its incident edges.
Every edge ancilla appears in two Gauss checks, so multiplying all Gauss checks cancels the edge factors and leaves $L$.
Completing the construction requires the deformed original checks, cycle checks and gauge initialization specified in the source.
The spatial distance minimizes the weight of nontrivial logical Pauli operators of the full deformed code, including original and auxiliary qubits.
Measuring one independent logical operator leaves $k-1$ logical qubits from an original $k$.
Our formalization starts from the resulting check matrices and proves their algebraic properties.
Section~\ref{sec:instances} computes the enlarged codes including their ancillas; the single-row argument above explains the loss of one logical degree of freedom on the original qubit set.

To isolate the role of repeated measurements, the temporal model assigns one measurement-fault bit to each of $T\ge1$ rounds.
Its fault class consists of independent measurement-outcome flips, detected by comparisons between adjacent rounds.
A nonzero fault that passes every comparison must occupy every round, so its weight is exactly $T$.
If the chosen linear readout returns one on this constant pattern, the temporal fault distance is $T$.
This is the repetition mechanism behind Supplementary Lemma~3 of the gauging paper, whose round count is $t_o-t_i$.
Theorem~\ref{thm:temporal} and Eq.~\eqref{eq:temporal} state this identity; Section~\ref{sec:sharp} separately studies models with persistent data errors.

\subsection{The spatial and temporal theorems}

\begin{theorem}[spatial $X$-type component]
\label{thm:spatial}
Let the auxiliary graph be loop-free and have expansion at least one.
Assume every vector in $\ker H_Z$ outside the row space of $H_X$ has weight at least $d$.
The weight of an operator is the number of qubits on which it acts.
Assume the deformed code has the following check-matrix structure.
The Gauss laws $A_v=X_v\prod_{e\ni v}X_e$ are $X$ checks, and the lifted original $X$ checks belong to the deformed $X$ check space.
Each original $Z$ check has a deformed partner that differs only on edge qubits.
The deformed $Z$ checks include a cycle basis of the auxiliary graph, and every Gauss law commutes with every deformed $Z$ check.
Let $F$ be an $X$-type logical operator of the deformed code: it lies in the kernel of its $Z$ checks and outside the row space of its $X$ checks.
Write $S$ for its vertex support and $R$ for its support outside the auxiliary graph.
Then its edge support is a cut $\partial U$ for some vertex set $U$ with $|U|\le|U^{\mathsf c}|$, and $d\le|S|+|\partial U|+|R|=\mathrm{wt}(F)$.

\end{theorem}

Theorem~\ref{thm:spatial} bounds physical $X$-type logical operators.
A bound on the full static CSS distance also requires the $Z$-type component.
The finite instances reported below check both components separately.
The spatial proof identifies the edge support, removes it with Gauss checks, and bounds the remaining logical operator's weight.
First, an undetectable edge pattern has even overlap with every cycle check.
Cut--cycle duality therefore makes its edge support a cut $E=\partial U$.
We can choose $U$ no larger than its complement.
Second, multiplying by the Gauss laws over $U$ removes this edge support: in binary support notation, their product is the mod-two sum of the $A_v$ vectors, with ones on $U$ and its crossing edges.

\begin{equation}
\sum_{v\in U}A_v \;=\; \bigl(\mathbf 1_U,\mathbf 1_E\bigr).
\label{eq:cleaning}
\end{equation}
Multiplying by stabilizer checks changes the operator's support while preserving its action on the encoded information.
The cleaned operator has support $S\triangle U$ on vertices and $R$ outside the graph, with no edge support.
The construction hypotheses make it a nontrivial logical operator of the original code, so

\begin{equation}
d \;\le\; \lvert S\triangle U\rvert + \lvert R\rvert \;\le\; \lvert S\rvert + \lvert U\rvert + \lvert R\rvert
\;\le\; \lvert S\rvert + \lvert\partial U\rvert + \lvert R\rvert.
\label{eq:cheeger}
\end{equation}
Section~\ref{app:counting} gives the coordinate proof of each step.

Expansion is used only in the last inequality of Eq.~\eqref{eq:cheeger}.
It bounds the number of vertex checks needed to remove the edge support by the number of edges already in that support.

\begin{theorem}[temporal component]
\label{thm:temporal}
Let one check be repeated over $T\ge1$ rounds, with a comparison between each pair of adjacent rounds.
Every nonzero binary fault pattern that passes all comparisons is the all-ones pattern and has weight $T$.

\end{theorem}

Write $z_t\in\Ztwo$ for the measurement-fault bit in round $t$.
Fault weight counts the independently flipped measurement bits.
Each comparison checks $z_t+z_{t+1}$, so passing all $T-1$ comparisons requires all bits to be equal.
A nonzero binary pattern must then be one in every round.
This pattern is a logical fault whenever the specified linear readout returns one on it, giving temporal fault distance $T$.
At $T=1$ there is no comparison, and the same identity holds:

\begin{equation}
z_t=1\ \text{for all } t,\qquad \lvert\mathrm{supp}\,F\rvert = T.
\label{eq:temporal}
\end{equation}

\begin{remark}[the assembly]
Once both static CSS components have lower bound $d$, a temporal distance $T\ge d$ places both the spatial and temporal distances at least $d$.
This final comparison uses the two component bounds.
Applying it to the source protocol also requires the assumptions used to establish those bounds, including the relevant boundary and detector conventions.
Section~\ref{app:assembly} gives the statement in the source's notation.

\end{remark}

\refstepcounter{smsection}\section*{\thesmsection\quad Checking the conditions}
\label{sec:decision}

C1 and C2 can be checked directly from the auxiliary graph, round count and target distance.
Section~\ref{sec:instances} gives exact values on two named codes, and Section~\ref{sec:sharp} examines what changes when a condition fails.
The two tests are

\begin{equation}
\lvert\partial S\rvert \ge \min\bigl(\lvert S\rvert,\lvert S^{\mathsf c}\rvert\bigr)
\ \text{for all } S
\qquad\text{and}\qquad
t_o-t_i \ge d ,
\label{eq:predicates}
\end{equation}
where $\partial S$ is the set of edges leaving $S$ and $t_o-t_i$ is the duration of the deformation in rounds.
C4 is also a finite predicate: it tests for linear dependencies among same-round checks.
C3 instead specifies a boundary convention.
We study its effect through related syndrome constraints in explicit fault models.

\subsection{The conditions, item by item}

We use C1--C4 to collect assumptions from different statements in the published supplement of Ref.~\cite{williamson2026gauging}.
Its item numbers differ from those in the arXiv version.
These are separate from the source's four requirements on the auxiliary graph.
The two lists share the expansion requirement.
The source also requires sparsity, constant-size matchings and a constant-weight cycle basis to keep the deformed code low-density.
The resource comparison below minimizes edge ancillas subject to expansion alone.

\begin{itemize}
\item \textbf{C1: graph expansion.} Every vertex set $S$ has at least $\min(|S|,|S^{\mathsf c}|)$ outgoing edges.
This is one of the two explicit hypotheses of the source's spacetime-fault-distance theorem.
Methods Remark~2 and Corollary~1 already preserve distance when the first layer has expansion at least one, even if cycle sparsification gives a final graph with expansion at most one half.

\item \textbf{C2: enough rounds.} The deformation lasts $t_o-t_i\ge d$ rounds.
Supplementary Lemma~3 already proves that the source's measurement-and-initialization fault distance is exactly $t_o-t_i$.
Supplementary Definition~3 defines spacetime fault distance, and Lemma~4 separates spatial and temporal contributions.
The spacetime-distance theorem uses the smaller contribution.
Theorem~\ref{thm:temporal} verifies the round-count identity in our measurement-fault model.
Our necessity statement allows any linear parity readout that evaluates to one on the all-ones pattern.
This readout condition states explicitly when the source's constant-pattern witness is a logical fault in our model.

\item \textbf{C3: perfect boundary measurements.} The source takes the first and last stabilizer-measurement rounds to be perfect.
Supplementary Remark~3 explains this convention using the $d$ rounds of error correction before and after gauging.
In that argument, a process involving both the gauging measurement and a boundary has distance greater than $d$.
Our protocol models impose an endpoint syndrome constraint: the syndrome at that endpoint must be zero.
Endpoint faults are allowed when their contributions to the syndrome cancel.
This defines a model without the source's surrounding error correction, whose noisy-boundary behavior requires a separate analysis.

\item \textbf{C4: no same-round detector.} A local detector is a nonempty set of one round's check rows whose sum is zero.
C4 is an assumption used to generate the source's detector set.
Supplementary Remark~2 explains that local relations, also called meta-checks, can produce such detectors.
The lemma can be extended to include them.
C4 holds exactly when the check rows are linearly independent, so matrix rank decides it.
A dependent set has a fixed combined ideal outcome; for example, two identical checks must have the same ideal outcome.
\end{itemize}

\subsection{Expansion and ancilla count}

Choosing a graph family with a proved expansion bound satisfies C1 for every family member.
The complete graph $K_m$ is one example.
If $|S|\le m/2$, any vertex in a nonempty $S$ has $|S^{\mathsf c}|$ edges to the complement, which is at least $|S|$.
For a larger set, apply the same argument to its complement.
Lean verifies this bound and supplies the expansion hypothesis of the spatial theorem.

With one ancilla per edge, $K_m$ uses $m(m-1)/2$ ancillas for a measured support of size $m$.
C2 can likewise be satisfied by choosing $d$ deformation rounds.
The schedule then meets C2 at the cost of $d$ measurement rounds.
For the toric family, constructed from two repetition codes~\cite{bravyi1998quantum}, the K\"unneth formula determines the code dimension from the homology of the factors~\cite{bravyi2014homological}.
Choosing a complete auxiliary graph and $d$ rounds then satisfies C1 and C2 throughout the family.
C1 concerns the auxiliary graph, and C2 concerns the schedule; neither is an intrinsic property of the original code.

The minimum ancilla count for expansion alone is linear in the support size.
A star on $m$ vertices satisfies C1 with $m-1$ edges, and this is the minimum possible edge count.
Adding edges preserves C1, so a star with $r$ additional edges between its leaves uses $m-1+r$ ancillas.
At four vertices and cycle rank three, this graph is $K_4$, and both constructions use six ancillas.
This is the auxiliary graph of the bivariate bicycle example in Section~\ref{sec:instances}.
At seven vertices, a star with three chords uses nine ancillas, whereas $K_7$ uses twenty-one.
Their cycle ranks are three and fifteen, respectively, so this is a comparison of two C1-satisfying graphs, not a comparison at fixed cycle rank.
A star has unbounded maximum degree as $m$ grows.
A low-density implementation additionally needs the source's degree, matching-size and cycle-weight bounds.

The supplement reports an exhaustive finite census over edge multisets for the overall minimum and over simple graphs for specified cycle ranks.
In the tested rows with a solution, the star-based witness uses fewer edges than the complete graph except when completeness is forced: three vertices with one cycle and four with three.
The supplement specifies the search range and reproduction script, and Lean verifies the reported witnesses.

The lower bound $m-1$ follows from rank and nullity.
Assign a bit to each vertex and map the assignment to the difference of the endpoint bits on every edge.
A nonconstant assignment in the kernel would split the vertices into two nonempty sets with no edge between them, contradicting C1.
The kernel therefore has dimension at most one, giving $m\le\varepsilon+1$ for $\varepsilon$ edges.
A connected graph with cycle rank $r$ has $\varepsilon=m-1+r$.
For fixed cycle rank three, the complete graph has this edge count only at $m=4$.
Separately, the formalized toric family has $k=2$ for every cycle length $m\ge2$, and measuring one independent logical operator reduces this to one logical qubit.

\refstepcounter{smsection}\section*{\thesmsection\quad Instances}
\label{sec:instances}

Two examples show how the spatial construction and the measurement model enter exact distance calculations.
The bivariate bicycle example computes the static code after adding the gauging ancillas.
The Bacon--Shor example isolates spatial and temporal behavior in two separately defined measurement models.

\subsection{A gauged bivariate bicycle code}

We use the bivariate bicycle code $[[18,4,4]]$ at shift parameters $\ell=m=3$~\cite{bravyi2024highthreshold}.
These shift parameters are distinct from the auxiliary-graph support size denoted by $m$ above.
Its check matrices come from the Lean-QEC development~\cite{ehatamm2026leanqec}.
Gauging a weight-four logical operator with auxiliary graph $K_4$ adds six ancillas and gives a $[[24,3,4]]$ code.
The construction reduces the number of logical qubits by one and preserves both static distance components at four.

Lean constructs the Gauss-law, pairing and cycle checks from the $K_4$ data and verifies their matrix entries.
The Gauss laws sum to the measured logical operator, placing it in the deformed check row space.
The code dimension and both distances are computed exactly.

Expansion follows from the complete-graph theorem, and choosing four rounds satisfies C2.
A Lean theorem combines C1 and C2 with the independently computed static distances to establish this instance's spatial bound.
The exact values $d_X=d_Z=4$ are certified separately; Section~\ref{app:conventions} defines each label by the check matrix whose kernel is used.
The temporal value is separately $T=4$ in the measurement-fault model with adjacent-round comparisons.
The temporal value follows from the repetition identity for independent measurement faults.

\subsection{A transversal measurement}

The second example is the Bacon--Shor subsystem code $[[9,1,3]]$ on a three-by-three grid~\cite{aliferis2007subsystem,bacon2006operator}.
It protects one logical qubit and also contains unprotected gauge degrees of freedom.
The distance three refers to the protected subsystem.
Treating the gauge degrees of freedom as additional logical qubits instead gives the associated stabilizer code distance two.
Thus the protected subsystem and the associated stabilizer code have different distance conventions.

In the static single-round model, every data qubit is measured in $X$, and two $X$-type stabilizers provide parity checks.
A bare $Z$-type logical operator of weight three is one minimum-weight undetected readout error.
Lean verifies this bare-logical witness and proves the lower bound over all undetected readout-flipping faults, including those that act on the gauge subsystem.
The temporal value comes from a different model: independent faults in the repeated support readout, detected by comparisons between adjacent rounds.
Its distance is $T$, so three rounds give temporal distance three.
The static value counts data-only readout errors, while the temporal value counts repeated measurement-outcome faults.

With only two rounds, the measurement-fault model has an explicit undetected readout error of weight two:

\begin{equation}
\begin{aligned}
d_{\text{temporal}}(T) &= T ,\\
d_{\text{temporal}}(d-1) &= 2 \;<\; 3 = d ,
\end{aligned}
\label{eq:c2tight}
\end{equation}
Equation~\eqref{eq:c2tight} makes the round requirement explicit: with independent measurement faults and fault-free data, $T\ge d$ is necessary.
In the persistent-fault readout model of main-text Sec.~\ref*{sec:readout}, data, check-result and readout faults are all allowed.
Its strict-majority distance is three at two rounds, using the predicate defined there.
The comparison changes the fault locations, detectors and readout criterion, as Table~\ref*{tab:models} makes explicit.

\refstepcounter{smsection}\section*{\thesmsection\quad How each condition affects distance}
\label{sec:sharp}

Each condition controls a different part of the distance argument.
C1 and C2 enter the assembled spatial and temporal statements.
C1 is sufficient under the construction hypotheses but does not determine every instance's distance.
C2 is necessary in the measurement-fault model.
C3 motivates the endpoint syndrome constraint in the persistent-fault readout model, and C4 tests for redundant same-round checks.
For redundancy, we first prove static distance invariance and then analyze the additional measurement in the protocol model.

\textbf{C1: distance below the expansion threshold depends on the construction.} We gauge the same weight-four logical operator of the smallest toric code using two different four-vertex paths (main-text Fig.~\ref*{fig:c1pair}).
Both paths violate C1 because some two-vertex set has only one outgoing edge.
Changing one edge changes the resulting code distance.
For the distance-losing path, $d_X^{\mathrm{phys}}=1$ and $d_Z^{\mathrm{phys}}=5$.
For the distance-preserving path, $d_X^{\mathrm{phys}}=2$, and neither side has a weight-one logical operator, so $d_Z^{\mathrm{phys}}\ge2$.
The total distances are therefore one and two, while the original code has distance two.

A zero column of the deformed $Z$-check matrix explains the weight-one logical operator in the first path.
For any CSS matrix pair, a unit vector $e_j$ is such an operator exactly when the $j$th $Z$-check column is zero and $e_j$ lies outside the $X$-check row space.
The zero column means that no $Z$ check detects $e_j$, and the row-space condition means that $e_j$ is not a product of $X$ stabilizers.
Applying this criterion on both CSS sides decides whether distance one occurs; larger distances require searching higher-weight vectors.

Each deformed $Z$ check is completed by an auxiliary-edge set whose mod-two boundary equals the check's support on the auxiliary graph~\cite{williamson2026gauging}.
The number of selected edges incident to a vertex is odd on that support and even elsewhere.
This allows three selected edges to meet at a support vertex, so it is more general than an ordinary perfect matching.
An auxiliary-edge column is zero exactly when the edge appears in no completed $Z$ check and no cycle check.
For the distance-losing path, we verify the zero column directly in the deformed matrix.

A second implementation enumerates every measured logical of support at most six and every connected auxiliary graph on that support for the toric codes $[[8,2,2]]$ and $[[18,2,3]]$.
Each deformed check uses a minimum-weight edge completion that commutes with every Gauss law, following the source's parity-based matching definition.
All enumerated C1-satisfying instances preserve the original distance.
Among the C1-violating instances, the distance ranges from one to the original distance.
The path pair and finite census therefore exhibit both distance loss and distance preservation below the expansion threshold using graphs without dummy vertices.
All enumerated instances have same-round detectors, placing this census in the regime where the source's detector construction must include local check relations.

The same-round detectors come from dependencies in the original code.
The four $X$ checks of the smallest toric code sum to zero, and their lifted checks retain this relation after gauging.
The bivariate bicycle instance also has a local relation.
Supplementary Remark~2 discusses precisely this case: the detector set must include the same-round relations.
By contrast, the eight generators of the Shor $[[9,1,3]]$ code are linearly independent and satisfy C4.

The construction also requires every deformed $Z$ row to commute with every Gauss law.
This is equivalent to requiring the row's edge completion to have odd-degree vertices exactly on the relevant support, the source's matching condition.
This condition constrains the edge completion, whereas C1 constrains the graph.
The construction must check the edge-completion condition and the graph-expansion condition separately.

\textbf{C2 is necessary in the measurement-fault model.} The Bacon--Shor example with $T=2$ has an undetected readout error of weight two, below $d=3$.
The general witness is the all-ones pattern: the same measurement bit is flipped in every round.
Adjacent-round comparisons see no change, and the fault weight is $T$.
Any linear parity readout $w$ with $w\cdot\mathbf 1=1$ reads this pattern as a logical error.
For every positive $T<d$, this gives a fault lighter than $d$.
Our result states the readout condition explicitly, while Supplementary Lemma~3 of the source gives the exact distance for its measurement-and-initialization fault class.

Throughout the protocol analysis, an undetectable logical fault fires no detector and satisfies the specified readout-error predicate.
The protocol distance is the minimum weight among these faults; the static distance instead minimizes over nontrivial logical operators in the check kernel outside the stabilizer row space.
Static distance tests all nontrivial logical operators, while protocol distance uses the selected logical readout and its decision rule.

\textbf{C3: a boundary syndrome constraint detects constant nonzero syndromes.} The persistent-fault readout model includes data, check-result and readout faults and requires the first syndrome to be zero.
At $T=1$, one data fault paired with a check-result fault that cancels its syndrome gives an undetected readout error of weight two.
Lean proves that no weight-one fault satisfies both the detector and readout conditions.
The one-round distance is therefore two, compared with three in the static data-only model and one in the measurement-fault model.
The fault class and detector set therefore change the one-round distance even for the same code.

For every $T\ge2$, $d_{\mathrm{maj}}(T)=3$.
This statement uses the predicate of Eq.~(\ref*{eq:majority-distance}).

For $T\ge4$, the lower bound follows from three observations.
First, a weight-at-most-two fault occupies at most two rounds.
Second, in a round with no new faults, the syndrome is $H$ times the accumulated data error.
The boundary syndrome constraint and adjacent comparisons force every syndrome to zero.
The accumulated error has weight at most two, so the single-round distance three prevents it from flipping that round's readout.
Third, only the at most two occupied rounds can have flipped readouts, which is not a strict majority when $T\ge4$.
Lean checks $T=2$ and $T=3$ by finite enumeration.
A weight-three logical data fault in the first round gives the matching upper bound for all $T$.

The general boundary-collapse construction is given in main-text Sec.~\ref*{sec:readout}; Lean certifies its Bacon--Shor specialization.

The one-round formula is the $q=1$ case of Eq.~(\ref*{eq:weighted-readout-distance}); its Bacon--Shor value is also certified in Lean.
Let $y$ be the merged data and readout error, let $H$ be the check matrix, and let $w$ be the logical parity support.
An undetected fault must use check-result error $Hy$ to cancel the syndrome, while a readout error requires $w\cdot y=1$.
Minimizing the total weight gives

\begin{equation}
d_{T=1} \;=\; \min\bigl\{\,\wt(y)+\wt(Hy)\;:\;w\cdot y=1\,\bigr\},
\label{eq:t1closed}
\end{equation}
For the Bacon--Shor model, the minimum is attained at $y=e_0$: one data error and one check-result error that cancels its syndrome.
The value is two.
For other check matrices and readout supports, Eq.~\eqref{eq:t1closed} gives the value directly.
Section~\ref{app:oneround} gives related lower bounds and endpoint variants.
The paired data and check-result errors count as two independent fault locations.

Figure~\ref*{fig:sharpness} of the main text compares the persistent-fault model, the measurement-fault model and the persistent-fault model without its boundary syndrome constraint.
A second implementation recomputes the figure, and all twelve of its points are exactly the proved ones, four on each curve.

\textbf{C4: check redundancy preserves static distance and can improve protocol distance.} A same-round detector is a nonzero linear combination of check rows that sums to zero.
Such a combination exists exactly when the row rank is smaller than the number of rows.
A redundant row is a binary sum of existing check rows, so adding it leaves both the check row space and the check kernel unchanged.
Both CSS distance components therefore remain unchanged.
A code with redundant generators represents the same static code after those generators are removed.

Measuring an extra check has a separate protocol effect.
In our persistent-fault readout model, duplicating the first check adds an independent check-result fault location.
Deleting the extra check-result fault location leaves the readout unchanged, preserves the original detector constraints and can only reduce or preserve the fault weight.
The duplicated-check distance is therefore at least the original distance for every round count.
At $T=1$, the distance rises from two to three: a single data fault that flips the logical parity must be hidden in both copies of the check using two independent check-result faults.
The improvement uses one extra check measurement and one extra independent check-result fault location per round.
The detector-generating lemma describes relations among measurement outcomes.
These results show how same-round detectors can be retained alongside the spatial and temporal analysis.

\refstepcounter{smsection}\section*{\thesmsection\quad Extensions to surgery and fault complexes}
\label{sec:open}

The same matrix-based approach yields further results for surgery, mapping cones and fault complexes.
The gauging operation verified here acts on check matrices.
The graph, ancillas, cycle checks and edge completions of Ref.~\cite{williamson2026gauging} enter through explicit construction hypotheses.
The named instances construct these data and verify the resulting matrix entries.
Connecting these results to a gate-level circuit requires identifying that circuit's checks, fault locations and error propagation with the specified matrices.

The check-matrix construction includes a useful commutation criterion.
A deformed check commutes with every Gauss law exactly when its edge completion has odd-degree vertices exactly on the check's relevant support.
This is the source's parity-based meaning of a matching.
Such a completion exists on a connected graph without self-loops when the support has even cardinality.
The parity condition ensures commutation, and expansion supplies the weight bound used in the spatial proof.

\textbf{The verified graph regime.} Lean verifies all three steps of Theorem~\ref{thm:spatial}.
Cut--cycle duality identifies edge sets orthogonal to every cycle with cuts.
The cycle space is the kernel of the edge-to-vertex incidence map and is spanned by graph cycles.
The formalized auxiliary graph is on all original code qubits.
Every vertex therefore carries a data qubit; adding the source's dummy vertices would require extending this construction to vertices that carry checks without data.
Thus the outside-graph term $R$ in the displayed counting argument is not instantiated by this formal construction.
The verified $X$-type spatial bound assumes loop-free graphs with $h(G)\ge1$; self-loops and $0<h(G)<1$ require additional arguments.
Section~\ref{app:conventions} defines the Cheeger constant $h(G)$.

The C4 results concern check dependencies and static distances, together with the specified duplicated-check protocol model.
The detector-generating statement uses a label model in which distinct labels can take arbitrary outcome values, while repeated occurrences of one label share an outcome.
In this model, a parity combination vanishes for every allowed outcome assignment exactly when each label occurs an even number of times.
Relations among dependent check operators must also be included when applying this model to their measurements.
Within this label model, adjacent comparisons leave precisely the repeated-label fault patterns undetected.
Initialization and readout contribute single-site detectors at the endpoints.
These sites define the spacetime check set, to which a circuit-level fault model can be attached by specifying how gate faults change the outcomes.

\textbf{Surgery and the coupling between complexes.} We verify a $Z$-type specialization of the distance conclusion of Theorem~7.7 of Ref.~\cite{logicalmeasurements2026} for the single-block model described below.
Its second proof case compares a deformed-code logical operator $\overline{\Lambda}_Z$ with an auxiliary-hypergraph vertex set $\kappa_\lambda$.
An explicit coupling is needed to compare the logical operator and the vertex set because they belong to different spaces.
We use the one-to-one data attachment of formula~(2), together with the chain-map construction in \S4 and \S5 of Ref.~\cite{logicalmeasurements2026}.

In this specialization, the port covers all original data qubits.
Lean proves that the kernel of the deformed $X$-check matrix $H'_X$ is contained in the row space of the deformed $Z$-check matrix $H'_Z$.
Thus every undetected physical $Z$-type operator in this specialization is a stabilizer; no logical operator remains on that side.
For a free coupling in our own model, complete cycle checks make all legal couplings agree on admissible vectors.
Cleaning removes the coupling contribution without requiring the coupling itself to vanish.

The source obtains its block-reading fault-distance guarantee from the gauging arguments in \S3.1.
A mapping cone puts the deformed-code complex, the auxiliary complex and the map connecting them into one complex, allowing us to analyze distance transfer in that combined object.
The connecting map, called the port map, has three linear components $F_1,F_2,F_3$.
In the specific small-code and triangular-auxiliary assembly studied here, the commuting-square condition requires $F_2$ to have three equal columns.
A second implementation enumerates the squares and confirms that these are exactly the admissible $F_2$ matrices.
For every admissible port in this assembly, the cone distance equals the small code's distance when the first column of $F_3$ is nonzero, and equals one when that column is zero.
The column test is an exact distance criterion for this small-code and triangular-auxiliary assembly.

A separate theorem gives the general algebraic criterion.
Let $D$ be the idle-volume complex and $C$ the auxiliary complex, and suppose $D$ is exact at the middle.
Exactness means that every cycle at that term is a boundary.
A pair $(u,v)$ is then an undetected nontrivial cone fault exactly when $D.d_2u+F_3v=0$ and $v\notin\operatorname{im}C.d_2$.
A nontrivial undetected fault must satisfy the linear equation and lie outside the specified image space.
Once the chain-map conditions hold, this fault criterion is determined by $F_3$ and the two complex differentials.
A weight-one cone fault exists exactly when a weight-one auxiliary vector $v$ lies outside $\operatorname{im}C.d_2$ and satisfies $F_3v=0$.
The first-column test above is this criterion specialized to the triangular assembly.

The exactness condition concerns the idle-volume complex $D$.
For $D$ built as a tensor product of a two-term factor and a code complex, the supplement gives a complete criterion for middle exactness.
Its criterion uses the homology of both factors.
A code complex with vanishing $H_0$ and $H_1$ gives middle exactness with every partner.

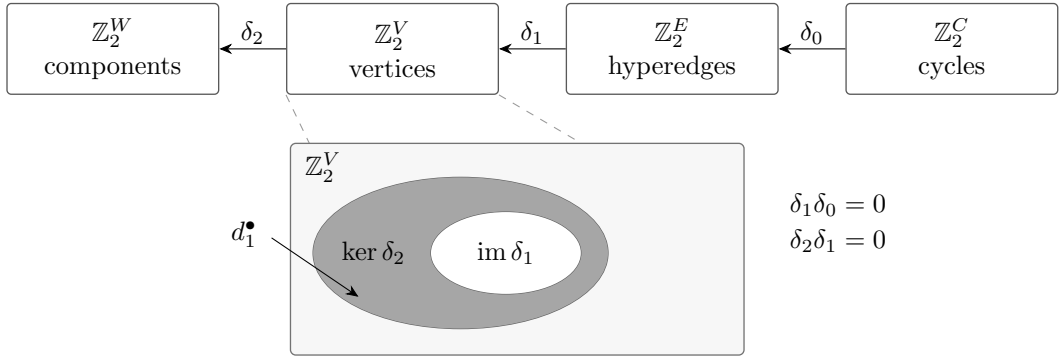
\begin{figure*}[t]
\centering
\begin{tikzpicture}[
    font=\normalsize,
    >={Stealth[length=1.7mm,width=1.4mm]},
    space/.style={draw=black!65, line width=0.5pt, rounded corners=1.6pt,
                  align=center, inner sep=3pt, minimum width=2.80cm,
                  minimum height=1.20cm},
    maplbl/.style={font=\normalsize, inner sep=1.5pt},
    host/.style={draw=black!55, line width=0.5pt, rounded corners=2pt,
                 fill=black!3},
    ker/.style={draw=black!60, line width=0.5pt, fill=black!16},
    im/.style={draw=black!60, line width=0.5pt, fill=white},
    sub/.style={font=\normalsize, inner sep=1pt},
    law/.style={font=\normalsize, align=left, inner sep=0pt},
  ]

  \node[space] (w) at (0, 0)      {$\Ztwo^{W}$ \\[1pt] components};
  \node[space] (v) at (3.70, 0)   {$\Ztwo^{V}$ \\[1pt] vertices};
  \node[space] (e) at (7.40, 0)   {$\Ztwo^{E}$ \\[1pt] hyperedges};
  \node[space] (c) at (11.10, 0)  {$\Ztwo^{C}$ \\[1pt] cycles};

  \draw[->] (v.west) -- (w.east) node[maplbl, midway, above] {$\delta_2$};
  \draw[->] (e.west) -- (v.east) node[maplbl, midway, above] {$\delta_1$};
  \draw[->] (c.west) -- (e.east) node[maplbl, midway, above] {$\delta_0$};

  \node[law, anchor=west] at (8.95, -2.30)
    {$\delta_1\delta_0 = 0$\\[2pt] $\delta_2\delta_1 = 0$};

  \draw[dashed, black!40, line width=0.4pt] (v.south west) -- (2.60, -1.25);
  \draw[dashed, black!40, line width=0.4pt] (v.south east) -- (6.20, -1.25);
  \draw[host] (2.35, -4.05) rectangle (8.35, -1.25);
  \node[sub, anchor=north west] at (2.50, -1.33) {$\Ztwo^{V}$};
  \draw[ker, fill=black!8] (4.60, -2.70) ellipse [x radius=1.95, y radius=1.00];
  \draw[im]  (5.20, -2.70) ellipse [x radius=1.00, y radius=0.55];
  \fill[even odd rule, fill=black!35]
    (4.60, -2.70) ellipse [x radius=1.95, y radius=1.00]
    (5.20, -2.70) ellipse [x radius=1.00, y radius=0.55];
  \node[sub] at (3.45, -2.70) {$\ker\delta_2$};
  \node[sub] at (5.20, -2.70) {$\mathrm{im}\,\delta_1$};
  \node[sub, anchor=east] at (1.95, -2.45) {$d_1^{\bullet}$};
  \draw[->] (2.10, -2.50) -- (3.25, -3.28);

\end{tikzpicture}
\caption{
The auxiliary four-term complex used in the surgery analysis of Ref.~\cite{logicalmeasurements2026}.
From left to right, the spaces are spanned by the chosen component family, vertices, hyperedges and cycles.
The maps $\delta_2,\delta_1,\delta_0$ point left, and consecutive compositions vanish.
Both $\ker\delta_2$ and $\operatorname{im}\delta_1$ therefore lie in the vertex space.
The shaded region represents $\ker\delta_2\setminus\operatorname{im}\delta_1$.
When this set is nonempty, $d_1^\bullet$ is its minimum vector weight.
The formal definition assigns zero when this set is empty.
OpenAI Codex assisted the drawing code and layout; see main-text Sec.~\ref*{sec:verification}.
}

\label{fig:complex}
\end{figure*}

\textbf{Cosystolic distance and its assumptions.} When $\ker\delta_2\setminus\operatorname{im}\delta_1$ is nonempty, the auxiliary quantity $d_1^\bullet$ is the minimum weight of its vectors (Fig.~\ref{fig:complex}).
A component is a vertex set that meets every hyperedge in an even number of vertices.
The component family $W$ is a chosen collection of these sets.
The vector must pass all component parity checks but cannot be generated by adding hyperedge vectors; over $\mathbb F_2$, both requirements are finite parity tests on its support.
When the set is nonempty, a lower bound excludes all lighter admissible supports, and an upper bound supplies one admissible support.
The characterization reduces distance calculation to a minimum-weight search subject to the two parity tests.
The problem is to find the lightest vector in a linear code outside a specified subcode.
The supplement gives a reduction from linear-code minimum distance, which is NP-hard~\cite{vardy1997intractability}.
The free-coupling problem has the same algebraic form: a linear system together with exclusion from a specified subspace.

Cosystolic bounds for simplicial complexes suggest ways to construct auxiliary systems~\cite{evra2022decodable,lubotzky2017expanders,kaufman2020highorder,dikstein2024coboundary}.
Using a specific bound here requires identifying the corresponding cochain complex, relating its weight convention to ours, and verifying all hypotheses of that theorem.
For three structured families, the supplement gives exact distance formulas.
These families have one nonempty hyperedge, so their cycle space is zero.
Applying them to sparse surgery requires bounds on incidence degrees and check weights as the family grows, together with the code and coupling hypotheses of the chosen surgery theorem.
The supplement identifies the entries whose weights grow in these examples.

The graph specialization illustrates why nonemptiness matters.
Each hyperedge is then an ordinary edge, and $\delta_1$ is the edge-to-vertex incidence map.
On a connected graph, leaving a vertex out of the chosen component family gives distance one.
Whether this occurs depends on the chosen component family.
If the whole vertex set is the single chosen component, $\ker\delta_2\setminus\operatorname{im}\delta_1$ is empty.
The formal definition assigns zero in this case because it uses the natural-number infimum, which equals zero for an empty set.
Here zero records an empty admissible set, so it represents the absence of a cosystolic vector.
Positive lower-bound criteria therefore require at least one admissible vector.

\textbf{Two foliations give different distance laws.} Foliation stacks a code complex along a time direction~\cite{bolt2016foliated}.
With repetition-code foliation, the matrix-indexed distance $d_Z$ is $T$ whenever the code complex has nonzero zeroth homology $H_0$.
Here $H_0$ is the cokernel of the bottom map.
With dual foliation, the corresponding distance equals the original code's matrix-indexed distance $d_Z$ for every positive $T$, provided that the original code has a nontrivial logical operator on that side.
For an empty logical set, the fault-complex infimum is zero, whereas the static finite-search function returns $n+1$ for an $n$-qubit code.
These values record the absence of candidates; neither is the weight of a logical operator.
The K\"unneth formula describes the homology~\cite{logicalmeasurements2026}; separate weight bounds and explicit faults attaining them give these distance laws.
Selecting the distance law for a physical circuit therefore requires identifying which foliation models that circuit.

The supplement lists the formal declarations supporting these results and states the assumptions needed to connect them to the source constructions.
It also verifies that each detector functional vanishes on the image of the fault-complex boundary map, as required by the complex condition.
To use this identity in the source's spacetime cleaning proof, one must identify the fault coordinates and establish the corresponding weight bound.
Related work describes fault complexes as graph states~\cite{hillmann2025faultcomplexes} and extends the construction through spacetime lifting~\cite{xu2026lifting}.

\subsection{Proof verification and numerical calculations}

Lean verifies the proof scripts, and separate numerical implementations check the reported finite instances.
DeepSeek-V4.1-Flash assisted the Lean derivations and numerical debugging; OpenAI Codex (GPT-6) assisted literature checks, formal-proof exploration and numerical cross-checks.
AI-generated proof code is checked by the Lean kernel, and numerical claims are checked against independently computed values.
OpenAI Codex also assisted the four TikZ schematics and the plotting code; their supports, operator relations and plotted distances were checked against the definitions and numerical data.

\refstepcounter{smsection}\section*{\thesmsection\quad Formal statements supporting the results}
\label{sec:sm-modules}

The development is a single \Lean\ project of 75 modules plus a root
module. This repository contains 19 modules and imports 56 from QECCertificates, which provides the GF(2) and Pauli layers and part of the \code{Codes} layer. The project also depends on mathlib and the Lean-QEC formalization at \code{QCL-SUAT/Lean-QEC}; full revision hashes pin all three dependencies. Table~\ref{tab:sm-map} and Table~\ref{tab:sm-map-continued} group the modules by layer and name the theorem or group of results supporting each claim.

The K\"unneth row summarizes several results, whose derivations are given below. The dimension identities cover all four K\"unneth degrees, including both a zero differential in the two-term factor and the general case. For repetition-code foliation, which stacks the code along time, the project also proves elementwise decoupling as described below.

\begin{table*}[t]
\caption{Modules and the results they support, part I: the machine, decision and first spatial layers. Every module is in the default build; Section~\ref{sec:sm-audit} reports the axiom dependencies of its supporting statements. Codes are written $[[n,k,d]]$, $d_1^\bullet$ is the $1$-cosystolic distance of the auxiliary hypergraph and $f_p$ is the port function.}
\label{tab:sm-map}
\centering
\begingroup
\footnotesize
\renewcommand{\arraystretch}{0.96}
\begin{ruledtabular}
\begin{tabular}{@{}p{60pt}p{175pt}p{247pt}@{}}
Layer & Module & \raggedright\arraybackslash Supporting statement \\
\hline
machine & \code{GF2.Basic} & \raggedright\arraybackslash \code{spanL\_map\_addSmul\_append}, \code{Matrix.rowSpace\_eq\_spanL\_ofFn} \\
machine & \code{GF2.RowReduce} & \raggedright\arraybackslash \code{spanL\_rowReduce}, \code{isReduced\_rowReduce} \\
 & \code{GF2.KernelBasis} & \raggedright\arraybackslash \code{kerL\_rowReduce} \\
 & \code{GF2.Canonical} & \raggedright\arraybackslash \code{rowReduce\_mem\_iff\_of\_spanL\_eq} \\
 & \code{GF2.RankCertificate} & \raggedright\arraybackslash \code{Matrix.rank\_eq\_length\_rowReduce} \\
 & \code{GF2.RankEchelon} & \raggedright\arraybackslash \code{rankEchelon\_eq\_length\_rowReduce} \\
 & \code{GF2.Witness} & \raggedright\arraybackslash \code{DualWitness.not\_mem\_rowSpace}, \code{eq\_minWeight\_of\_bounds} \\
 & \code{GF2.Membership} & \raggedright\arraybackslash \code{inSpanB\_iff}, \code{inKerB\_iff} \\
 & \code{GF2.LowerBound} & \raggedright\arraybackslash \code{eq\_minWeight\_of\_decide}, \code{lightCand\_length\_le} \\
 & \code{GF2.WeightEnum} & \raggedright\arraybackslash \code{mem\_lightVecs} \\
 & \code{GF2.HGP}, \code{GF2.HGPCompression} & \raggedright\arraybackslash \code{hgp\_orthogonal}, \code{hgpHX\_mulVec\_eq\_zero\_iff} \\
 & \code{GF2.HGPCleaning}, \code{GF2.HGPCleaningDual} & \raggedright\arraybackslash cleaning lemmas, two-sided \\
 & \code{GF2.KunnethCore}, \code{GF2.HGPKunneth} & \raggedright\arraybackslash \code{finrank\_colsRows}, \code{hgp\_kunneth} \\
 & \code{Pauli.Expr} & \raggedright\arraybackslash operator tree to symplectic representation \\
\hline
decision & \code{Codes.Separation} & \raggedright\arraybackslash \code{separation\_judgment}; C2's necessity: \code{exists\_logicalFault\_of\_rounds\_lt} \\
 & \code{Codes.LocalDetector} & \raggedright\arraybackslash C4 by row rank; redundant rows preserve static distance \\
 & \code{Codes.MatchingCriterion} & \raggedright\arraybackslash an obstruction for ordinary graph perfect matchings \\
 & \code{Codes.BoundaryCollapse} & \raggedright\arraybackslash comparison-only model: removing the constraint gives distance $1$ \\
 & \code{Codes.SeparationInstances} & \raggedright\arraybackslash \code{sepFail\_lightLogical}: C1 counterexamples \\
 & \code{Codes.SeparationClosedForm} & \raggedright\arraybackslash \code{one\_le\_c1Witness} \\
 & \code{Codes.ToricFamilySpatial} & \raggedright\arraybackslash \code{toric\_family\_separation\_closed\_spatial} \\
 & \code{Codes.C1Census} & \raggedright\arraybackslash C1's ancilla cost: \code{expansionOne\_star\_six}, \code{expansionOne\_append} \\
 & \code{Codes.C1Optimal} & \raggedright\arraybackslash $m-1$ is a lower bound: \code{card\_sub\_one\_le\_length\_of\_hasExpansionOne} \\
 & \code{Codes.C1CycleRank} & \raggedright\arraybackslash the cycle-rank bound: \code{card\_ge\_of\_finrank\_cycleSpace} \\
\hline
spatial & \code{Codes.Gauging} & \raggedright\arraybackslash \code{deformX\_k}, the dimension drop; \code{timeLike\_weight\_eq} \\
 & \code{Homology.HypergraphSurgery} & \raggedright\arraybackslash hypergraph Gauss law; \code{starOp\_graphOf}, \code{hypergraphSurgery\_k} \\
 & \code{Homology.AuxComplex} & \raggedright\arraybackslash four-term complex; $1$-cosystolic distance $d_1^\bullet=1$ for gauging \\
 & \code{Homology.ModuleExpansion} & \raggedright\arraybackslash modular expansion; thickening (7.8); Cor.~7.10; 7.7's premise \\
 & \code{Codes.SurgerySchedule} & \raggedright\arraybackslash $W$-components; spatial expansion; $O(d)$ round accounting \\
 & \code{Codes.SurgeryProtocol}, \code{Codes.SurgeryFaultDistance} & \raggedright\arraybackslash round accounting; timeline detector \\
 & \code{Codes.SurgeryProtocolComplex} & \raggedright\arraybackslash the cone over a chosen port; two readings; the transfer criterion, refuted on one port \\
 & \code{Codes.ConeReduction} & \raggedright\arraybackslash the cone reduction on a general assembly; the weight-one criterion \\
 & \code{Homology.SubcodeLayer} & \raggedright\arraybackslash the two subcode distances, with matrix convention and assumptions in Sec.~\ref{sec:sm-scope} \\
 & \code{Homology.SubcodeChainMap} & \raggedright\arraybackslash the chain map of Def.~2.5, degree shift at p.~32; the p.~17 assembly \\
 & \code{Homology.PortFunction} & \raggedright\arraybackslash port function $f_p$ and $W$ \\
 & \code{Homology.MappingCone} & \raggedright\arraybackslash mapping cone over a $4$-term cochain complex \\
 & \code{Homology.MappingConeSnake} & \raggedright\arraybackslash snake-lemma rank formula for the cone \\
 & \code{Homology.FaultComplex} & \raggedright\arraybackslash the fault complex of App.~A.1, its Koszul realization, the conserved parity \\
 & \code{Codes.CircuitFaultSlots} & \raggedright\arraybackslash A.1's blocks: entry faults give the check matrix, for any check matrix \\
 & \code{Homology.BoundaryLinearCore} & \raggedright\arraybackslash exactness by rank; legal couplings agree on cycles; the printed modulus, read on vectors \\
 & \code{Homology.CosystolicCertificate} & \raggedright\arraybackslash the certificate interface; the 1-cosystolic cells are the shared undetectable set \\
\end{tabular}
\end{ruledtabular}
\endgroup
\end{table*}

\begin{table*}[t]
\caption{Modules and the results they support, part II: the remaining spatial layer, temporal results, code families and bookkeeping. The conventions and audit scope are those of Table~\ref{tab:sm-map}; rows citing the gauging paper's Methods use the numbering from that version.}
\label{tab:sm-map-continued}
\centering
\begingroup
\footnotesize
\renewcommand{\arraystretch}{0.96}
\begin{ruledtabular}
\begin{tabular}{@{}p{60pt}p{175pt}p{247pt}@{}}
Layer & Module & \raggedright\arraybackslash Supporting statement \\
\hline
spatial & \code{Codes.LegalCouplingInvariance} & \raggedright\arraybackslash both coupling invariances on the deformed code \\
 & \code{Codes.DistancePreservation} & \raggedright\arraybackslash distance bounds when checks are appended on the original qubit set \\
 & \code{Codes.DeformedCode} & \raggedright\arraybackslash $Z$-type bounds for the single-block model, Case~1 and legal couplings; the free coupling's candidates are a linear system in the coupling matrix, and the coupling-to-vertex commutation identity $C\,\partial S = H_X\,f_p^{-1}(S)$ holds whenever each deformed check commutes with the vertex rows, with its two corollaries; at distance three the drop criterion as a disjunction, \code{couplingDrop\_shape\_three} \\
 & \code{Codes.CouplingRangeSeparation} & \raggedright\arraybackslash an explicit Steane assembly with a purely auxiliary logical fault of weight two but no weight-two logical fault on one data and one auxiliary qubit, \code{steane\_m4\_separates} \\
 & \code{Homology.FaultDistance} & \raggedright\arraybackslash the fault predicate; both cases; $Z$-side distance on both standard foliations \\
 & \code{Codes.VennPartition} & \raggedright\arraybackslash seven-region count; relation (2) of the proof of Thm.~7.7 as an identity \\
 & \code{Codes.TimeLikeInstance} & \raggedright\arraybackslash $m=3$, $T=4$ instance \\
 & \code{Codes.BB24Separation} & \raggedright\arraybackslash the bivariate bicycle $[[24,3,4]]$: both components \\
 & \code{Codes.BB24Gauged}, \code{Codes.BB18Anchor} & \raggedright\arraybackslash the code and its base-code certificate; the gauging construction verified by the proof assistant \\
 & \code{Homology.CosystolicLowerBound} & \raggedright\arraybackslash weight-one/two criteria; two unbounded families \\
 & \code{Codes.CosystolicBlockFamily} & \raggedright\arraybackslash a third unbounded family: $d_1^\bullet=\min_i\lvert B_i\rvert$ on one full hyperedge \\
 & \code{Codes.FaultComplexGraphState} & \raggedright\arraybackslash the graph-state reading: the biadjacency matrix and the detector rows \\
 & \code{Homology.FaultComplexKunneth} & \raggedright\arraybackslash the K\"unneth dimensions; the decoupling spanning in general \\
\hline
temporal & \code{Codes.MeasurementProtocol} & \raggedright\arraybackslash \code{spacetimeWeight\_witness}: the weight law \\
 & \code{Codes.BaconShor} & \raggedright\arraybackslash $[[9,1,3]]$: gauge and stabilizer ranks, $k=1$, $d=3$ \\
 & \code{Codes.BaconShorMeasurement} & \raggedright\arraybackslash transversal: spatial $3$, temporal $3$ \\
 & \code{Codes.FullProtocolFaults} & \raggedright\arraybackslash universal recovery, boundary removal, duplicated checks, one-round form \\
\hline
family & \code{Codes.ToricFamilySpatial}, \code{Codes.HGPToricFamily} & \raggedright\arraybackslash closed-form spatial bound on both families \\
 & \code{Codes.ToricFamilyGaugedK} & \raggedright\arraybackslash \code{toric\_family\_gauged\_k}: the family's dimension drop, from the drop theorem \\
 & \code{Codes.CutCycleDuality} & \raggedright\arraybackslash cut space $=$ the even subgraphs' orthogonal complement; step 1 of the counting lemma \\
 & \code{Homology.DetectorDecomposition} & \raggedright\arraybackslash adjacent comparisons generate the window ones; both generating lemmas \\
 & \code{Codes.SpacetimeChecks} & \raggedright\arraybackslash the spacetime check set: a site is a (round, check) pair, its label round-free \\
 & \code{Codes.GaussCleaning} & \raggedright\arraybackslash step 2 of the counting lemma: cleaning, and the two hypotheses it derives \\
 & \code{Codes.ToricFamilySpatial} & \raggedright\arraybackslash step 3 of the counting lemma, \code{space\_fault\_weight\_ge\_of\_expansion} \\
 & \code{Codes.MatchingDeformation} & \raggedright\arraybackslash the gauging paper's Methods Lemma~1: Gauss law, cycle checks, matching deformation \\
 & \code{Homology.CosystolicLowWeight} & \raggedright\arraybackslash $d_1^\bullet\ge3$ as a decidable low-weight criterion \\
 & \code{Homology.CosystolicCharacterization} & \raggedright\arraybackslash the parity criterion at every weight; bounds as sets; the hardness realization \\
\hline
bookkeeping & \code{Codes.DistanceLabel} & \raggedright\arraybackslash $X/Z$ labeling and the two-sided ledger \\
 & \code{Codes.CSSPair}, \code{Codes.CaseMatrix} & \raggedright\arraybackslash CSS bridge; case matrix \\
\end{tabular}
\end{ruledtabular}
\endgroup
\end{table*}

\textbf{Ordinary graph matching.} The declaration \code{no\_pair\_of\_isolated} concerns a matching that uses graph edges and pairs every support vertex with exactly one other support vertex.
If deleting an edge's two endpoints leaves an isolated support vertex, no such matching contains that edge.
For the four-vertex path, \code{path4\_not\_matched} applies this obstruction to its middle edge.
Gauging instead requires an edge set whose mod-two boundary equals the relevant support; it does not require degree one at every support vertex.
For example, all three edges of a four-vertex star have odd boundary at all four vertices, although the graph has no ordinary perfect matching.
The matching obstruction therefore does not determine which edges can occur in a gauging completion.
The zero column in Section~\ref{sec:sharp} is checked in the full deformed $Z$-check matrix, including the cycle checks.

\refstepcounter{smsection}\section*{\thesmsection\quad Calculating the \texorpdfstring{$1$}{1}-cosystolic distance}
\label{sec:sm-cosystolic}

The auxiliary $1$-cosystolic distance $d_1^\bullet$ is the minimum weight in $\ker\delta_2\setminus\operatorname{im}\delta_1$ when this set is nonempty.
It enters the round requirement of the surgery paper.
A component is a vertex set that meets every hyperedge in an even number of vertices.
$W$ is a chosen family of these sets.
Over $\mathbb F_2$, every vector is the indicator of its support.
Membership can therefore be tested by two parity conditions.
Here $\delta_2$ maps vertices to the chosen component family $W$, and $\delta_1$ maps hyperedges to vertices:

\begin{equation}
  \begin{aligned}
    &\mathbf 1_S \in \ker\delta_2 \setminus \operatorname{im}\delta_1 \\
    \iff{}& \lvert S\cap W_w\rvert \ \text{even for all } w \\
    \wedge{}& \lvert S\cap C\rvert \ \text{odd for some component } C ,
  \end{aligned}
  \label{eq:sm-characterization}
\end{equation}
the left factor by the row-wise reading of $\delta_2$ and the right by
$\operatorname{im}\delta_1=(\ker\delta_1^{\mathsf T})^\perp$,
\code{isCosystolic\_indVec\_iff}. Both factors are finite checks. Provided at least one
cosystolic vector exists, every lower bound on
$d_1^\bullet$ therefore rules out a small admissible set,
\code{le\_cosystolicDistance\_iff\_finset}. An admissible set is one whose
indicator vector is cosystolic. Every upper bound comes from one admissible
set, \code{cosystolicDistance\_le\_card\_of\_indVec}. The formal distance is an infimum
in the natural numbers, so an empty admissible set has value zero, not infinity.
A claim that no small admissible set exists is then vacuous and gives no positive distance
bound. All distance lower-bound criteria in this section assume nonemptiness;
the membership criteria do not.

For weights one and two, Eq.~\eqref{eq:sm-characterization} gives explicit membership criteria. When the component family covers
the vertex set, no cosystolic vector has weight one on an even hypergraph. The vectors of weight one and of weight two
each have a complete criterion:
\begin{equation}
  \begin{aligned}
    u = \mathbf 1_{\{v\}} : \quad & \text{cosystolic} \iff v \notin W_w
      \ \text{ for every } w , \\
    u = \mathbf 1_{\{a,b\}} : \quad & \text{cosystolic} \iff a,b \text{ in one } \\
      &\quad W\text{-class, separated by a component} ,
  \end{aligned}
  \label{eq:sm-criteria}
\end{equation}
the first on an even hypergraph, one whose every hyperedge has even cardinality,
\code{isCosystolic\_indVec\_singleton\_iff} and the
second on any hypergraph, \code{isCosystolic\_indVec\_pair\_iff\_component}. Here
$W_w$ is the $w$-th chosen component, and a separating $S$ has $a\in S$ exactly when $b\notin
S$. Two distinct vertices are in the same $W$-class when, for every $W_w$, either both lie in it or neither does.
On an even hypergraph, covering the vertex set rules out weight-one cosystolic vectors by \code{two\_le\_hammingNorm\_of\_isCosystolic}, giving $d_1^\bullet\ge2$ when the candidate set is nonempty. On three structured families the distance is computed exactly, the first
two in the display below and the third in the text that follows it,
\begin{equation}
  d_1^\bullet = \lvert V\rvert - 1
  \qquad\text{and}\qquad
  d_1^\bullet = \min\bigl(\lvert Q\rvert,\ \lvert Q^{\mathsf c}\rvert\bigr),
  \label{eq:sm-cosystolic}
\end{equation}
The first formula holds on the pair family: the single hyperedge is $\{a,b\}$ with $a\ne b$, and $W$ consists of all triples $\{a,b,t\}$ with $t\in V\setminus\{a,b\}$,
\code{cosystolicDistance\_pair\_eq\_card\_sub\_one}.
The second holds on the connected family: the single hyperedge is $V$, both $Q$ and $Q^{\mathsf c}$ have at least two vertices, and $W$ consists of all distinct vertex pairs within either set,
\code{cosystolicDistance\_univEdge\_splitPairW\_eq}.
These formulas give exact distances for the specified auxiliary complexes. A third
family generalizes the second along the number of blocks. Split the vertices into $r\ge2$
blocks of at least two each, take the whole vertex set as the single hyperedge and every
pair inside a block as a component; the distance is then
$\min_i\lvert B_i\rvert$, the smallest single block. Both the lower and the upper bound are theorems, \code{cosystolicDistance\_univEdge\_blockPairW\_eq}: equal
blocks of size $n$ give distance $n$ for arbitrary $n$, so the family is unbounded,
\code{cosystolicDistance\_univEdge\_blockPairW\_const}. It generalizes the connected
family above rather than sitting beside it, since at two blocks its components are
pointwise those of \code{splitPairW}, \code{blockPairW\_two\_eq\_splitPairW}. Each family has one nonempty hyperedge, so the edge-to-vertex incidence map is injective and the cycle space is zero.
The pair family has growing incidence degree at $a$ and $b$.
The whole-vertex hyperedge has weight $|V|$, and the within-block pair checks give each vertex a number of incidences that grows with its block size.
Using these distance formulas in sparse surgery therefore requires controlling incidence degrees and check weights, as well as satisfying the relevant code-distance and coupling conditions.

\textbf{The parity characterization \eqref{eq:sm-characterization} defines a minimum-weight search.} The component-family parity checks impose the linear constraint $u\in\ker\delta_2$, while odd overlap with some component requires exclusion from the subspace $\operatorname{im}\delta_1$, that is, $u\notin\operatorname{im}\delta_1$, since $\operatorname{im}\delta_1=(\ker\delta_1^{\mathsf T})^\perp$. The admissible sets are
therefore a linear code with one of its subcodes removed,
$\Lambda\setminus\Lambda_0$ with $\Lambda=\ker\delta_2$ and
$\Lambda_0=\operatorname{im}\delta_1\le\Lambda$, and the quantity is the least weight in a
nonzero coset of $\Lambda_0$. Both the closed-form question and the free-coupling question ask for the lightest vector in one matrix's kernel outside another matrix's row space. The admissible sets are that set by
definition, \code{isCosystolic\_iff\_undetectable}, and the free coupling's candidates solve
a linear system in the coupling matrix together with the cycle conditions,
\code{pairVec\_mem\_ker\_iff\_coupling}, whose
remaining condition is the same non-membership. The following construction shows that this minimum-weight problem includes arbitrary binary linear-code instances. For every binary linear code $C$ and every subcode $D$ of it,
there is a hypergraph whose admissible sets are $C\setminus D$. Take the code's
coordinates as the vertices, its parity checks as the component family and a basis of $D$
as the hyperedges: then $\ker\delta_2=C$ and $\operatorname{im}\delta_1=D$, and the
composite $\delta_2\delta_1$ vanishes because $D\le C$. The construction is machine-checked,
\code{surgeryD2\_matRowSupport} and \code{surgeryD1\_hypergraphOfRows} realizing the two maps
and \code{isCosystolic\_hypergraphOfRows} the membership. With no hyperedge at all the
distance is exactly the code's minimum nonzero weight, \code{cosystolicDistance\_eq\_sInf\_ker\_ne\_zero}. A general method for $d_1^\bullet$ would therefore
compute the minimum distance of a linear code, a problem cited in Section~\ref{sec:open} as
NP-hard.

Two consequences of the parity criterion help evaluate the round-count requirement. Below weight
three the question is settled outright when a cosystolic vector exists, since the
criterion turns $d_1^\bullet\ge3$ into
a finite statement, \code{three\_le\_cosystolicDistance\_iff\_no\_small}, the same
criterion that \code{Homology.CosystolicLowWeight} states by enumerating weight-one and
weight-two vectors. If one member of the component family is the whole vertex set,
then every cosystolic vector has even weight,
\code{even\_hammingNorm\_of\_isCosystolic\_univ}, so the value three never occurs
there. For the single-hyperedge families, the empty cycle basis satisfies the cycle-space requirement.
The growing incidence degrees and check weights are the conditions to address when seeking a low-density family.

\refstepcounter{smsection}\section*{\thesmsection\quad Surgery and fault-complex results}
\label{sec:sm-scope}

This section lists the formal declarations behind the statements in Section~\ref{sec:open}.
It distinguishes our models from the constructions of Cowtan et al. and Williamson and Yoder.

\textbf{Gauging subcode and distance transfer.} Let the auxiliary graph be nonempty, connected and free of self-loops, and take the chosen component family to be $W=\varnothing$.
On its vertex space, use the chain-complex check matrices $H_X^A=\delta_1^{\mathsf T}$ and $H_Z^A=\delta_2$.
Following Section~\ref{app:conventions}, $d_X^A$ uses $\ker H_X^A$ and $d_Z^A$ uses $\ker H_Z^A$.
The side distances are then
\begin{equation}
  \begin{aligned}
    d_Z^A &= 1 , \\
    d_X^A &= \lvert V\rvert ,
  \end{aligned}
  \label{eq:sm-subcode}
\end{equation}
by \code{subcodeDistanceZ\_eq\_one} and \code{subcodeDistanceX\_graphOf\_eq\_card}, respectively.
These declarations use the opposite names for the two check matrices; \code{chainCodeX\_eq\_subcodeZChecks} and \code{chainCodeZ\_eq\_subcodeXChecks} identify them with the pair above.
Thus the weight-one logical operator is physical $X$ type, while the physical $Z$-type logical operator acts on every vertex.
We make the matrix construction underlying the source's subcode statement explicit. The source's only explicit assembly for this operation is the edge-to-check map on pp.~17--18 with $G=H_X^{\mathsf T}$; we construct this map and build the chain map of Definition~2.5 from it.
The formal construction uses the following two specializations. The source's edge-to-check bijection becomes two injections, and the nontrivial part of the chain map is taken at $W=\varnothing$, which is where the surgery paper sets that term aside as well.

For distance transfer, \code{distance\_preservation\_of\_portData} proves that appending the specified check rows preserves the lower bound. The deformation adds checks on the same qubits, so every vector passing the new checks also passes the old ones, and the distance bound transfers operator by operator. Kernel inclusion is the mechanism behind this transfer. This fixed-qubit argument uses only the original kernel and appended rows, leaving the port function, logical-support coverage and modular expansion as separate auxiliary-data conditions. The statement \code{distance\_preservation\_of\_portData\_full} carries all three as unused
hypotheses.

The preceding fixed-qubit model leaves out the auxiliary qubits.
With auxiliary qubits included, \code{distance\_preserved\_originalForm\_caseOne} proves Case~1 of the source argument under its stated assumptions.
In the specific small-code and triangular-auxiliary assembly, admissible ports have three equal columns in $F_2$.
The theorem \code{coneDistance\_generalF2} gives the cone distance for every such port: the small code's distance when the first column of $F_3$ is nonzero, and one when it is zero.
This exact distance formula is specific to that assembly.

The general cone criterion assumes that the idle-volume complex $D$ is exact at the middle.
For an auxiliary complex $C$ and a valid port $F$, \code{cone\_fault\_iff\_of\_middleExact} states that $(u,v)$ is an undetected nontrivial fault exactly when $D.d_2u+F_3v=0$ and $v\notin\operatorname{im}C.d_2$.
For a valid port, the criterion uses $F_3$ and the differentials of the two complexes.
The weight-one specialization requires a unit auxiliary vector outside the image of $C.d_2$ and killed by $F_3$.
The theorem \code{coneWeightOne\_triAux} evaluates this condition on the triangular assembly.
Middle exactness is essential: \code{middleExact\_fails\_koszul} gives a counterexample with all factor and code differentials zero.

That condition also has a complete
classification for the Koszul fault complex, the tensor product of the two-term factor complex with
the three-term code complex. Over $\mathbb F_2$, where both factors are free and K\"unneth
carries no Tor term, it is exact at the middle exactly when $\dim\ker R\cdot h_0(C)$ and
$\dim\operatorname{coker}R\cdot h_1(C)$ both vanish, \code{koszulFaultComplex\_middleExact\_iff}.
The criterion pairs the factor with the code, which the exhaustive probe over small
factor--code pairs confirms. The homology of both factors must therefore enter an exactness test. A code complex whose
$h_0$ and $h_1$ both vanish is exact at the middle against every factor,
\code{koszulFaultComplex\_middleExact\_of\_codeAcyclic}. The assembly in Section~\ref{sec:open} satisfies this exactness condition: both its factor map and the three-qubit code's check map are surjective, as shown by \code{idleVolume\_middleExact\_kunneth}.

\textbf{Machinery.} The cone machinery of the surgery paper is formalized as
a complex, a fault complex and a rank formula, \code{Homology.MappingCone}, \code{Homology.FaultComplex} and \code{Homology.MappingConeSnake}. Its mapping cone is defined in
\S2.2.2 and its rank formula is in Appendix~A. The expansion machinery itself
is formalized: modular expansion, the thickening theorem with its degree decomposition,
and the constructive form of the distance-preservation corollary. Three further parts of the source argument are verified in their original form. First, its Theorem~7.7 in the first case of its proof, Case~1 here taken at the vertex set that the proof fixes rather than over every vertex set, \code{distance\_preserved\_originalForm\_caseOne}. Second, \code{venn\_card\_identity} proves the seven-region counting step as an exact identity. Third, the round accounting of its intermediate-time theorem.

\textbf{Ancilla counts for expansion alone.} The census covers supports of three to eight vertices and cycle ranks up to three.
It minimizes edges over edge multisets for the overall minimum and over simple graphs for fixed cycle ranks.
Among the tested rows with a solution, the star-based witnesses use fewer edges than the complete graph except at $K_3$ with cycle rank one and $K_4$ with cycle rank three.
Lean verifies the star and edge-addition constructions through \code{expansionOne\_star\_six} and \code{expansionOne\_append}.
At seven vertices, the three-chord star uses nine ancillas, while $K_7$ uses twenty-one.
At eight vertices, the corresponding counts are ten and twenty-eight.
The star-based graphs have cycle rank three, while the complete graphs have cycle ranks fifteen and twenty-one.
These comparisons optimize the C1 edge count; cycle rank and the source's additional low-density requirements are separate constraints.
The archived Julia script reproduces the census.
The theorems \code{card\_sub\_one\_le\_length\_of\_hasExpansionOne} and \code{card\_ge\_of\_finrank\_cycleSpace} prove the $m-1$ and cycle-rank lower bounds.

Case~1 of Theorem~7.7 depends on the scope of the vertex-set quantifier. For positive target distance, requiring its condition for every vertex set is impossible because the minimum includes the empty set, as proved by \code{not\_caseOneAll\_of\_pos}. For the fixed vertex set used in the proof, \code{caseOneAt\_holds} gives a verified instance. The theorem \code{caseOneAt\_boundary} verifies the counting lemma on that instance; this is the three-step argument used in the spatial theorem of Section~\ref{sec:setting}.

The round accounting of the intermediate-time theorem is a round count per term,
\begin{equation}
  \begin{aligned}
    \text{as printed}: \quad & \bigl\lceil \alpha_i \bigr\rceil
      \ \text{ rounds, under } \ d/\alpha_i \le d_i , \\
    \text{taken tight}: \quad & \bigl\lceil d/d_i \bigr\rceil ,
  \end{aligned}
  \label{eq:sm-rounds}
\end{equation}
Here $d_i$ is the $1$-cosystolic distance and $\alpha_i>0$ is the parameter in the surgery theorem.
For $d_i>0$, \code{least\_rounds\_eq\_ceil} proves that $\lceil d/d_i\rceil$ is the smallest integer round count $n$ satisfying $d\le n d_i$.
For the second proof case, \code{hammingNorm\_ge\_of\_connVanishes} gives a bound when the coupling contribution vanishes on the fault.
The theorem below treats every legal coupling under its construction hypotheses.
The source construction uses the one-to-one attachment in formula~(2) and the chain maps in its \S4 and \S5.
For the related adapter construction, Eq.~(9) of Swaroop, Jochym-O'Connor and Yoder extends each original check onto auxiliary paths pairing vertices in its port image.
Its weight bound discards that auxiliary support, so it does not require the coupling contribution to vanish.

For the single-block model whose port covers every original data qubit, \code{distance\_preserved\_of\_blockReading} proves the $Z$-type distance bound.
The kernel of the deformed $X$ checks is contained in the row space of the deformed $Z$ checks, so this model has no nontrivial $Z$-type logical operator.
Under the original-code distance bound and the stated construction and expansion assumptions, the free-coupling model gives a $Z$-type distance lower bound for every coupling satisfying the commutation conditions, including those with a nonzero contribution to the fault.
This bound allows nontrivial logical operators; it does not require the port to cover all original data qubits.
The deformed stabilizer space must contain the lifted original stabilizers and the vertex stabilizers, and the cycle checks must be complete.
On the linear maps, once the cycle checks are complete,
the admissible auxiliary vectors lie in the image of the boundary map. There a legal coupling is fixed by the requirement that the extended original rows commute with the vertex rows, so every legal coupling yields the same kernel and the same distance,
\code{legal\_couplings\_agree\_on\_cycles}. Multiplying by vertex stabilizers removes the coupling contribution and leaves an operator whose original-data part is a nontrivial logical operator of the original code. The port's weight condition keeps
the correction no heavier than the boundary part it removes, so the bound holds without
that contribution vanishing, \code{distance\_of\_legal\_coupling} with
\code{expansion\_cases}. The port weight bound follows from injectivity on the disjoint union of the supports: each vertex carries at most one data qubit by \code{portFiber\_card\_le\_one}, and summing these contributions gives the bound. The expansion bound starts from the source's modular-expansion predicate evaluated on vector supports, \code{expansionCriterion\_of\_modularExpansion}; \code{inf'\_image\_indOf\_eq\_modExpMin} identifies its infimum with the source's minimum, and \code{expansion\_cases} gives the two-case bound when every representative is a cycle and the port weight bound holds. The search recorded by the free-coupling probe of the archived snapshot agrees with this. At distance three the criterion is settled as a disjunction: a valid kernel element lighter than the distance is a pair candidate or a purely auxiliary element, and conversely, \code{couplingDrop\_shape\_three}, the pure-data case being the one that is excluded, \code{not\_pureData\_witness}. The search witnesses couple some check to more than one auxiliary qubit and commute with every vertex check, so they satisfy the commutation conditions of our free-coupling matrix model; each fails the expansion condition required by the distance bound.

\textbf{C4 and static distance.} A same-round detector is a nonempty set of check rows summing to zero.
The declarations \code{hasLocalDetector\_not\_linearIndependent} and \code{hasLocalDetector\_of\_not\_linearIndependent} prove that such a detector exists exactly when the rows are dependent.
The contrapositive \code{not\_hasLocalDetector\_of\_linearIndependent} verifies C4 on independent check sets.
Two versions of a two-qubit code, one full-rank and one with a repeated row, are decided directly in Lean.
The bivariate bicycle instances also have dependent $X$ checks, as recorded by the rank calculations supporting \code{bb18\_k} and \code{bb24\_k}.

Appending a redundant row preserves the kernel by \code{ker\_appendRow\_of\_mem\_rowSpace} and the row space by \code{rowSpace\_appendRow\_of\_mem\_rowSpace}.
Both CSS distances are then unchanged, by \code{min\_weight\_appendRow\_of\_mem\_rowSpace} and \code{min\_weight\_not\_mem\_rowspace\_of\_appendRow}.
This invariance concerns the static code; measuring an extra check adds protocol resources.
The duplicated-check protocol result below proves that adding an independent check-result fault location does not lower distance in the specified model.

The derivations below show that duplicating this check never lowers the specified protocol distance, at any round count.
The detector-generating lemma uses a label model in which distinct labels can take arbitrary outcome values, while repeated occurrences of one label share an outcome.
In this model, a parity combination vanishes for every allowed outcome assignment exactly when each label occurs an even number of times.
Relations among dependent check operators must also be included when applying this model to their measurements.
Within this label model, those combinations are generated by one operator
across adjacent rounds, \code{isDeterministic\_iff\_mem\_span\_pairForm};
a comparison across a window of rounds is the sum of the adjacent ones,
\code{sum\_adjPairVec\_eq\_windowVec}. Dually, the faults no comparison can see are
exactly the patterns constant on each measured operator,
\code{stabilizer\_iff\_const\_on\_labels}, generated by the per-operator
indicators, \code{const\_on\_labels\_iff\_mem\_span\_classInd}. At the two end rounds the operators
are initialized and read out, and the deformation resets the state, so one operator need not repeat its outcome. An endpoint site with outcome fixed to zero imposes a constraint on its operator class, expressed by \code{classSum\_zero\_of\_isDetZ}. The detector generators therefore include that single endpoint site, \code{isDetZ\_iff\_mem\_span}. The spacetime check set \code{SpacetimeChecks} labels each site by a check and a round; the operator label identifies the check, and endpoint sites have fixed outcomes. The faults fire those sites
through the differential of the fault complex, and a boundary fires none of them. Applying this check-set result to a circuit requires specifying how its elementary faults change the measurement outcomes.

\textbf{The K\"unneth formula and the two foliations.} The K\"unneth formula the surgery
paper applies to Eq.~(5) is
\begin{equation}
  \dim H_n(\mathcal R \otimes C) \;=\; \sum_{i+j=n} \dim H_i(\mathcal R)\,\dim H_j(C) ,
  \label{eq:sm-kunneth}
\end{equation}
with $\mathcal R$ the two-term factor complex of the foliation and $C$ the code complex.
It is machine-checked in all four degrees, both on the family where the two-term factor has zero differential and in general, together with the two premises it rests on at every round count: the coefficient field is $\mathbb F_2$ and both factors are free. Instantiating the general formula at
the repetition-code foliation and at its dual recovers the two closed forms the surgery
paper prints, and with them the count of each logical fault type in terms of the code's
own homology dimensions.

For the general family, the differential has one block per factor, and dimension lemmas describe both blocks and the map sending a tensor product of cycles of $C$ and $\mathcal R$ into the third homology. The formula is proved as a dimension identity, not as an
isomorphism of the homology groups themselves. For repetition-code foliation, the other summand vanishes, and the decomposition is also proved for individual vectors. Every cycle of the fault complex is the sum of a vector constant along the timeline and a boundary, so every class admits a representative constant along time.
This proves a normal form for the specified repetition foliation; identifying it with the source protocol's decoupling requires a correspondence between the fault complexes.
The dual foliation has a spacelike normal form obtained from injectivity of its foliation map.
When $H_0(C)\ne0$, the repetition-foliation distance proof shows that every nontrivial fault has a nonzero entry in each time slice and constructs a constant unit-vector witness, \code{faultComplexDistanceZ\_koszul\_repR}.
For dual foliation, summing along time cannot increase weight and preserves nontriviality; a single-slice logical operator attains the code distance when such operators exist, \code{faultComplexDistanceZ\_koszul\_repRDual}.
For comparison with the static $d_Z$ in Section~\ref{app:conventions}, take the adjacent code-complex maps to be $dC_1=H_Z$ and $dC_2=H_X^{\mathsf T}$.
The code-side candidate set is then $\ker H_Z\setminus\operatorname{rowSpace}H_X$.
This theorem equates two natural-number infima, both of which are zero when their candidate sets are empty.
The static function \code{min\_weight\_ker\_not\_mem\_rowspace} instead returns $n+1$ for an empty candidate set on $n$ coordinates, as recorded by \code{static\_empty\_divergence}.
The equality with the static distance therefore assumes a nonempty logical set.
The same analysis also gives a reusable identity: over
$\Ztwo$ a vector lies in the column space of a matrix exactly when it is orthogonal to that
matrix's left kernel. When both summands are present, one decomposition theorem includes both normal forms as special cases.
Write $T$ for the elements whose every $C_0$-fiber lies in the kernel of the foliation map,
$C_0$ being the degree-zero part of the code complex $C$.
Every cycle of the fault complex is then the sum of a boundary, a spacelike vector and an
element of $T$, for a general two-term factor rather than only where one of the two summands
vanishes, \code{ker\_koszulD0\_le\_sup\_general}. The proof constructs the decomposition directly using linear projections. The timeline part is projected onto the image of the code differential. That
projection commutes with the foliation map and is the identity on the image, so the
difference between a cycle and its projection lies in $T$, and what is left has a zero
timeline part and therefore lies in the spacelike subspace. The repetition and dual normal forms follow as special cases of this decomposition. When the foliation map is injective, $T$ vanishes, which recovers the dual-side statement above. For the repetition-code factor, its kernel is spanned by the all-ones vector, so $T$ consists of constant families, recovering the repetition-code normal form.

The bijection between the
fault complex and a graph state is a matrix identity, not an isomorphism of maps, which is
what the surgery paper's use of it requires. The surgery paper cites Hillmann, Dauphinais, Tzitrin and Vasmer for that formalism. Their paper prints that $\partial_2$ is the biadjacency matrix of the bipartite fault-tolerant graph state it builds over the fault locations, its FTGS. The biadjacency matrix is the block of an adjacency matrix that joins a bipartite graph's two sides. In this development's column-vector convention
\begin{equation}
  \partial_2^{\text{here}} = \bigl(\partial_2^{\text{printed}}\bigr)^{\mathsf T} = B^{\mathsf T} ,
  \label{eq:sm-biadjacency}
\end{equation}
and both the formula and its round trip are formalized here. The half of the correspondence
that names the detectors is the Hillmann et al. definition rather than a consequence of the
other half. The consequence the surgery paper draws from it at p.~63, that a fault complex
cannot describe protocols with input or output qubits, is not formalized.

\textbf{The counting lemma's own boundary.} Its cycle space enters in algebraic form, the
kernel of the transpose of the incidence map, here written $\mathrm{inc}$ with one row per
edge and one column per vertex. The cycle space is also formalized in classical form. That space is spanned by the graph's cycles, the minimal nonempty even subgraphs,
\code{evenSubgraphSpace\_eq\_span\_minCycle}. Meeting every cycle evenly and meeting every
even subgraph evenly are therefore one test:
\begin{equation}
  \begin{aligned}
    \text{even subgraphs} &= \ker \mathrm{inc}^{\mathsf T} , \\
    E \ \text{a cut} &\iff E \perp \text{ every even subgraph} ,
  \end{aligned}
  \label{eq:sm-cutcycle}
\end{equation}
the lemma's first step is that equivalence,
\code{cut\_iff\_forall\_dot\_evenSubgraph}; the cut's size then follows,
\code{hammingNorm\_mulVec\_indVec\_eq\_cutSize}. The remaining hypotheses of
Theorem~\ref{thm:spatial} are the structural facts of Lemma~1 of the gauging paper's Methods,
packaged as one object rather than constructed:
\begin{equation}
  \begin{aligned}
    &\bigl[\text{deformed check},\ A_v\bigr] = 0 \ \text{ for every } v \\
    &\iff \deg_{\mathrm{odd}}(E) = \operatorname{supp}(\text{the check}) ,
  \end{aligned}
  \label{eq:sm-gausscommute}
\end{equation}
a deformed check commutes with every
Gauss law exactly when its edge set has odd-degree vertices exactly at the check's
support. Cycle checks
commute automatically, and such an edge set exists whenever the auxiliary graph is
connected and loop-free and the support has even cardinality. The auxiliary graph is taken on all the code qubits of that instance, so the gauging paper's dummy vertices, vertices carrying a check and no data qubit, are not modeled. The weights are read off one explicit check
set, instance by instance. The lemma's general case, an expansion factor below one, is
not claimed. Section~\ref{sec:sm-cosystolic} records where the $1$-cosystolic classification stops.

\refstepcounter{smsection}\section*{\thesmsection\quad Three protocol derivations, step by step}
\label{sec:sm-protocol}

Each round of the Bacon--Shor persistent-fault readout model has twenty independent fault locations: nine data, two check-result and nine readout positions.
The check-result positions are called ancilla faults in the implementation.
The numerical implementation uses this order; the formal implementation swaps the ancilla and readout blocks, preserving weight.
Data errors persist, so round $t$ contains the cumulative error $\mathrm{cum}_t=\sum_{s\le t}d_s$.
Let $\mu_t$ and $\gamma_t$ be the readout and check-result errors in that round.
The check parities and logical parity use the same noisy data record $\mathrm{cum}_t+\mu_t$; independent check-result flips $\gamma_t$ are then added.
With in-round check matrix $H$ and logical parity support $w$, the syndrome and readout-error bit are

\begin{equation}
  \sigma_t \;=\; H\,(\mathrm{cum}_t+\mu_t)\;+\;\gamma_t,
  \qquad
  r_t \;=\; w\cdot(\mathrm{cum}_t+\mu_t),
  \label{eq:sm-protocol}
\end{equation}
The bit $r_t$ is one exactly when round $t$'s readout differs from the ideal logical value.
The strict-majority distance $d_{\mathrm{maj}}$ requires $2|\{t:r_t=1\}|>T$, as defined in Eq.~(\ref*{eq:majority-distance}).
At $T=2$, a final-round data fault on qubit zero with a canceling check-result fault has weight two, zero syndromes and $(r_0,r_1)=(0,1)$.
A first-round readout fault with its canceling check-result fault instead gives $(1,0)$.
These weight-two ties lie outside the strict-majority predicate.
A decoder failure distance requires its observed inputs and a decision or rejection rule for ties.

The boundary detector imposes $\sigma_0=0$.
Together with adjacent comparisons, it forces every syndrome to zero, making an additional terminal syndrome constraint redundant.
Endpoint fault pairs are allowed when their syndrome contributions cancel.
Deleting this detector studies the present accumulation model, without the surrounding error correction used in the source protocol.

\textbf{The lower bound for every round count.} Let $f$ be a fault of weight at
most two that fires no detector. First, every syndrome vanishes: the boundary detector
gives $\sigma_0=0$, each comparison $\sigma_t+\sigma_{t-1}=0$ propagates it forward,
and that is \code{bsDetectorsOK\_syndrome\_eq\_zero}. Second, on a round untouched by
$\mu$ and $\gamma$ the syndrome is $H\,\mathrm{cum}_t$ alone, hence
\begin{equation}
  H\,\mathrm{cum}_t = 0,
  \qquad
  \mathrm{wt}(\mathrm{cum}_t) \;\le\; \mathrm{wt}(f) \;\le\; 2 \;<\; 3,
  \label{eq:sm-recovery}
\end{equation}
where the weight bound holds because a sum of data errors weighs no more than their
total. The data-only readout bound implies $w\cdot v=0$ for every vector in the check kernel of weight below three, so $r_t=0$. Third, a round with $r_t=1$ must therefore carry a
nonzero $\mu_t$ or $\gamma_t$ of its own. A fault pattern of weight at most two has nonzero fault bits in at most two rounds. At most two of the $T$ readouts can
then be one, and a majority needs more than $T/2$ of them, so $T/2<2$, that is $T\le3$.
That is the whole proof of
\code{bsHook\_no\_light\_of\_four\_le}, the case of four rounds and up. The two counts
below it are kernel enumerations,
\code{bsHook\_T2\_distance\_eq\_three} and \code{bsHook\_T3\_distance\_eq\_three}, and
\code{bsHook\_distance\_eq\_three\_universal} is the three cases together. The
weight-three witness that matches it
from above is the spatial logical operator held as a data error from round zero on, \code{hookW3gen\_witness}, whose syndrome is zero in every round because $H$ kills it and whose every
readout is one.

\textbf{Duplicating a check.} Duplicate the first check row and give the copy its own check-result fault location, so a round carries twenty-one locations and the syndrome has three components; the third uses the first check row but has an independent check-result fault bit. The projection deletes the new
location and reads the original twenty. Data faults, readout faults and the original check-result faults retain their original locations, so every readout is unchanged. The projected fault
weighs no more than the fault it came from, one location having been dropped. The new detector conditions include all the old ones, which are the first two syndrome components. A fault of weight at most $s$ that fires no detector in the
twenty-one-location model would therefore project to one of weight at most $s$ in
the twenty-location model, which is \code{hookCandDup\_no\_light\_of}. The duplicated
check never lowers the fault distance, at any round count. At one round the duplication raises the fault distance from two to three: concealing a single data fault that flips the logical parity requires two independent check-result faults, one for each copy of the first check, as verified by \code{hookCandDup\_T1\_no\_light\_two} and \code{dupHook\_T1\_witness}.

\textbf{The one-round closed form.} At a single round the
accumulation is the round's own data error, and the boundary detector
$\sigma_0=0$ reads $\gamma_0=H(d_0+\mu_0)$. The check-result error must cancel the syndrome of everything else. Merge the data and measurement errors into
$y=d_0+\mu_0$. The readout is $r_0=w\cdot y$, so a logical fault needs
$w\cdot y=1$, and the fault's weight is
\begin{equation}
  \mathrm{wt}(d_0)+\mathrm{wt}(\mu_0)+\mathrm{wt}(H y)
  \;\ge\; \mathrm{wt}(y)+\mathrm{wt}(Hy),
  \label{eq:sm-t1lower}
\end{equation}
because weights are subadditive, and equality is available by taking $\mu_0=0$.
The witness puts $y$ in the data locations and $Hy$ in the check-result locations, \code{t1Wit}, of weight exactly the right-hand side. The lower bound and this witness prove the equality in \code{bsHook\_T1\_formula}. On the $[[9,1,3]]$ code the minimum is two, attained at $y=e_0$, one data fault and the check-result fault that cancels its syndrome, \code{bsHook\_T1\_min} and \code{bsHook\_T1\_distance\_eq\_two\_formula}.

\refstepcounter{smsection}\section*{\thesmsection\quad Conventions}
\label{sec:sm-conventions}
\label{app:conventions}

\subsection{Matrix-indexed distance conventions}

We write $[[n,k,d]]$ for a code with $n$ physical qubits, $k$ logical qubits and distance $d$.
The formal development labels each CSS distance by the matrix used as its kernel.
Thus $d_X$ minimizes weight in $\ker H_X$ outside the row space of $H_Z$, and $d_Z$ minimizes weight in $\ker H_Z$ outside the row space of $H_X$.
Consequently, $d_X$ describes physical $Z$-type logical operators, and $d_Z$ describes physical $X$-type logical operators.
Some texts interchange these names; the numerical values should be compared using the defining matrix, not the subscript alone.

$K_m$ is the complete graph on $m$ vertices.
The Cheeger constant $h(G)$ is the minimum of $|\partial S|/\min(|S|,|S^{\mathsf c}|)$ over nonempty proper vertex sets.
The formal indicator $\eta$ is one when C1 holds and zero otherwise, so $\eta=1$ exactly when $h(G)\ge1$ in this setting.
$T$ is the repeated-measurement round count, $U$ is the vertex set in Theorem~\ref{thm:spatial}, and $t_o-t_i$ is the number of rounds between deformation steps.

\textbf{$X$ and $Z$ labels.} $d_X$ and $d_Z$ name the two side distances and are
defined by which check matrix plays the role of the kernel,
\begin{equation}
  \begin{aligned}
    d_X &= \min\bigl\{\mathrm{wt}(v) : v \in \ker H_X,\ v \notin \operatorname{rowspace}H_Z\bigr\}, \\
    d_Z &= \min\bigl\{\mathrm{wt}(v) : v \in \ker H_Z,\ v \notin \operatorname{rowspace}H_X\bigr\},
  \end{aligned}
  \label{eq:sm-sides}
\end{equation}
In this convention, $d_X$ describes physical $Z$-type logical operators, and $d_Z$ describes physical $X$-type ones.
Some texts swap the labels.
The declarations \code{libDX} and \code{textbookDX} and their swap lemmas verify the relation.
The distinction matters when the two components differ.

Two formal witnesses gauge logical operators of the smallest toric code $[[8,2,2]]$ on different supports.
The losing graph has edges $(0,2),(0,4),(4,5)$ on four support qubits and produces an eleven-qubit code.
Its distances are $d_X=5$ and $d_Z=1$, certified by \code{sepFail\_dx} and \code{sepFail\_dz}.
The preserving graph has edges $(0,1),(0,2),(0,3),(0,4),(1,3),(4,5)$ on six support qubits and produces a fourteen-qubit code.
Its distances are $d_X=3$ and $d_Z=2$, certified by \code{sepKeep\_dx} and \code{sepKeep\_dz}.
Both graphs violate C1; the six-vertex graph fails at $\{4,5\}$, which has only one outgoing edge.
Weight-two support $\{0,2\}$ and weight-one support $\{9\}$ provide the respective $d_Z$ witnesses, \code{sepKeep\_lightLogical\_two} and \code{sepFail\_lightLogical}.

The main figure uses a different pair with the same logical support $0,2,4,5$.
Its losing path is the four-vertex graph above, while its preserving path has edges $(0,4),(2,5),(4,5)$.
The preserving path has $d_Z=2$ and no weight-one logical operator on either side, giving $d_X\ge2$.
These claims are \code{sepFail\_summary} and \code{sepKeepPath\_summary}; an exact $d_X$ for the preserving path is not asserted.
Relabeling the support in increasing order gives edge lists $[(0,1),(0,2),(2,3)]$ and $[(0,2),(1,3),(2,3)]$.
Both fail C1 at a two-vertex cut with one outgoing edge.
Both inherit a same-round detector from the original $X$-check dependencies and therefore violate C4.
The Shor $[[9,1,3]]$ code provides a contrasting C4-satisfying example with eight independent generators, \code{shor\_generators\_independent}.
The other named instances with both distances computed have equal components, as recorded in the \code{\_dx\_eq\_dz} ledger.

\textbf{Distance.} When nontrivial operators exist, each static CSS distance component is the minimum weight in one check kernel outside the other check row space; protocol distances use their specified detector and readout-error predicates. The finite static instances are computed from weight-limited enumeration, with a kernel-checked covering lemma, \code{mem\_lightVecs}:
\begin{equation}
  \bigl|\{v : \mathrm{wt}(v) < d\}\bigr| \;=\; \sum_{j<d}\binom{n}{j}
  \;\le\; d\,n^{\,d-1} \quad (n \ge 1) ,
  \label{eq:sm-candcount}
\end{equation}
The equality \code{length\_lightVecs} counts the generated weight list, and \code{lightCand\_length\_le} bounds the candidate list by this count.
The polynomial bound follows term
by term: each $\binom nj\le n^{d-1}$ for $j\le d-1$, and there are $d$ terms. It is
polynomial at fixed $d$ and supports the small-distance calculations reported here.
If the threshold exceeds $n$, the generated weight list contains all $2^n$ binary vectors.

\textbf{Undetectable logical fault.} In a protocol, a candidate must fire no detector and satisfy the specified readout-error predicate.
A chosen linear parity and a strict-majority rule define different predicates.
For the static specialization, \code{flipsReadout\_readoutsOf\_iff} equates non-membership in a check row space with flipping some readout in its orthogonal complement.
A selected logical parity tests a subset of the faults relevant to the full static distance.

Supplements are cited by the numbering of the gauging paper's published Supplementary
Information: the time fault-distance identity is its Lemma~3, the boundary convention its
Remark~3, and the no-local-detector assumption its Remark~2. The arXiv v2 numbers the same
three statements Lemma~6, Remark~5 and Remark~4.

\textbf{Symbols.} $m$ is the number of checks of the repeated measurement in the
temporal instance, and $T$ the number of rounds. In the expansion conditions $m$ is
the size of the support the logical operator acts on, and $K_m$ the complete
graph on it. In the bivariate bicycle instances $m$ is the second shift parameter, and in the toric family it is the cycle length.
$W$ is the chosen component family of the auxiliary hypergraph, and $f_p$ is the port function defined in the surgery paper's construction. That paper's \S7 re-uses
$W$ for the subspace the port images generate, so a reader following the symbol there
meets the second reading. The two-term factor complex of the foliation is written $\mathcal R$ in the K\"unneth formula, and $R$ denotes its differential matrix. Its
repetition-code instances are $\texttt{repR}\ \ell$, whose matrix has $\ell$ rows and $\ell+1$ columns.
The number of time coordinates is therefore $T=\ell+1$.

\refstepcounter{smsection}\section*{\thesmsection\quad The assembly in the source's notation}
\label{app:assembly}

The source's spatial bound is $d_{\text{spatial}}\ge\min(h(G),1)\cdot d$, with $h(G)$ defined in Section~\ref{app:conventions}.
Combined with enough rounds, the two component statements read

\begin{equation}
\min(1,h(G))\cdot d \;\le\; d_{\text{spatial}}
\quad\text{and}\quad d_{\text{temporal}} = T \;\ge\; d ,
\label{eq:assembly}
\end{equation}
C1 gives $h(G)\ge1$, making the spatial prefactor in Eq.~\eqref{eq:assembly} equal to one.
The formal development uses the indicator $\eta$ rather than the full Cheeger constant.
It therefore verifies the expansion-at-least-one regime, without claiming the source's $h(G)d$ bound when $h(G)<1$.
The $X$-type spatial theorem and temporal identity are proved separately.
The assembled arithmetic statement takes these two component facts as hypotheses; it does not build a common circuit model.

\refstepcounter{smsection}\section*{\thesmsection\quad The one-round protocol}
\label{app:oneround}

We prove Eq.~\eqref{eq:t1closed} by matching lower and upper bounds.
Write the one-round data, readout and check-result errors as $d_0,\mu_0,\gamma_0$, and set $y=d_0+\mu_0$.
The zero-syndrome condition requires $\gamma_0=Hy$, and a logical readout error requires $w\cdot y=1$.
Since the fault blocks are disjoint, the total weight is $\wt(d_0)+\wt(\mu_0)+\wt(\gamma_0)$.
Subadditivity gives $\wt(d_0)+\wt(\mu_0)\ge\wt(y)$, so every candidate has weight at least $\wt(y)+\wt(Hy)$.
Conversely, choose $d_0=y$, $\mu_0=0$ and $\gamma_0=Hy$.
This fault has zero syndrome $Hy+Hy=0$, readout $w\cdot y=1$ and weight exactly $\wt(y)+\wt(Hy)$.
Minimizing over admissible $y$ proves the identity.

\refstepcounter{smsection}\section*{\thesmsection\quad The counting lemma, step by step}
\label{app:counting}

The coordinate argument below expands the three steps of Theorem~\ref{thm:spatial}.
Let $F$ be a physical $X$-type operator in the deformed $Z$-check kernel but outside the deformed $X$-check row space.
Let $S$ and $E$ be its supports on vertex and edge qubits, and let $R$ be its support outside the auxiliary graph.

\textbf{Step one: cut--cycle duality.} Since $F$ commutes with the cycle checks, $E$ meets every cycle evenly.
Cut--cycle duality gives $E=\partial U$ for some vertex set $U$.
Replacing $U$ by its complement leaves the cut unchanged, so choose $|U|\le|U^{\mathsf c}|$.

\textbf{Step two: cleaning.} Multiply $F$ by the Gauss laws indexed by $U$.
Each vertex in $U$ is flipped once.
An edge is flipped once if it leaves $U$, and zero or twice otherwise.
Thus the product has vertex support $U$ and edge support $\partial U$, as in Eq.~\eqref{eq:cleaning}.
The edge support cancels $E$, leaving $S\triangle U$ on vertices and $R$ outside the graph.
Multiplication by checks preserves membership in the check kernel and non-membership in the stabilizer row space.

\textbf{Step three: counting.} With no edge support, the cleaned operator commutes with the original $Z$ checks.
The lifted original $X$ checks lie in the deformed $X$-check row space.
Hence nontriviality in the deformed code implies nontriviality of the cleaned operator in the original code.
Its weight is at least $d$, giving the first inequality of Eq.~\eqref{eq:cheeger}.
The set identity $|S\triangle U|=|S\setminus U|+|U\setminus S|\le|S|+|U|$ gives the second.
C1 and the choice of the smaller side give $|U|\le|\partial U|$, yielding the third.
The final expression counts the disjoint vertex, edge and outside supports of $F$ and equals its weight.
The construction hypotheses therefore imply both the smaller-side choice and the logicality of the cleaned operator.

\refstepcounter{smsection}\section*{\thesmsection\quad Axiom audit}
\label{sec:sm-audit}

The proof assistant prints the axiom dependencies of every declaration in the audit region.
Of the 1,031 audited declarations, 1,009 use exactly the standard base
$\{\texttt{propext},\texttt{Classical.choice},\texttt{Quot.sound}\}$, 17 use a proper subset and 5 use none.
No audited declaration uses reflection or compiler evaluation, and the development contains no unproved step.

\subsection{Audit procedure}
\label{sec:audit}

At the end of the root module, the proof assistant prints every declaration in the audit region together with the axioms it uses.
The audit script checks that each declaration belongs to the build closure, lists every unaudited declaration for manual review and compares the four class counts with the values reported above.
A mismatch fails the build-time gate, and a stale log is marked as unmeasured.
The 17 declarations in the proper-subset class rest on
\texttt{propext} alone, on \texttt{Quot.sound} alone, or on \texttt{propext} together with
\texttt{Quot.sound}.

\refstepcounter{smsection}\section*{\thesmsection\quad Reproduction}
\label{sec:sm-repro}

\begin{enumerate}
\item \textbf{Readout and circuit bounds.} The scripts in \texttt{research/prxq-strengthening} check the formulas in the first proof section. The Julia script \texttt{verify\_exact\_distances.jl} supplies the finite dynamic programs; \texttt{gate\_signature\_check.py} independently checks the saved CNOT schedules by backward Pauli propagation. The script \texttt{circuit\_schedule\_bound.py} verifies every eligible constructive witness on those schedules, on two serial schedules and on 128 additional interleavings. It also checks all 120 local orderings of a five-interaction example. These finite checks accompany the direct proofs; they establish no optimum over all circuits.
\item \textbf{Build.} \texttt{lake build}, run in the development directory of the
  archived snapshot named in the paper's data-availability statement,
  fetches every dependency from its pinned revision, so no machine-local path is
  configured. One of those dependencies is the QECCertificates package, which ships the
  modules named throughout this supplement and is pinned like the rest. On a machine that has not built mathlib before, \texttt{lake exe cache get}
  is run first. The build ends with the audit lines summarized in
  Sec.~\ref{sec:sm-audit} of this supplement.
\item \textbf{Audit.} The classification is read off those lines and then checked
  against the build log. A declaration whose list is not a subset of the three standard
  axioms, or one that mentions a reflection axiom, fails the counting gate, which
  compares the transcribed numbers with what the build printed.
\item \textbf{Numbers.} Every instance parameter and all three families are recomputed by the
  kernel during the build: the instance parameters by a \texttt{by decide} evaluation
  over finite $\Ztwo$ data, the three families by general theorems; every point of the data
  figure is a kernel assertion, and the second implementation below recomputes all of
  them independently. The instance table covers seven code families and sixteen parameter sets. Each parameter set has its check matrices written out, and both its dimension and its exact distance are settled by \texttt{by decide}.
\item \textbf{Enumerations.} Several computations are Julia scripts in that snapshot, run from its code directory as \texttt{julia --project=<code-directory> <script>}. They are the enumerations behind the data figure; the classification of the expansion condition over every measured logical of support at most six and every connected auxiliary graph on its support, on the two smallest members of the toric family; and the condition-cost census. Each carries assertions on the values this paper quotes from
  it, and the figure's script asserts every value it shares with a kernel-checked
  distance. The classification covers $107{,}428$ pairs on the $[[8,2,2]]$ code and $1{,}388{,}136$ on the $[[18,2,3]]$ one; the support-size cap excludes no logical operator on the first code, where none is heavier than six, and restricts the search to supports of size at most six on the second. Each
  deformed check is completed by a minimum-weight edge set satisfying the source's parity condition.
  Minimum weight is a choice of this numerical implementation, not an additional requirement of the source's definition.
  On both codes every instance meeting the condition keeps the code's distance. The $X$-side reading is recorded in the same snapshot, and no instance on either code carries an $X$-side logical below the code distance.
\item \textbf{A control model.} The data files also carry a further fault model, over the
  same detector set with the criterion requiring a strict majority of wrong readouts replaced by a flip of the
  last round's readout. Its distances at $T=1,2,3,4$ are $2,2,2,2$, where the strict-majority model gives $2,3,3,3$: a final-round data fault and its canceling check-result fault suffice at every round count. This numerical control is not among the twelve kernel-certified
  points drawn in the figure. Its regression tests distinguish a last-round flip
  from flipping every round and explicitly check the two-round ties.
\item \textbf{Cross-checks.} The independent probes used as further cross-checks are
  two Julia scripts for the two foliation
  distances, over $202$ and $3564$ small instances respectively, with no counterexample
  to the closed form each tests. The dual one was run beside two competing closed forms,
  $(\ell+1)\,d_Z$ and $\ell\,d_Z$, with $\ell$ the length of the repetition factor, and
  refutes each of them, on $2286$ and $1524$ instances. The
  census carries its own checker beside it. A third probe, in the snapshot's own tools directory, tests the K\"unneth identity of the foliated construction for general factor differentials $R$ over $136$ random pairs of chain complexes, with no counterexample to the formula, and uses the zero-differential case $R=0$ as a control.
  The remaining probes in that directory are cross-checks, not sources of quoted numbers.
\item \textbf{The gauging step.} The step itself is carried inside the kernel. The development builds the Gauss laws, pairing edges and cycle checks from the $K_4$ data and checks them entrywise against the matrices of the modules listed in Table~\ref{tab:sm-map}. It states two facts as theorems: the measured operator's promotion to a check, \code{bb24L\_mem\_rowSpace}, and the dimension drop, \code{bb18\_gauged\_drop}. The numerical prototype is the independent second path of the step.
\end{enumerate}

\end{document}